\pdfoutput=1
\documentclass[10pt]{jfm}
\usepackage{graphicx}
\usepackage{multirow}

\newcommand{\enma}[1]   {\ensuremath{#1}}

\newcommand{\beq}{\begin{equation}}
\newcommand{\eeq}{\end{equation}}
\newcommand{\bseq}{\begin{subequations}}
\newcommand{\eseq}{\end{subequations}}
\newcommand{\beqn}{\begin{eqnarray}}
\newcommand{\eeqn}{\end{eqnarray}}
\newcommand{\ba}{\begin{array}}
\newcommand{\ea}{\end{array}}
\newcommand{\bct}{\begin{center}}
\newcommand{\ect}{\end{center}}
\newcommand{\btmz}{\begin{itemize}}
\newcommand{\etmz}{\end{itemize}}
\newcommand{\benum}{\begin{enumerate}}
\newcommand{\eenum}{\end{enumerate}}

\newcommand{\diag}      {\enma{\mathrm{diag}}}

\newcommand{\trace}     {\enma{\mathrm{trace}}}

\newcommand{\col}       {\enma{\mathrm{col}}}

\newcommand{\bv}{{\bf v}}

\newcommand{\matbegin}{
        \left[
}
\newcommand{\matend}{
        \right]
}

\newcommand{\tbt}[4]{
  \matbegin \begin{array}{cc}
       #1 & #2 \\ #3 & #4
       \end{array} \matend }
\newcommand{\thbt}[6]{
  \matbegin \begin{array}{cc}
       #1 & #2 \\ #3 & #4 \\ #5 & #6
       \end{array} \matend }
\newcommand{\tbth}[6]{
  \matbegin \begin{array}{ccc}
       #1 & #2 & #3\\ #4 & #5 & #6
       \end{array} \matend }

\newcommand{\ca}{{\cal A}}

\newcommand{\be}{\begin{equation}}
\newcommand{\ee}{\end{equation}}

\newcommand{\cplxs}{ C\kern -.35em \rule{0.03 em}{.7 ex}~   }

\def\complex{\hbox{C\kern -.45em \rule{0.03 em}{1.5 ex}}~}

\newcommand{\bi}{\begin{itemize}}
\newcommand{\ei}{\end{itemize}}

\usepackage{amsfonts,amssymb,latexsym,color,amsmath,pifont,epsfig,mathdots,mathtools,natbib}
\usepackage{setspace,subfigure}

\newcommand{\cx}{{\cal X}}
\newcommand{\cb}{{\cal B}}
\newcommand{\csq}{{\cal S}}

\newcommand{\bbZ}{\mathbb{Z}}

\newcommand{\bbR}{\mathbb{R}}

\newcommand{\inprod}[2]{\left< #1 , #2 \right>}

\newcommand{\non}{\nonumber}
\newcommand{\ds}{\displaystyle}

\newcommand{\mrd}{\mathrm{d}}
\newcommand{\mre}{\mathrm{e}}
\newcommand{\mri}{\mathrm{i}}

\newcommand{\bd}{{\bf d}}

\newcommand{\bu}{{\bf u}}
\newcommand{\bF}{{\bf F}}

\newcommand{\bpsi}{\mbox{\boldmath$\psi$}}

\newcommand{\bnabla}{\mbox{\boldmath$\nabla$}}

\newcommand{\mrP}{\mathrm{P}}

\newcommand{\py}{\partial_y}

\newcommand{\pt}{\partial_t}
\newcommand{\pyy}{\partial_{yy}}

\newcommand{\toep}{\mbox{toep}}

\newcommand{\bmrE}{\bar{\mathrm{E}}}
\newcommand{\tmrE}{\tilde{\mathrm{E}}}

\title{Controlling the onset of turbulence by streamwise traveling waves. \\
Part~1: Receptivity analysis}

\shorttitle{Controlling the onset of turbulence by traveling waves: receptivity analysis}

\author{Rashad Moarref \and Mihailo R.\ Jovanovi\'c}

\affiliation{Department of Electrical and Computer Engineering,
University of Minnesota, Minneapolis, MN 55455, USA}

\shortauthor{R.\ Moarref \& M.\ R.\ Jovanovi\'c}

\begin{document}

    \maketitle

    \begin{abstract}
We examine the efficacy of streamwise traveling waves generated by a zero-net-mass-flux surface blowing and suction for controlling the onset of turbulence in a channel flow. For small amplitude actuation, we utilize weakly nonlinear analysis to determine base flow modifications and to assess the resulting net power balance. Receptivity analysis of the velocity fluctuations around this base flow is then employed to design the traveling waves. Our simulation-free approach reveals that, relative to the flow with no control, the downstream traveling waves with properly designed speed and frequency can significantly reduce receptivity which makes them well-suited for controlling the onset of turbulence. In contrast, the velocity fluctuations around the upstream traveling waves exhibit larger receptivity to disturbances. Our theoretical predictions, obtained by perturbation analysis (in the wave amplitude) of the linearized Navier-Stokes equations with spatially periodic coefficients, are verified using full-scale simulations of the nonlinear flow dynamics in companion paper,~\citet*{liemoajov10}.
    \end{abstract}

\section{Introduction}
    \label{sec.intro}

The problem of turbulence suppression in a channel flow using feedback control with wall-mounted arrays of sensors and actuators has recently received a significant attention. This problem is viewed as a benchmark for turbulence suppression in a variety of geometries, including boundary layers. Also, there has been mounting evidence that the linearized Navier-Stokes (NS) equations represent a good control-oriented model for the dynamics of transition. Recent research suggests that, in wall-bounded shear flows, one must account for modeling imperfections in the linearized NS equations since they are exceedingly sensitive to external excitations and unmodelled dynamics~\citep[see, for example,][]{tretrereddri93,farioa93,jovbamJFM05,sch07}. This has motivated several research groups to use the linearized NS equations for model-based design of estimators and controllers in a channel flow~\citep{bewliu98,leecorkimspe01,kim03,hogbewhen03,hogbewhen03b,hoechebewhen05,chehoebewhen06,kimbew07,vazkrs07,vazkrs07-book,cockrs09}. These results suggest that the proper turbulence suppression design paradigm is that of disturbance attenuation or robust stabilization rather than modal stabilization.

An alternative approach to feedback flow control relies on the understanding of the basic flow physics and the open-loop implementation of controls (i.e., without measurement of the relevant flow quantities and disturbances). Examples of sensorless strategies include: wall geometry deformation such as riblets, transverse wall oscillations, and control of conductive fluids using the Lorentz force. Although several numerical and experimental studies show that properly designed sensorless strategies may yield significant drag reduction, an obstacle to fully utilizing these physics-based approaches is the absence of a theoretical framework for their design and optimization.

An enormous potential of sensorless strategies was exemplified by~\cite{minsunspekim06}, where direct numerical simulations (DNS) were used to show that a surface blowing and suction in the form of an upstream traveling wave (UTW) results in a sustained sub-laminar drag in a fully developed turbulent channel flow. The underlying mechanism for obtaining drag smaller than in a laminar flow is the generation of the wall region Reynolds shear stresses of the opposite signs compared to what is expected based on the mean shear. By assuming that a wall actuation {\em only\/} influences the velocity fluctuations,~\cite{minsunspekim06} determined an explicit solution to the two dimensional NS equations linearized around parabolic profile; they further used an expression for skin-friction drag in fully developed channel flows~\citep*{fukiwakas02,bewaam04}, and showed that the drag is increased with the downstream traveling waves (DTWs) and decreased with the upstream traveling waves.

A comparison of laminar and turbulent channel flows with and without control was presented by~\cite*{marjosmah07}, where a criterion for achieving sub-laminar drag was derived. This study considered effectiveness of streamwise traveling waves at high Reynolds numbers and discussed why such controls can achieve sub-laminar drag. Another recent study,~\citet{hoefuk09}, emphasized that the UTWs introduce a larger flux compared to the uncontrolled flow which motivated the authors to characterize the observed mechanism as a {\em pumping\/} rather than as a drag reduction. It was shown that, even with no driving pressure gradient, blowing and suction along the walls induces pumping action in a direction opposite to that of the wave propagation. By considering flows in the absence of velocity fluctuations~\cite{hoefuk09} showed that it costs more to drive a fixed flux with wall-transpiration type of actuation than with standard pressure gradient type of actuation. A fundamental limitation on the balance of power in a channel flow was recently examined by~\cite{bew09}; this study showed that any transpiration-based control strategy that results in a sub-laminar drag necessarily has negative net efficiency compared to the laminar flow with no control. Furthermore,~\cite*{fuksugkas09} showed that a lower bound on the net driving power in a duct flow with arbitrary constant streamline curvature is determined by the power required to drive the Stokes flow. It was thus concluded that the flow has to be relaminarized in order to be driven with the smallest net power. However, since the difference between the turbulent and laminar drag coefficients grows quadratically with the Reynolds number,~\cite{marjosmah07} argued that relaminarization may not be possible in strongly inertial flows. An alternative approach is to design a controller that reduces skin-friction drag in turbulent flows; provided that the control power is less than the saved power, a positive net efficiency can still be achieved.

In this paper, we show that a positive net efficiency can be achieved in a channel flow subject to streamwise traveling waves if the controlled flow stays laminar while the uncontrolled flow becomes turbulent. Starting from this observation, we develop a framework for design of the traveling waves that are capable of (i) improving dynamical properties of the flow; and (ii) achieving positive net efficiency. We quantify receptivity of the NS equations linearized around UTWs and DTWs to stochastic disturbances by computing the {\em ensemble average energy density\/} of the statistical steady-state. Motivated by our desire to have low cost of control we confine our study to small amplitude blowing and suction along the walls. This also facilitates derivation of an explicit formula for energy amplification (in flows with control) using perturbation analysis techniques. Our simulation-free design reveals that the UTWs are poor candidates for preventing transition; conversely, we demonstrate that {\em properly designed\/} DTWs are capable of substantially reducing receptivity of three dimensional fluctuations (including streamwise streaks and Tollmien-Schlichting (TS) waves). This indicates that the DTWs can be used as an effective means for controlling the onset of turbulence. Moreover, we show the existence of DTWs that result in a positive net efficiency compared to the uncontrolled flow that becomes turbulent. Our theoretical predictions are verified in Part 2 of this paper~\citep{liemoajov10} using DNS of the NS equations. Thus, our work (i) demonstrates that the theory developed for the linearized equations with uncertainty has considerable ability to capture full-scale phenomena; and (ii) exhibits the predictive power of the proposed perturbation-analysis-based method for designing traveling waves.



This paper represents an outgrowth of the study performed during the 2006 Center for Turbulence Research Summer Program~\citep*{jovmoayouCTR06}. While~\citet{jovmoayouCTR06} only focused on receptivity of UTWs with large wavelength, our current study does a comprehensive analysis of the influence of both UTWs and DTWs on the fluctuations' kinetic energy and the overall efficiency. We also note that linear stability and transient growth of traveling waves were recently examined by~\citet*{leeminkim08}. For selected values of parameters, it was shown that the UTWs destabilize the laminar flow for control amplitudes as small as $1.5 \, \%$ of the centerline velocity; on the other hand, the DTWs with phase speeds larger than the centerline velocity remain stable even for large wave amplitudes. Moreover, the UTWs (DTWs) exhibit larger (smaller) transient growth relative to the uncontrolled flow. Our study confirms all of these observations; it also extends them at several different levels. First, we pay close attention to a net efficiency by computing the net power gained (positive efficiency) or lost (negative efficiency) in the presence of wall-actuation. Second, we conduct much more detailed study of the influence of traveling waves on velocity fluctuations; this is done by a thorough analysis of the influence of the wave speed, frequency, and amplitude on receptivity of full three dimensional fluctuations. Third, we confirm all of our theoretical predictions in Part~2 of this study, and highlight remaining research challenges.

Our presentation is organized as follows: in~\S~\ref{sec.problem}, we formulate the governing equations in the presence of traveling wave wall-actuation. The influence of control on the nominal bulk flux and the nominal net efficiency is also discussed in this section. A frequency representation of the NS equations linearized around base velocity induced by traveling waves is presented in~\S~\ref{sec.vel-fluc}. We further discuss a notion of the ensemble average energy density of the statistical steady-state and describe an efficient method for determining this quantity in flows subject to small amplitude traveling waves. In~\S~\ref{sec.energy-amp}, we employ perturbation analysis to derive an explicit formula for energy amplification. This formula is used to identify the values of wave frequency and speed that reduce receptivity of the linearized NS equations; we show that the essential trends are captured by perturbation analysis up to a second order in traveling wave amplitude. We also discuss influence of amplitude on energy of velocity fluctuations and reveal physical mechanisms for energy amplification. A brief summary of the main results along with an overview of remaining research challenges is provided in~\S~\ref{sec.concl}.

\section{Steady-state analysis}
    \label{sec.problem}


\subsection{Governing equations}
    \label{sec.gov-eq}

Consider a channel flow governed by the non-dimensional incompressible NS equations
    \beq
    \bu_{\bar{t}}
    \; = \,
    - \left( \bu \cdot \bnabla \right) \bu
    \; - \;
    \bnabla \mrP
    \; + \;
    ({1}/{R_c}) \Delta \bu
    \; + \;
    \bF,
    ~~
    0
    \; = \;
    \bnabla {\bf \cdot} \bu,
    \label{eq.NScts}
    \eeq
with the Reynolds number defined in terms of the centerline velocity of the parabolic laminar profile $U_c$ and channel half-height $\delta$, $R_c \, = \, U_c \, \delta / \nu$. The kinematic viscosity is denoted by $\nu$, the velocity vector is given by $\bu$, $\mrP$ is the pressure, $\bF$ is the body force, $\bnabla$ is the gradient, and $\Delta = \bnabla \cdot \bnabla$ is the Laplacian. The spatial coordinates and time are represented by $(\bar{x}, \bar{y}, \bar{z})$ and $\bar{t}$, respectively.

    \begin{figure}
    \begin{center}
    \includegraphics[width=0.6\textwidth]
    {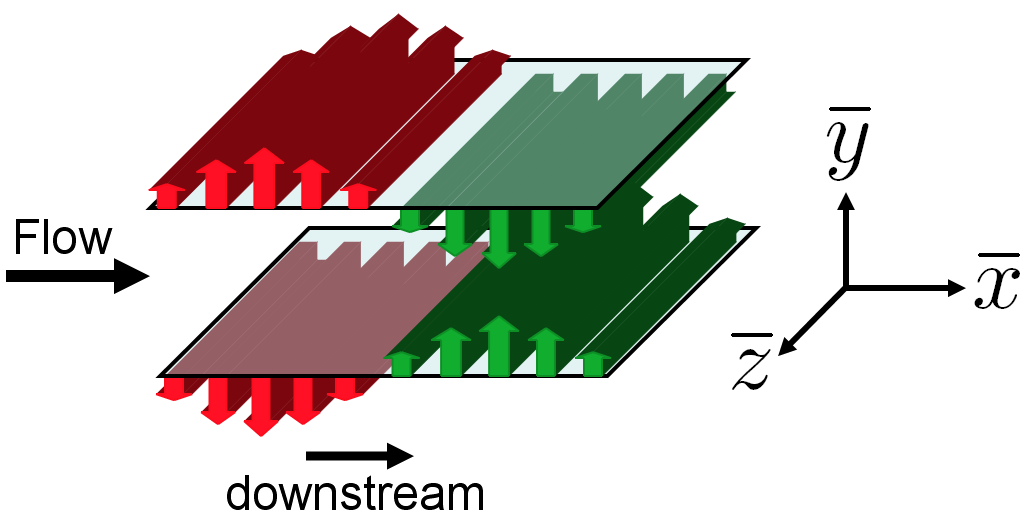}
    \end{center}
    \caption{
    A pressure driven channel flow with blowing and suction along the walls.
    }
    \label{fig.channel}
    \end{figure}

In addition to a constant pressure gradient, $P_{\bar{x}}$, the flow is exposed to a zero-net-mass-flux surface blowing and suction in the form of a streamwise traveling wave (see figure~\ref{fig.channel} for illustration). In the absence of the nominal body force, $\bar{\bF} \equiv 0$, base velocity ${\bf u}_b = ({U}, \, {V}, \, {W})$ represents the steady-state solution to~(\ref{eq.NScts}) subject to
    \beq
    \ba{c}
    V(\bar{y} \, = \, \pm 1)
    \; = \;
    \mp 2 \, \alpha \,
    \cos \, (\omega_x (\bar{x} \, - \, c \, \bar{t})),
    ~~
    \bar{\bF}
    \,
    \equiv
    \,
    0,
    \\[0.1cm]
    U(\pm 1)
    \; = \;
    V_{\bar{y}}(\pm 1)
    \; = \;
    W(\pm 1)
    \; = \;
    0,
    ~~
    P_{\bar{x}}
    \; = \, -\, 2/R_c,
    \ea
    \label{eq.BCorig}
    \eeq
where $\omega_x$, $c$, and $\alpha$, respectively, identify frequency, speed, and amplitude of the traveling wave. Positive values of $c$ define a DTW, whereas negative values of $c$ define a UTW. The time dependence in $V(\pm 1)$ can be eliminated by the Galilean transformation,
    $
    (
    x = \bar{x} - c \bar{t},
    \,
    y = \bar{y},
    \,
    z = \bar{z},
    \,
    t = \bar{t}
    ).
    $
This change of coordinates does not influence the spatial differential operators, but it transforms the time derivative to
    $
    \partial_{\bar{t}}
    =
    \partial_{t}
    -
    c \, \partial_{x},
    $
which adds an additional convective term to the NS equations
    \beq
    \bu_{t}
    \; = \;
    c \bu_x
    \; - \;
    \left( \bu \cdot \bnabla \right) \bu
    \; - \;
    \bnabla \mrP
    \; + \;
    ({1}/{R_c}) \Delta \bu
    \; + \;
    \bF,
    ~~
    0
    \; = \;
    \bnabla {\bf \cdot} \bu.
    \label{eq.NScts-trans}
    \eeq
In new coordinates, i.e.\ in the frame of reference that travels with the wave, the wall-actuation~(\ref{eq.BCorig}) induces a two dimensional base velocity,
    $
    {\bf u}_b
    =
    ({U(x,y)}, \, {V(x,y)}, \, {0}),
    $
which represents the steady-state solution to~(\ref{eq.NScts-trans}). Note that the spatially periodic wall actuation,
    $
    V(y = \pm 1)
    =
    \mp 2 \, \alpha \,
    \cos \, (\omega_x x),
    $
induces base velocity which is periodic in $x$.

The equations describing dynamics (up to a first order) of velocity fluctuations $\bv = ({u}, \, {v}, \, {w})$ around base velocity, $\bu_b$, are obtained by decomposing each field in~(\ref{eq.NScts-trans}) into the sum of base and fluctuating parts, i.e., $\{ \bu = {\bf u}_b + \bv$, $\mrP = P + p$, $\bF = 0 + \bd \}$, and by neglecting the quadratic term in $\bv$
    \beq
    \bv_{t}
    \; = \;
    c \bv_x
    \; - \;
    \left( \bu_b \cdot \bnabla \right) \bv
    \; - \;
    \left( \bv \cdot \bnabla \right) \bu_b
    \; - \;
    \bnabla p
    \; + \;
    ({1}/{R_c}) \Delta \bv
    \; + \;
    \bd,
    ~~
    0
    \; = \;
    \bnabla {\bf \cdot} \bv.
    \label{eq.linear-NS}
    \eeq
Note that the boundary conditions~(\ref{eq.BCorig}) are satisfied by base velocity and, thus, velocity fluctuations acquire homogeneous Dirichlet boundary conditions.

\subsection{Base flow}
    \label{sec.base-flow}

Let us first consider a surface blowing and suction of a small amplitude $\alpha$. In this case, a weakly nonlinear analysis can be employed to  solve~(\ref{eq.NScts-trans}) subject to~(\ref{eq.BCorig}) and determine the corrections to base parabolic profile; similar approach was previously used by~\citet{jovmoayouCTR06,hoefuk09}. Up to a second order in control amplitude $\alpha$, $U(x,y)$ and $V(x,y)$ can be represented as
    \be
    \ba{rcl}
    U(x,y)
    \!\! & = & \!\!
    U_{0}(y)
    \, + \,
    \alpha \, U_{1}(x,y)
    \, + \,
    \alpha^2 \, U_{2}(x,y)
    \, + \,
    {\cal O}(\alpha^3),
    \\[0.1cm]
    V(x,y)
    \!\! & = & \!\!
    \alpha \, V_{1}(x,y)
    \, + \,
    \alpha^2 \, V_{2}(x,y)
    \, + \,
    {\cal O}(\alpha^3),
    \ea
    \non
    \ee
where $U_{0}(y) = 1 - y^2$ denotes base velocity in Poiseuille flow and (see Appendix~\ref{app.detail-nom-velocity})
    \be
    \ba{rcl}
    U_{1}(x,y)
    & \!\! = \!\! &
    U_{1,-1}(y) \, \mre^{-\mri \omega_x x}
    \, + \,
    U_{1,1}(y) \, \mre^{\mri \omega_x x},
    \\[0.1cm]
    V_{1}(x,y)
    & \!\! = \!\! &
    V_{1,-1}(y) \, \mre^{-\mri \omega_x x}
    \, + \,
    V_{1,1}(y) \, \mre^{\mri \omega_x x},
    \\[0.1cm]
    U_{2}(x,y)
    & \!\! = \!\! &
    U_{2,0}(y)
    \, + \,
    U_{2,-2}(y) \, \mre^{-2 \mri \omega_x x}
    \, + \,
    U_{2,2}(y) \, \mre^{2 \mri \omega_x x},
    \\[0.1cm]
    V_{2}(x,y)
    & \!\! = \!\! &
    V_{2,-2}(y) \, \mre^{-2 \mri \omega_x x}
    \, + \,
    V_{2,2}(y) \, \mre^{2 \mri \omega_x x}.
    \ea
    \label{eq.base-vel-pert}
    \ee

    \begin{figure}
    \begin{center}
    \begin{tabular}{cc}
    \subfigure[]{\includegraphics[height=1.9in,width=2.5in]
    {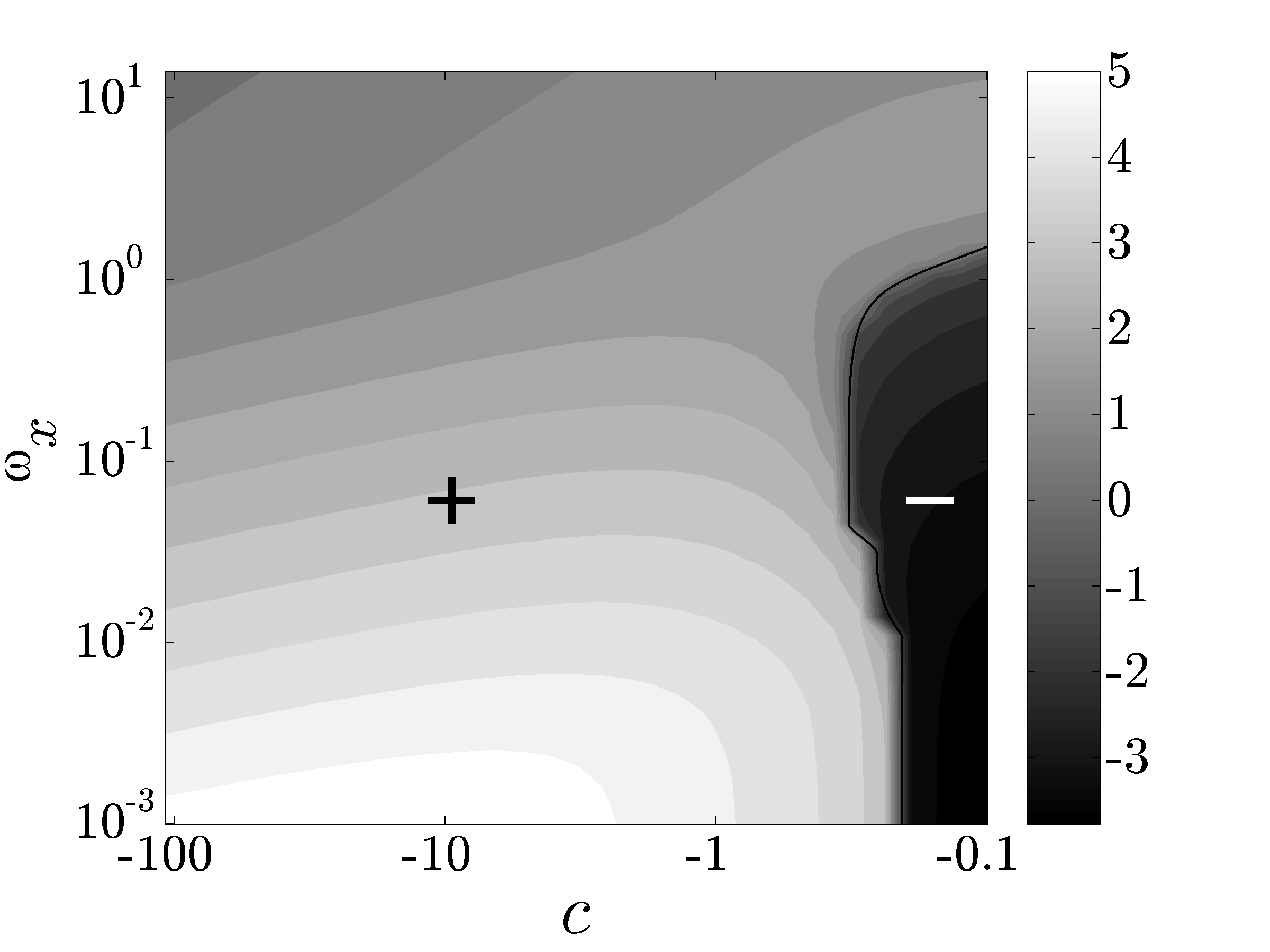}
    \label{fig.UB2-cm}}
    &
    \subfigure[]{\includegraphics[height=1.9in,width=2.5in]
    {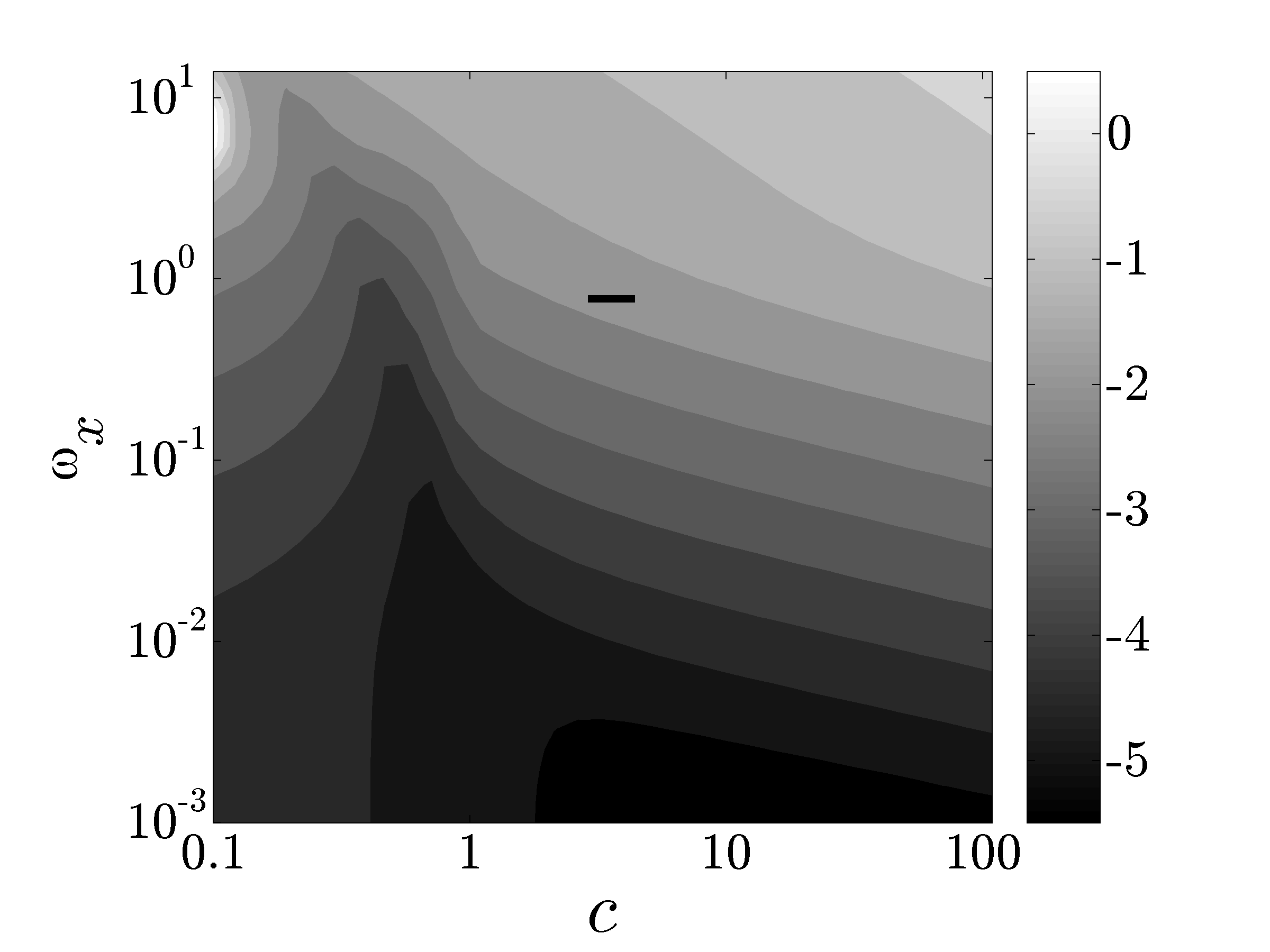}
    \label{fig.UB2-cp}}
    \end{tabular}
    \end{center}
    \caption{
    Second order correction to the nominal flux, $U_{B,2}(c, \omega_x)$, for (a) upstream waves; and (b) downstream waves in Poiseuille flow with $R_c = 2000$. Note: the level sets are obtained using a sign-preserving logarithmic scale; e.g., $5$ and $-3$ should be interpreted as $U_{B,2} = 10^5$ and $U_{B,2} = -10^3$, respectively.
    }
    \label{fig.UB2-freq}
    \end{figure}

\citet{hoefuk09} recently showed that, in the absence of driving pressure gradient, the traveling waves induce nominal bulk flux (i.e., {\em pumping\/}) in the direction opposite to the direction in which the wave travels. While the first order of correction to the base velocity is purely oscillatory, the quadratic interactions in the NS equations introduce mean flow correction $U_{2,0}(y)$ at the level of $\alpha^2$. The nominal bulk flux is determined by $U_B = (1/2)\int_{-1}^{1} \overline{U} (y) \, \mrd y $ where the overline denotes averaging over horizontal directions. In the presence of a pressure gradient, the nominal flux in flow with no control is $U_{B,0} =  (1/2)\int_{-1}^{1} U_0 (y) \, \mrd y = 2/3$, and the second order correction (in $\alpha$) to $U_B$ is given by $U_{B,2} = (1/2) \int_{-1}^{1} U_{2,0} (y) \, \mrd y$. Figure~\ref{fig.UB2-freq} shows $U_{B,2}$ as a function of wave frequency, $\omega_x$, and wave speed, $c$, in Poiseuille flow with $R_c = 2000$. Except for a narrow region in the vicinity of $c = 0$, the upstream and downstream waves increase and reduce the nominal flux, respectively. Furthermore, for a given wave speed $c$, the magnitude of the induced flux increases as the wave frequency is decreased.

    \begin{figure}
    \begin{center}
    \begin{tabular}{cc}
    \subfigure[]{\includegraphics[height=1.9in,width=2.5in]
    {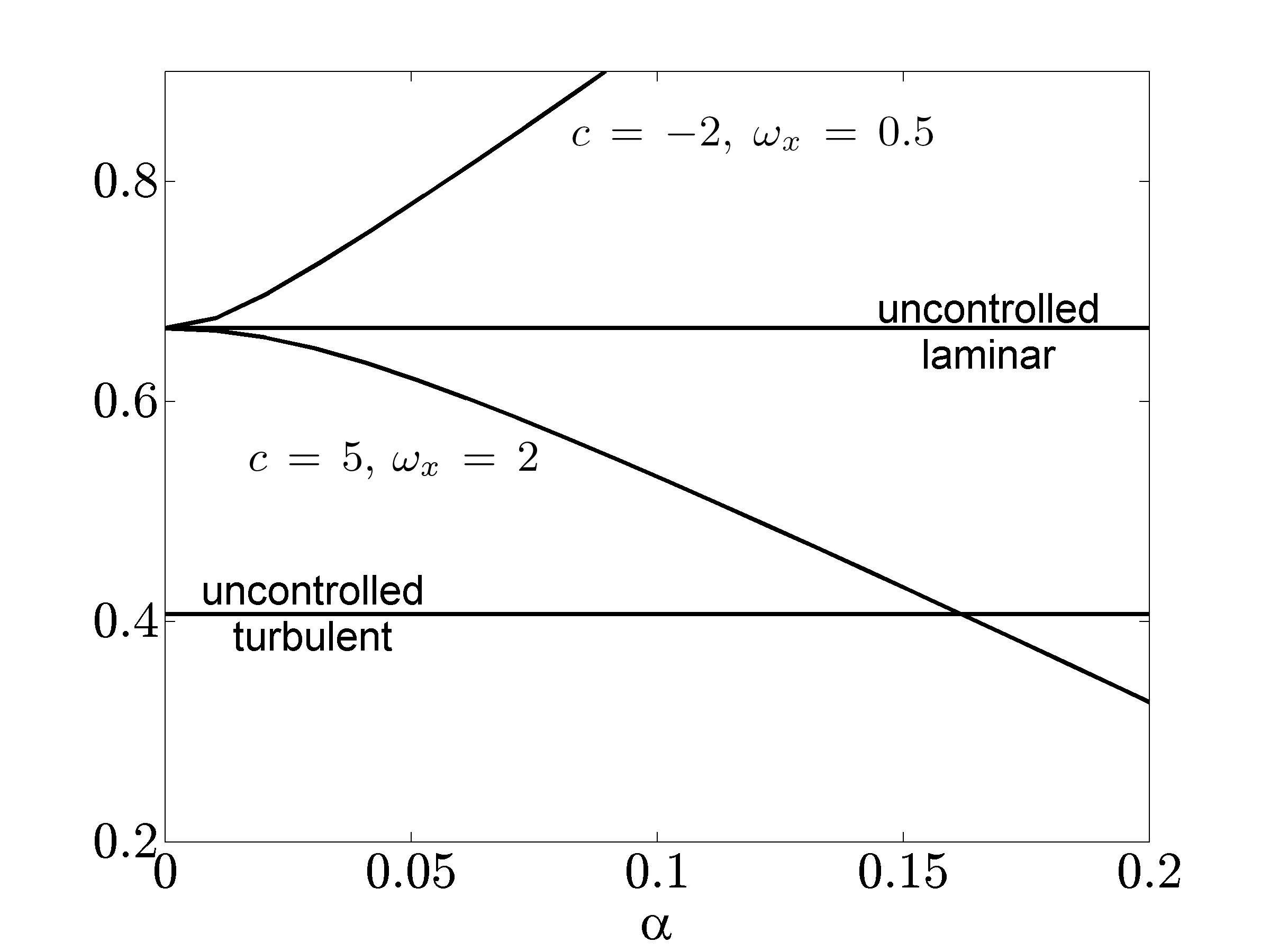}
    \label{fig.UB-nom-Newton}}
    &
    \subfigure[]{\includegraphics[height=1.9in,width=2.5in]
    {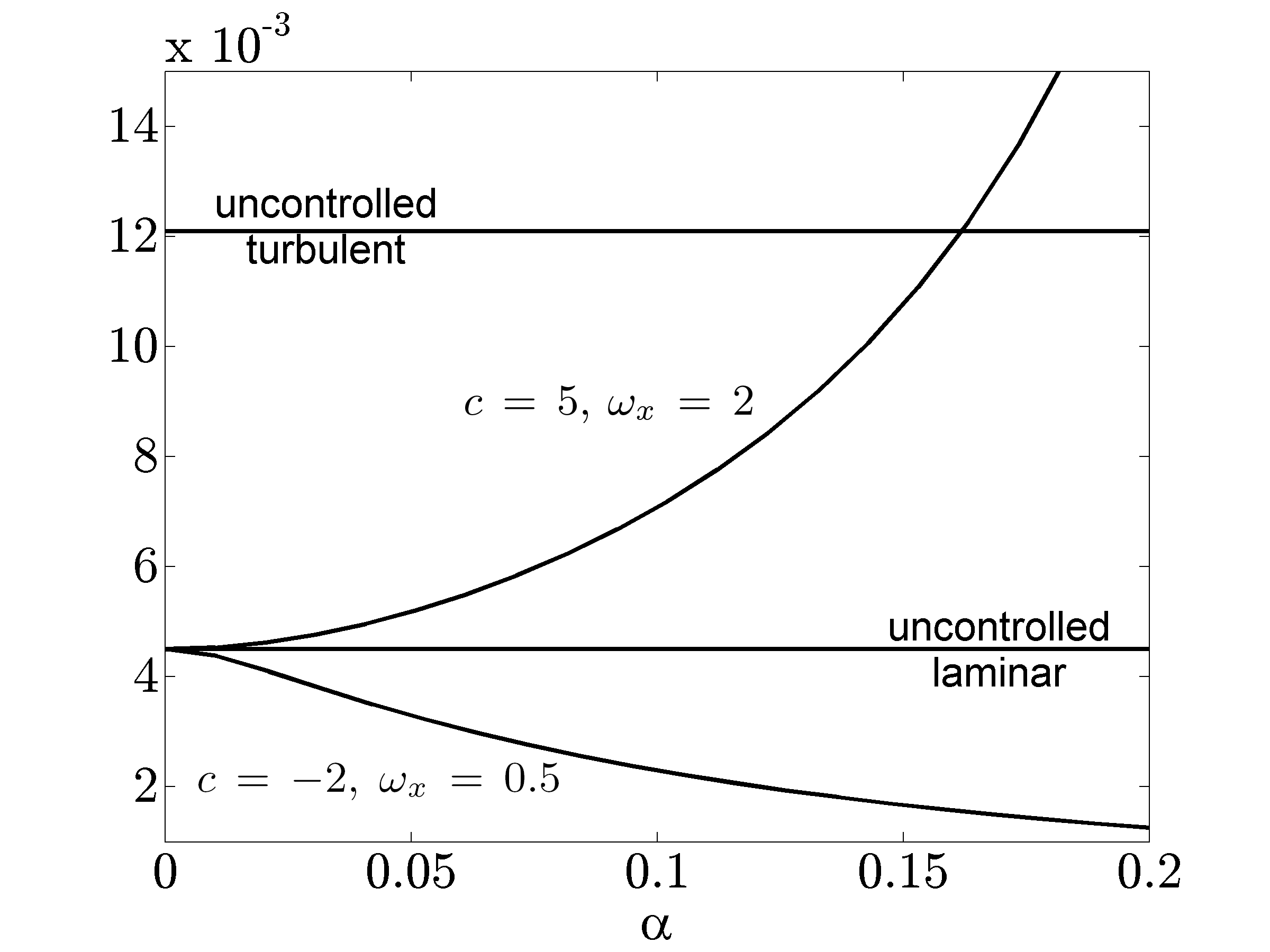}
    \label{fig.Cf-nom-Newton}}
    \end{tabular}
    \end{center}
    \caption{
    (a) The nominal flux, $U_B(\alpha)$;
    and
    (b) the nominal skin-friction drag coefficient, $C_f(\alpha)$,
    for a pair of UTWs and a pair of DTWs in Poiseuille flow with $R_c = 2000$.
    The results are obtained by solving~(\ref{eq.NScts-trans}) subject to~(\ref{eq.BCorig}), in the steady-state, using Newton's method; $U_B$ and $C_f$ of the uncontrolled laminar and turbulent flows are also shown for comparison.
    }
    \label{fig.UB-Cf-nom-Newton}
    \end{figure}

Figure~\ref{fig.UB-Cf-nom-Newton} is obtained by finding the steady-state solution of~(\ref{eq.NScts-trans}) subject to~(\ref{eq.BCorig}) using Newton's method. Originally, we have used base flow resulting from the weakly nonlinear analysis to initialize Newton iterations; robustness of our computations is confirmed using initialization with many different incompressible base flow conditions. The nominal flux and its associated nominal drag coefficient for a UTW with $c = -2$ and $\omega_x = 0.5$, and a DTW with $c = 5$ and $\omega_x = 2$ are shown in this figure. The flux and drag coefficient of both laminar and turbulent flows with no control are also given for comparison. The nominal skin-friction drag coefficient is defined as~\citep{mccomb91}
    \be
    C_f
    \; = \;
    {2 \, \overline{\tau}_w}/{U_B^2}
    \; = \;
    {-2 \, P_x}/{U_B^2},
    \non
    \ee
where $\overline{\tau}_w$ is the nondimensional average wall-shear stress. For the fixed pressure gradient, $P_x = -2/R_c$, the nominal skin-friction drag coefficient is inversely proportional to square of the nominal flux and, in uncontrolled laminar flow with $R_c = 2000$, we have $C_f = 4.5 \times 10^{-3}$. The UTWs produce larger nominal flux (and, consequently, smaller nominal drag coefficient) compared to both laminar and turbulent uncontrolled flows. On the other hand, the DTWs yield smaller nominal flux (and, consequently, larger nominal drag coefficient) compared to uncontrolled laminar flow. In situations where flow with no control becomes turbulent, however, the DTWs with amplitudes smaller than a certain threshold value may have lower nominal drag coefficient than the uncontrolled turbulent flow; e.g., for a DTW with $c = 5$ and $\omega_x = 2$ this threshold value is given by $\alpha = 0.16$ (cf.\ figure~\ref{fig.Cf-nom-Newton}).

\subsection{Nominal net efficiency}
    \label{sec.nom-eff}

For the fixed pressure gradient, the difference between the flux of the controlled and the uncontrolled flows results in production of a driving power (per unit horizontal area of the channel)
    \be
    \Pi_{prod}
    \; = \;
    -2 P_x \, (U_{B,c} \, - \, U_{B,u}),
    \non
    \ee
where $U_{B,c}$ and $U_{B,u}$ are the nominal flux of the controlled and uncontrolled flows, respectively. On the other hand, the required control power exerted at the walls (per unit horizontal area of the channel) is given by~\citep{cur03}
    \be
    \Pi_{req}
    \; = \;
    \left.
    \overline{V P}
    \right|_{y \, = \, -1}
    \, - \,
    \left.
    \overline{V P}
    \right|_{y \, = \, 1}.
    \label{eq.Preq}
    \ee
The control net efficiency is determined by the difference of the produced and required powers~\citep{quaric04}
    \be
    \Pi_{net}
    \; = \;
    \Pi_{prod}
    \, - \,
    \Pi_{req},
    \non
    \ee
where $\Pi_{net}$ signifies the net power gained (positive $\Pi_{net}$) or lost (negative $\Pi_{net}$), in the presence of wall-actuation.

%


For small control amplitudes, the produced power can be represented as
    \be
    \Pi_{prod}
    \; = \;
    \Pi_{prod,0}
    \, + \,
    \alpha^2 \, \Pi_{prod,2}
    \, + \,
    {\cal O}(\alpha^4),
    \non
    \ee
where
    \be
    \Pi_{prod,0}
    \; = \;
    -2 P_x \, (U_{B,0} \, - \, U_{B,u}),
    ~~
    \Pi_{prod,2}
    \; = \;
    -2 P_x \, U_{B,2}.
    \non
    \ee
The nominal required control power can be determined from~(\ref{eq.Preq}) by evaluating the horizontal average of the product between base pressure, $P$, and base wall-normal velocity, $V$, at the walls. Since, at the walls, the nonzero component of $V$ contains {\em only\/} first harmonic in $x$ (cf.\ (\ref{eq.base-vel-pert})), we need to determine the first harmonic (in $x$) of $P$ to compute $\Pi_{req}$. Base pressure can be obtained by solving the two dimensional Poisson equation
    \be
    P_{xx}
    \, + \,
    P_{yy}
    \, = \,
    -\left(
    U_x \, U_x
    \, + \,
    2 V_x \, U_y
    \, + \,
    V_y \, V_y
    \right),
    \label{eq.Poisson}
    \ee
where $P$ satisfies the following Neumann boundary conditions
    \be
    \left.
    P_y
    \right|_{y \, = \, \pm1}
    \, = \,
    \left.
    \left(
    \left(
    V_{xx}
    \, + \,
    V_{yy}
    \right)/R_c
    \, + \,
    c \, V_x
    \right)
    \right|_{y \, = \, \pm1}.
    \non
    \ee
These are determined by evaluating the $y$-momentum equation at the walls. For small values of $\alpha$, weakly nonlinear analysis, in conjunction with the expressions for $U$ and $V$ given in \S~\ref{sec.base-flow}, can be employed to solve~(\ref{eq.Poisson}) for base pressure
    \be
    \ba{rcl}
    P(x,y)
    \!\! & = & \!\!
    \alpha \, P_{1}(x,y)
    \, + \,
    {\cal O}(\alpha^2),
    \\[0.1cm]
    P_{1}(x,y)
    \!\! & = & \!\!
    P_{1,-1}(y) \, \mre^{-\mri \omega_x x}
    \, + \,
    P_{1,1}(y) \, \mre^{\mri \omega_x x},
    \ea
    \non
    \ee
where $P_{1,-1}$ and $P_{1,1}$ are determined from
    \be
    \ba{rcl}
    P_{1,\pm1}^{\prime\prime}(y)
    \, - \,
    \omega_x^2 \, P_{1,\pm1}(y)
    \!\! & = & \!\!
    \mp2 \, \mri \, \omega_x V_{1,\pm1}(y) \, U'_0(y),
    \\[0.1cm]
    P_{1,-1}'(\pm1)
    \!\! & = & \!\!
    (V_{1,-1}^{\prime\prime}(\pm1)
    \, - \,
    \omega_x^2 \, V_{1,-1}(\pm1))/R_c
    \, + \,
    c \, \mri \, \omega_x V_{1,-1}(\pm1),
    \\[0.1cm]
    P_{1,1}'(\pm1)
    \!\! & = & \!\!
    (V_{1,1}^{\prime\prime}(\pm1)
    \, - \,
    \omega_x^2 \, V_{1,1}(\pm1))/R_c
    \, - \,
    c \, \mri \, \omega_x V_{1,1}(\pm1).
    \ea
    \non
    \ee
Here, the prime denotes the partial derivative with respect to $y$, and the required power can be represented as
    \be
    \ba{rcl}
    \Pi_{req}
    \!\! & = & \!\!
    \alpha^2 \, \Pi_{req,2}
    \, + \,
    {\cal O}(\alpha^4),
    \\[0.1cm]
    \Pi_{req,2}
    \!\! & = & \!\!
    \left.
    \left(
    P_{1,-1} V_{1,1} + P_{1,1} V_{1,-1}
    \right)
    \right|_{y \, = \, -1}
    \, - \,
    \left.
    \left(
    P_{1,-1} V_{1,1} + P_{1,1} V_{1,-1}
    \right)
    \right|_{y \, = \, 1}.
    \ea
    \non
    \ee

Since the second order correction to the nominal produced power, $\Pi_{prod,2}$, is directly proportional to $U_{B,2}$, $\Pi_{prod,2}$ is positive for UTWs and negative for DTWs. It turns out that smaller choices of $\omega_x$ result in larger produced (for UTWs) or lost (for DTWs) power. One of the main points of this paper, however, is to show that it may be misleading to rely on the produced power as the {\em only\/} criterion for selection of control parameters; in what follows, we demonstrate that the required control power as well as the dynamics of velocity fluctuations need to be taken into account when designing the traveling waves.

    \begin{figure}
    \begin{center}
    \begin{tabular}{cc}
    \subfigure[]
    {
    \includegraphics[height=1.92in,width=2.5in]
    {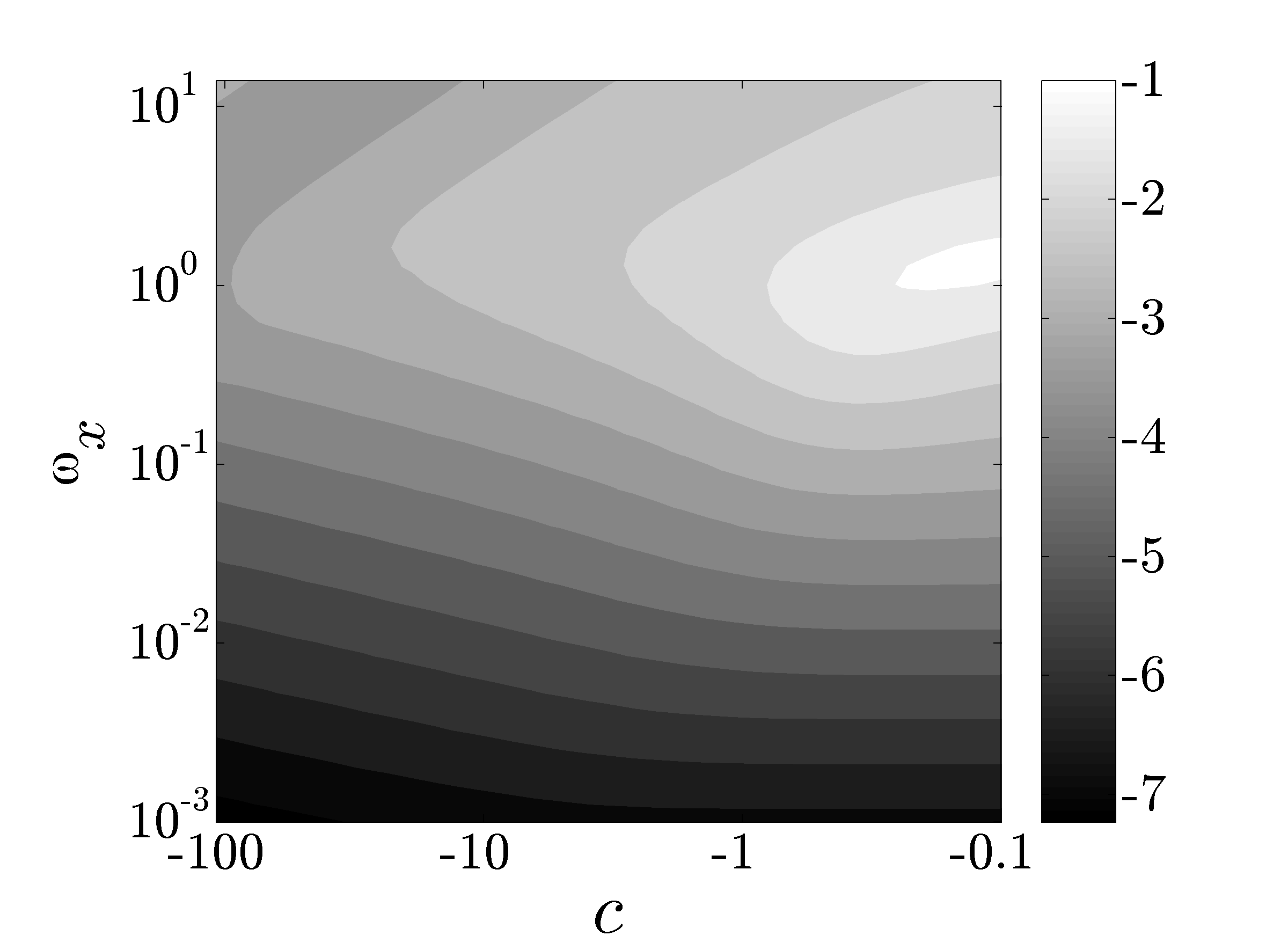}
    \label{fig.Pnet2-cm}}
    &
    \subfigure[]
    {
    \includegraphics[height=1.9in,width=2.5in]
    {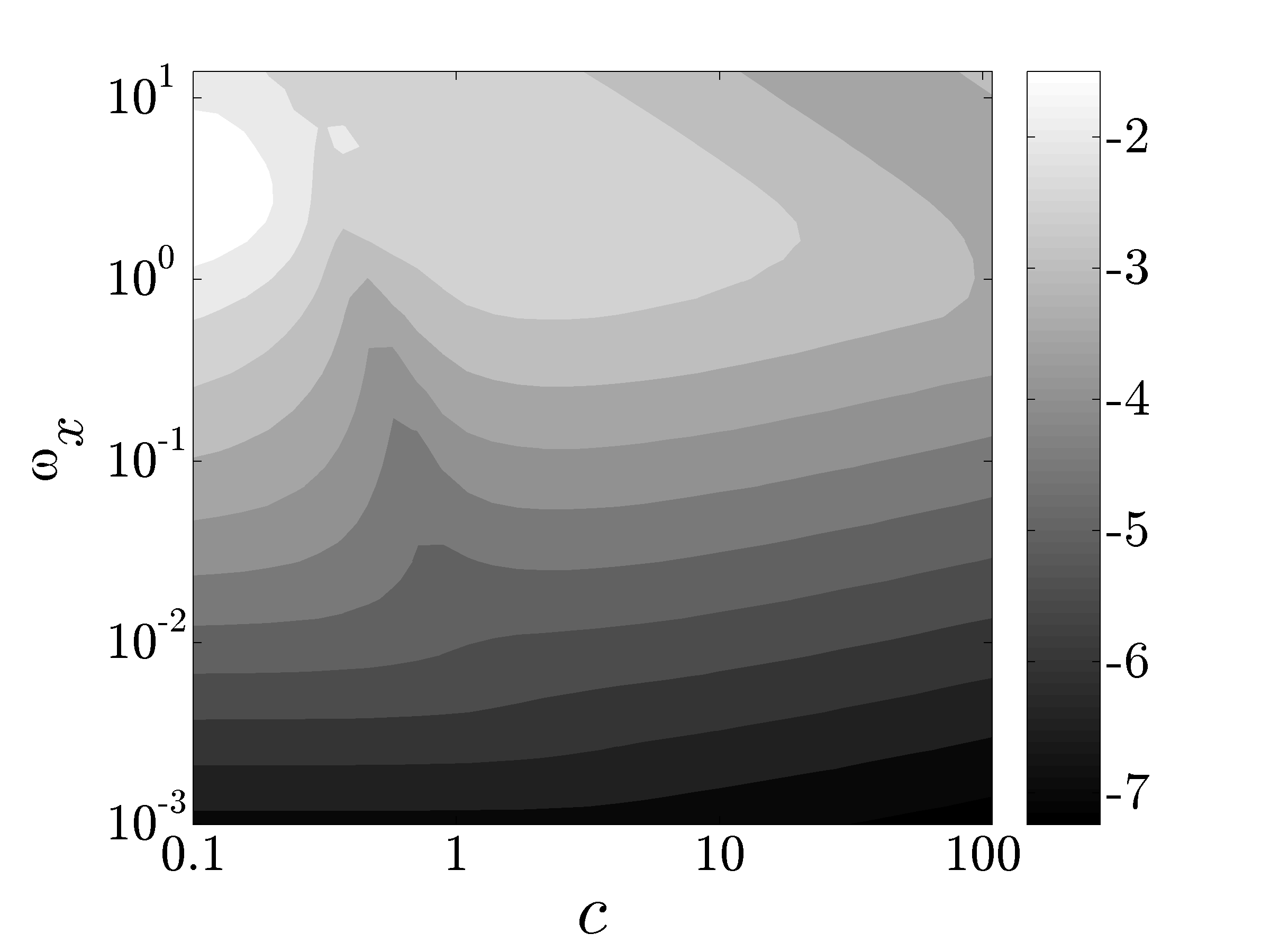}
    \label{fig.Pnet2-cp}}
    \end{tabular}
    \end{center}
    \caption{
    Second order correction to the nominal net efficiency, $\pi_2(c, \omega_x)$, for (a) upstream waves; and (b) downstream waves in Poiseuille flow with $R_c = 2000$. Note: the level sets are obtained using a sign-preserving logarithmic scale; e.g., $-4$ should be interpreted as $\pi_2 = -10^4$.
    }
    \label{fig.h2-nom}
    \end{figure}

    \begin{figure}
    \begin{center}
    \begin{tabular}{cc}
    \subfigure[]
    {
    \includegraphics[height=1.9in,width=2.5in]
    {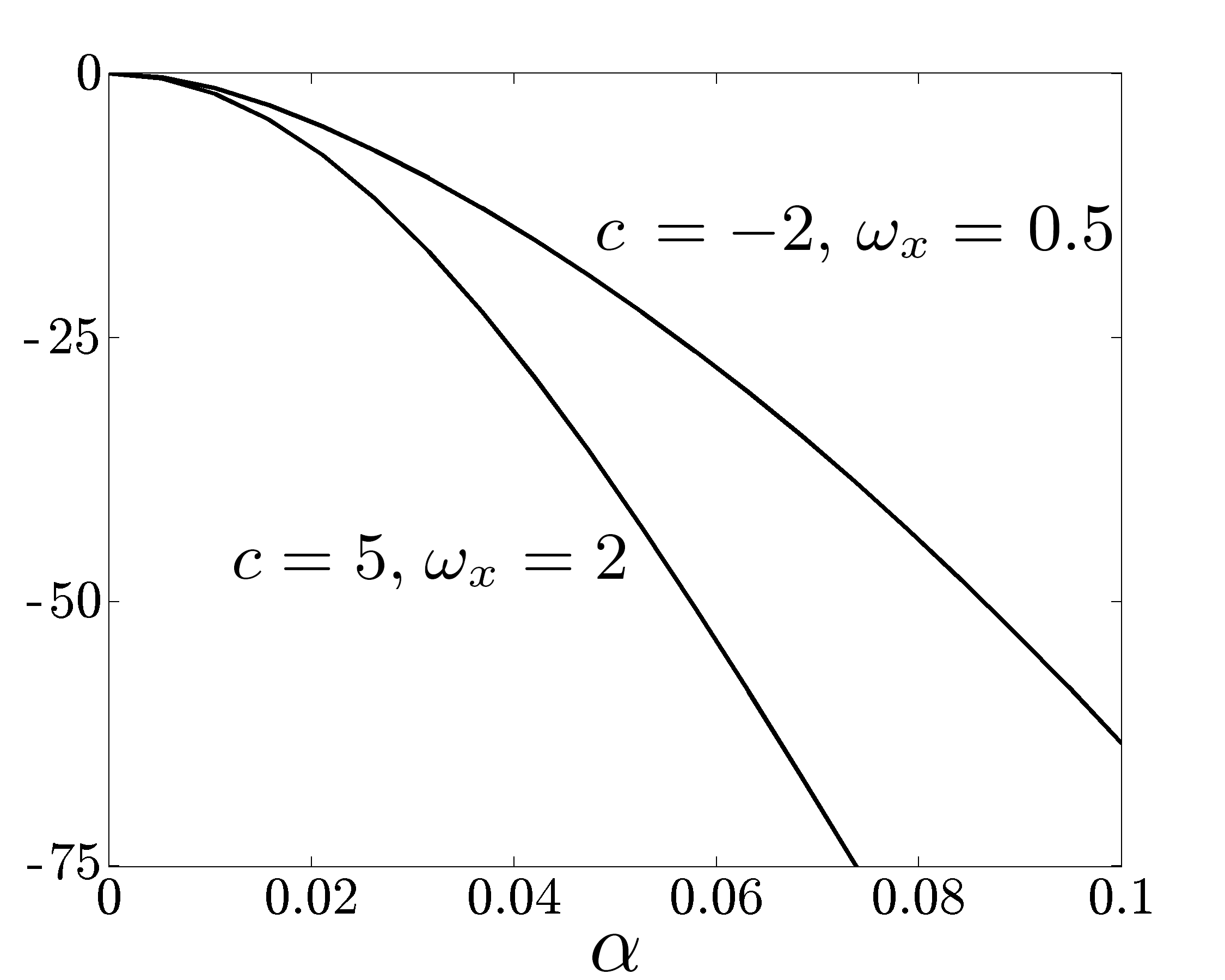}
    \label{fig.alpha-Pnet-lam-nom}}
    &
    \subfigure[]
    {
    \includegraphics[height=1.9in,width=2.5in]
    {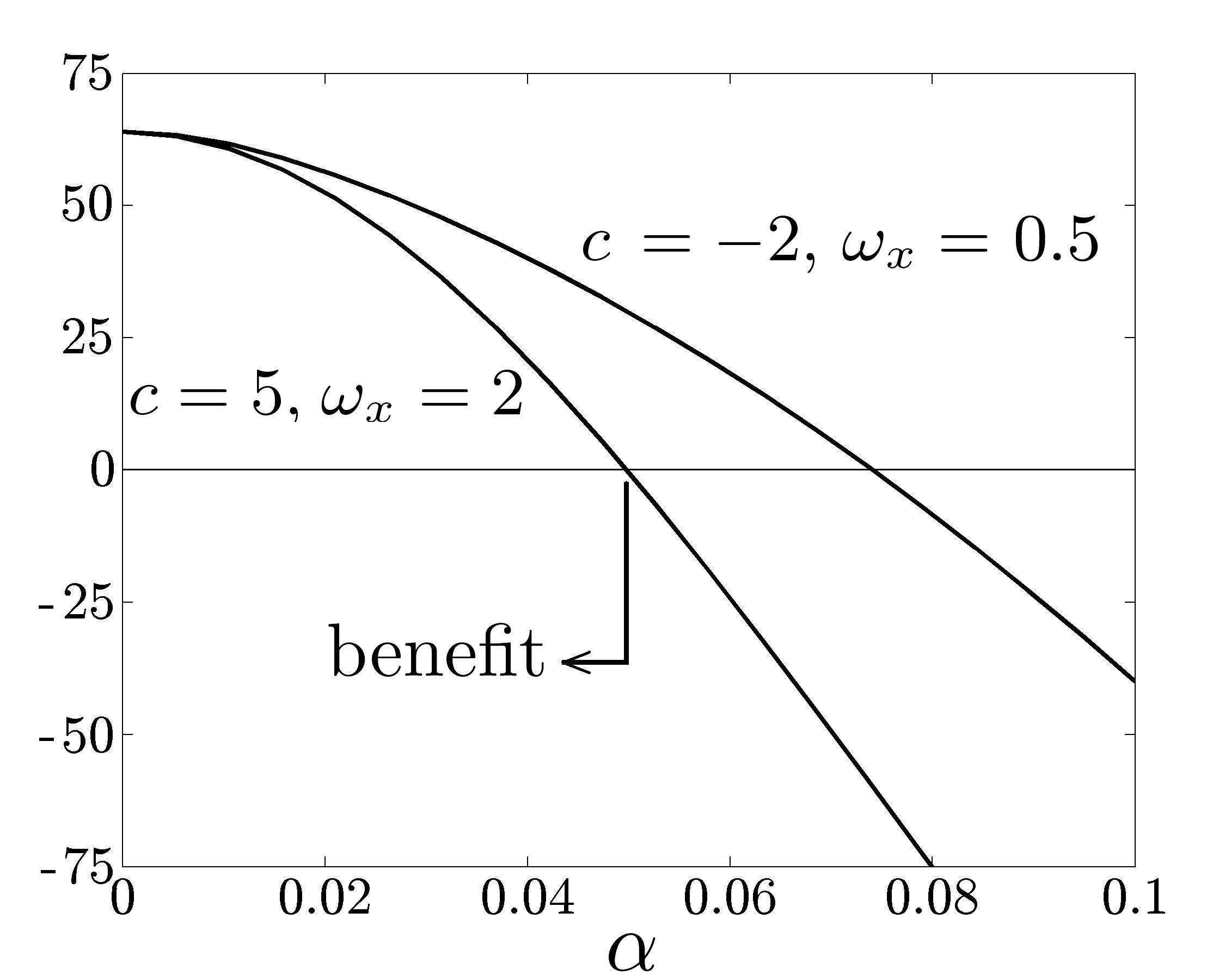}
    \label{fig.alpha-Pnet-turb-nom}}
    \end{tabular}
    \end{center}
    \caption{
    The steady-state net efficiency, $\% \Pi_{net}$, of \emph{laminar} controlled flows as a function of control amplitude $\alpha$ for a UTW with $( c = -2,$ $\omega_x = 0.5 )$ and a DTW with $( c = 5,$ $\omega_x = 2 )$ at $R_c = 2000$.
    The results are obtained by {\em assuming\/} that the uncontrolled flow
    (a) remains laminar; and
    (b) becomes turbulent.
    }
    \label{fig.alpha-Pnet-turb-lam-nom}
    \end{figure}


\subsection{Nominal efficiency of laminar controlled flows}
    \label{sec.nom-eff-laminar}

We next examine the nominal efficiency of {\em laminar\/} controlled flows.
Since we are interested in expressing the nominal efficiency relative to the power required to drive flow with no control, we provide comparison with both laminar and turbulent uncontrolled flows. The net efficiency in fraction of the power required to drive the uncontrolled laminar flow is determined by
    \be
    \% \Pi_{net}
    \; = \;
    \Pi_{net}/\Pi_{0}
    \, = \,
    -\alpha^2 \, |\pi_2 (R_c; c, \omega_x)|
    \, + \,
    {\cal O}(\alpha^4),
    \label{eq.net-lam}
    \ee
where $\Pi_{0} =  -2 P_x \, U_{B,0}$ and $\pi_2 = (\Pi_{prod,2} - \Pi_{req,2})/\Pi_{0}$. It can be shown that the second order correction to $\% \Pi_{net}$, $\pi_2$, is negative for all choices of $c$ and $\omega_x$ (see figure~\ref{fig.h2-nom}). This is because the required power for maintaining the traveling wave grows faster than the produced power as $\alpha$ is increased. In addition, figure~\ref{fig.h2-nom} shows that $|\pi_2|$ is minimized for small wave speeds and for $\omega_x \in (1, \, 4)$. Formula~(\ref{eq.net-lam}) demonstrates that the control net efficiency is negative whenever the uncontrolled flow stays laminar (cf.\ figure~\ref{fig.alpha-Pnet-lam-nom}). This is a special case of more general results by~\citet{bew09} and \citet{fuksugkas09} which have established that any transpiration-based control strategy necessarily has negative net efficiency compared to the laminar uncontrolled flow.


On the other hand, the net efficiency of the laminar controlled flow in fraction of the power required to drive the uncontrolled turbulent flow is determined by
    \be
    \% \Pi_{net}
    \; = \;
    \frac{\Pi_{net}}{\Pi_{turb}}
    \, = \,
    \frac{U_{B,0}}{U_{B,turb}} \,
    \left(
    \underset{> 0}{\underbrace{1 - \frac{U_{B,turb}}{U_{B,0}}}}
    \, - \,
    \alpha^2 \, |\pi_2 (R_c, c, \omega_x)|
    \right)
    \, + \,
    {\cal O}(\alpha^4),
    \label{eq.net-turb}
    \ee
where $\Pi_{turb} =  -2 P_x \, U_{B,turb}$. Since the bulk flux of the uncontrolled turbulent flow is smaller than that of the uncontrolled laminar flow (i.e., $U_{B,turb} < U_{B,0}$), it is possible to obtain a positive net efficiency for sufficiently small values of $\alpha$. Note that formula~(\ref{eq.net-turb}) is derived under the assumption that the controlled flow stays laminar while the
uncontrolled flow becomes turbulent. Clearly, this formula represents an idealization since it assumes that laminar flow can be maintained by both UTWs and DTWs even with infinitesimal control amplitudes. It also indicates that increasing the control amplitude always decreases the nominal net efficiency. In a nutshell, the control amplitude needs to be large enough to maintain a laminar flow but increasing the control amplitude beyond certain value brings the efficiency down and eventually leads to negative efficiency. If the efficiency is negative, maintaining a laminar flow does not lead to any net benefit in the presence of control. This is further illustrated in figure~\ref{fig.alpha-Pnet-turb-nom} where Newton's method is used to show that a positive net efficiency can be achieved for control amplitudes smaller than a certain threshold value (e.g., $\alpha < 0.05$ for the DTW with $c = 5$ and $\omega_x = 2$). In addition, the net efficiency monotonically decreases as $\alpha$ is increased, as predicted by the weakly nonlinear analysis up to a second order in $\alpha$ (cf.\ (\ref{eq.net-turb})).


    \begin{figure}
   \begin{center}
   \begin{tabular}{cc}
   \includegraphics[height=1.9in,width=2.5in]
   {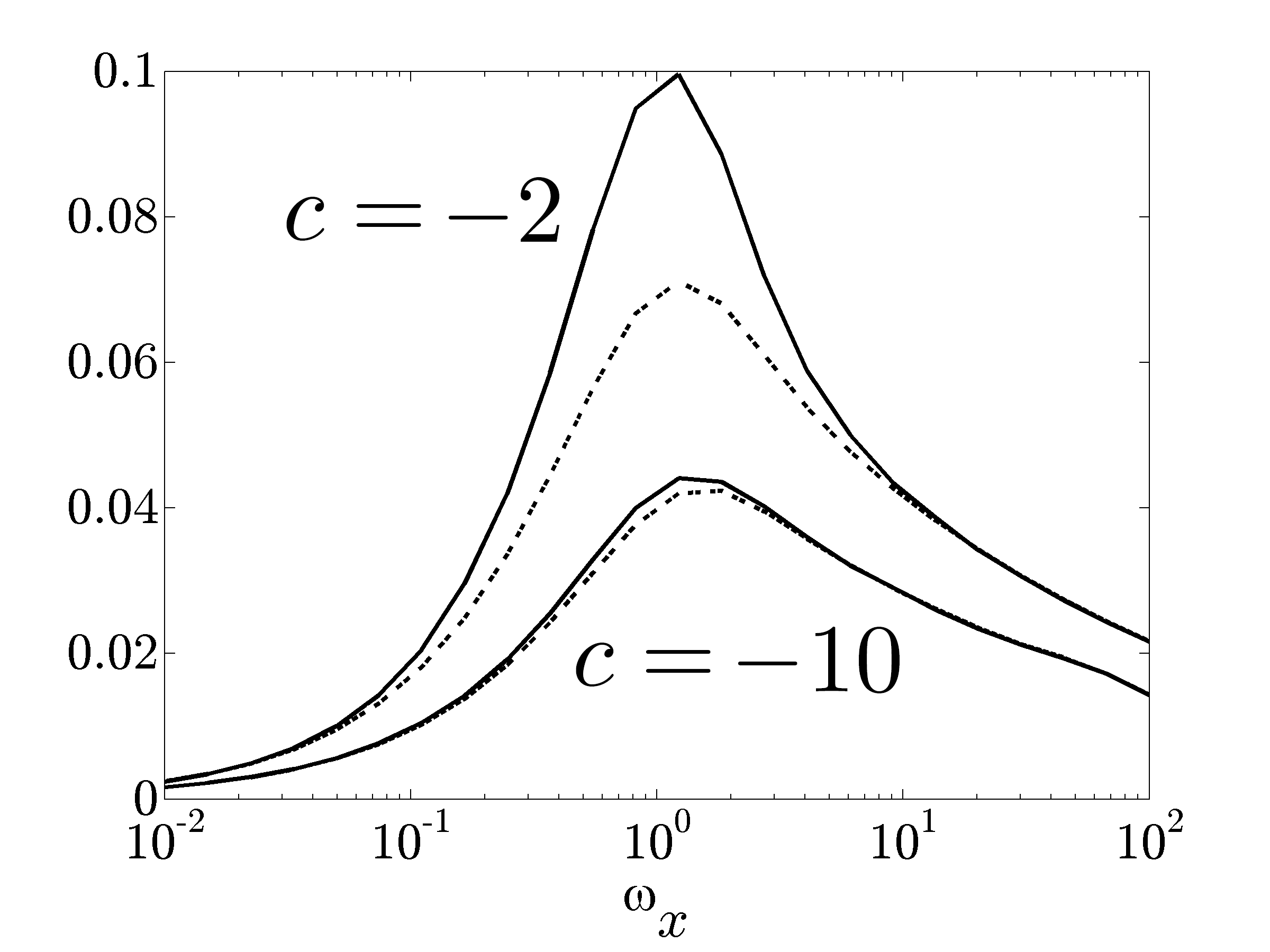}
   &
   \includegraphics[height=1.9in,width=2.5in]
   {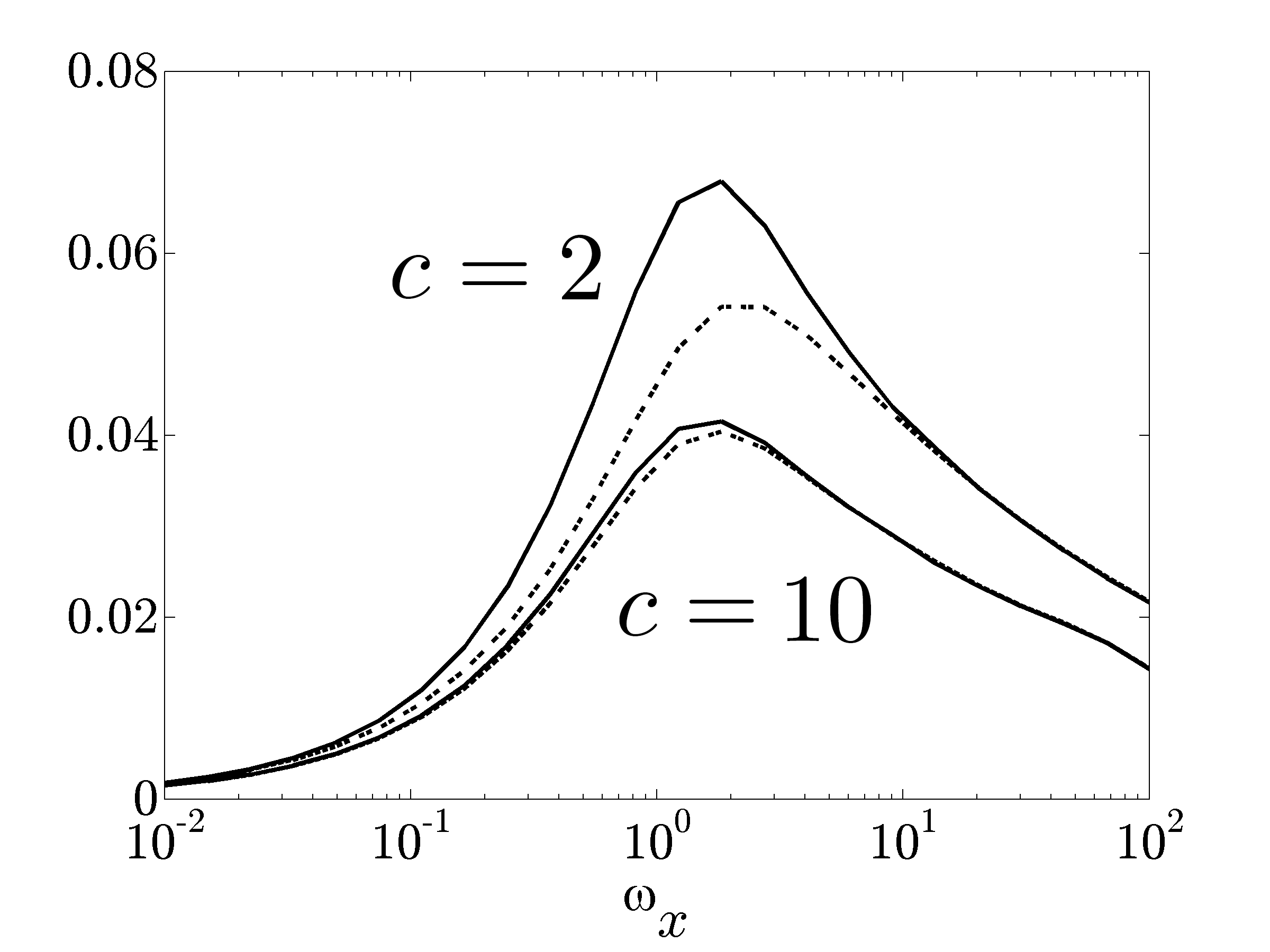}
   \end{tabular}
   \end{center}
   \caption{The wave amplitude, $\alpha_{\max}$, for which the nominal net efficiency, $\% \Pi_{net}$,
   is positive at $R_c = 2000$ for
   (a) a pair of UTWs; and
   (b) a pair of DTWs.
   The solid curves are computed using Newton's method, and the dotted curves are computed using~(\ref{eq.alpha_max}). The results are obtained by {\em assuming\/} that the controlled flow stays laminar while the uncontrolled flow becomes turbulent.
   }
   \label{fig.alpha-Pnet-nom}
   \end{figure}

An estimate for the maximum value of $\alpha$ for which a positive net efficiency is attainable can be obtained by solving the following equation (obtained using weakly nonlinear analysis)
    \beq
    (1 - U_{B,turb}/U_{B,0})
    \, - \,
    \alpha_{\max}^2 \, |\pi_2 (R_c, c, \omega_x)|
    \, = \, 0.
    \label{eq.alpha_max}
    \eeq
Figure~\ref{fig.alpha-Pnet-nom} shows $\alpha_{\max}$ as a function of $\omega_x$ for different values of $c$. The dotted curves denote the approximation for $\alpha_{\max}$ obtained using~(\ref{eq.alpha_max}). The values of $\alpha_{\max}$ (solid curves) obtained using Newton's method are also shown for comparison; we see that the predictions based on the second order correction capture the essential trends and provide good estimates for $\alpha_{\max}$ (especially for large wave speeds and for wave frequencies between $0.1$ and $10$). Figures~\ref{fig.alpha-Pnet-turb-lam-nom} and~\ref{fig.alpha-Pnet-nom} are obtained by assuming that the flow with control stays laminar while the flow with no control becomes turbulent. Whether or not the traveling waves can control the onset of turbulence depends on the velocity fluctuations; addressing this question requires analysis of the dynamics, which is a topic of~\S~\ref{sec.vel-fluc} and~\S~\ref{sec.energy-amp}, where we examine receptivity of velocity fluctuations around UTWs and DTWs to stochastic disturbances.

\section{Dynamics of fluctuations around traveling waves}
    \label{sec.vel-fluc}


\subsection{Evolution model with forcing}
    \label{sec.lnmodel}

A standard conversion of~(\ref{eq.linear-NS}) to the wall-normal velocity ($v$)/vorticity ($\eta$) formulation removes the pressure from the equations and yields the following evolution model with {\em forcing\/}
    \be
    \ba{rcl}
    E \, \bpsi_t(x,y,z,t)
    & \!\! = \!\! &
    F \, \bpsi(x,y,z,t)
    \; + \;
    G \, \bd(x,y,z,t),
    \\[0.1cm]
    \bv(x,y,z,t)
    & \!\! = \!\! &
    C \, \bpsi(x,y,z,t).
    \label{eq.LNSE}
    \ea
    \ee
This model is driven by the body force fluctuation vector $\bd = (d_1, \, d_2, \, d_3)$, which can account for flow disturbances. We refer the reader to a recent review article~\citep{sch07} and a monograph~\citep{schhen01} for a comprehensive discussion explaining why it is relevant to study influence of these excitations on velocity fluctuations. The internal state of~(\ref{eq.LNSE}) is determined by $\bpsi = (v, \, \eta)$, with Cauchy (both Dirichlet and Neumann) boundary conditions on $v$ and Dirichlet boundary conditions on $\eta$. All operators in~(\ref{eq.LNSE}) are matrices of differential operators in three coordinate directions $x$, $y$, and $z$. Operator $C$ in~(\ref{eq.LNSE}) captures a kinematic relation between $\bpsi$ and $\bv$, operator $G$ describes how forcing enters into the evolution model, whereas operators $E$ and $F$ determine internal properties of the linearized NS equations (e.g., modal stability). While operators $E$, $G$, and $C$ do not depend on base velocity, operator $F$ is base-velocity-dependent and, hence, it determines changes in the dynamics owing to changes in ${\bf u}_b$ (see Appendix~\ref{app.ABC-theta}). Moreover, for base velocity of~\S~\ref{sec.base-flow}, $F$ inherits spatial periodicity in $x$ from ${\bf u}_b$ and it can be represented as
    \be
    F
    \; = \;
    F_0
    \; + \;
    \sum_{l \, = \, 1}^{\infty}
    \alpha^l
    \sum_{r \, \overset{2}{=} \, - l}^{l}
    \mre^{\mri r \omega_x x}
    F_{l,r},
    \non
    \ee
where $F_0$ and $F_{l,r}$ are spatially invariant operators in the streamwise and spanwise directions and $\sum_{r \, \overset{2}{=} \, -l}^{l}$ signifies that $r$ takes the values $\{-l, -l+2, \ldots, l-2, l\}$. This expansion isolates spatially invariant and spatially periodic parts of operator $F$, which is well-suited for representation of~(\ref{eq.LNSE}) in the frequency domain.


    \begin{figure}
    \begin{center}
    \includegraphics[height=1.9in]
    {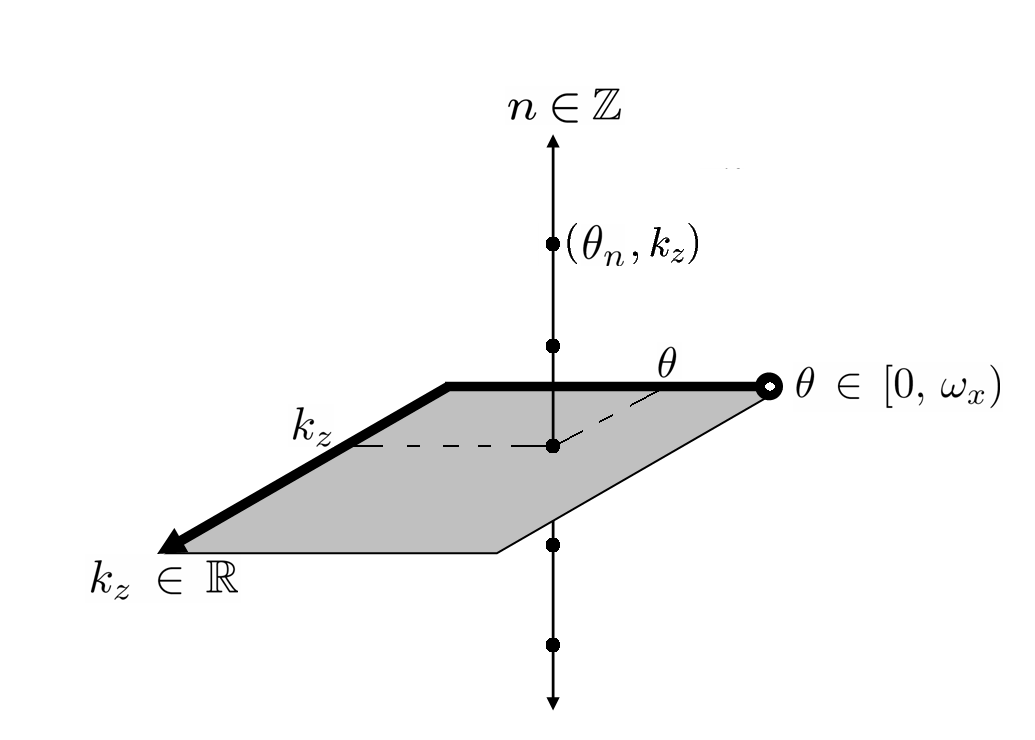}
    \end{center}
    \caption{
    A Bloch wave $\bd(x,y,z,t)$ defined in~(\ref{eq.NM}) is obtained by the superposition of weighted Fourier exponentials with frequencies $\left(\theta_n, k_z\right)|_{n \, \in \, \bbZ}$, with weights determined by $\bar{\bd}_n(y,k_z,t)$.
    }
    \label{fig.theta-kz-space}
    \end{figure}

\subsection{Frequency representation of the linearized model}
    \label{sec.freq3D3C}

Owing to the structure of the linearized NS equations, the differential operators $E$, $G$, and $C$ are invariant with respect to translations in horizontal directions. On the other hand, operator $F$ is invariant in $z$ and periodic in $x$. Thus, the Fourier transform in $z$ can be applied to algebraize the spanwise differential operators. In other words, the normal modes in $z$ are the spanwise waves, $\mre^{\mri k_z z}$, where $k_z$ denotes the spanwise wavenumber. On the other hand, the appropriate normal modes in $x$ are given by the so-called {\em Bloch waves}~\citep{odekel64,benliopap78}, which are determined by a product of $\mre^{\mri \theta x}$ and the $2 \pi/\omega_x$ periodic function in $x$, with $\theta \in [0, \, \omega_x)$. Based on the above, each signal in~(\ref{eq.LNSE}) (for example, $\bd$) can be expressed as
    \be
    \left.
    \ba{rcl}
    \bd(x,y,z,t)
    & = &
    \mre^{\mri k_z z}
    \mre^{\mri \theta x}
    \,
    \bar{\bd}(x,y,k_z,t)
    \\[0.1cm]
    \bar{\bd}(x,y,k_z,t)
    & = &
    \bar{\bd}(x + 2 \pi/\omega_x,y,k_z,t)
    \ea
    \right\}
    ~~
    k_z \, \in \, \bbR,
    ~~
    \theta \, \in \, [0, \, \omega_x),
    \non
    \ee
where only real parts are to be used for representation of physical quantities. Expressing $\bar{\bd}(x,y,k_z,t)$ in Fourier series yields (see figure~\ref{fig.theta-kz-space} for an illustration)
    \be
    \bd(x,y,z,t)
    \; = \;
    \ds{\sum_{n \, = \, - \infty}^{\infty}}
    \bar{\bd}_n(y,k_z,t)
    \,
    \mre^{\mri (\theta_n x \, + \, k_z z)},
    ~~
    \ba{l}
    \theta_n
    \; = \;
    \theta \, + \, n \omega_x,
    \\[0.0cm]
    k_z
    \, \in \,
    \bbR,
    ~
    \theta
    \, \in \,
    [0, \, \omega_x),
    \ea
    \label{eq.NM}
    \ee
where $\{ \bar{\bd}_n(y,k_z,t) \}_{n \, \in \, \bbZ}$ are the coefficients in the Fourier series expansions of $\bar{\bd}(x,y,k_z,t)$.

The frequency representation of the linearized NS equations is obtained by substituting (\ref{eq.NM}) into~(\ref{eq.LNSE})
    \be
    \ba{rcl}
    \pt \bpsi_{\theta}(y,k_z,t)
    & \!\! = \!\! &
    {\cal A}_\theta (k_z) \, \bpsi_{\theta}(y,k_z,t)
    ~+~
    {\cal B}_\theta (k_z) \, \bd_{\theta}(y,k_z,t),
    \\[0.1cm]
    \bv_\theta(y,k_z,t)
    & \!\! = \!\! &
    {\cal C}_\theta (k_z) \, \bpsi_{\theta}(y,k_z,t).
    \label{eq.FR}
    \ea
    \ee
This representation is parameterized by $k_z$ and $\theta$ and $\bpsi_\theta (y,k_z,t)$ denotes a bi-infinite column vector,
    $
    \bpsi_\theta (y,k_z,t)
    =
    \col \, \{\bpsi (\theta_n,y,k_z,t)\}_{n \, \in \, \bbZ}.
    $
The same definition applies to $\bd_{\theta}(y,k_z,t)$ and $\bv_\theta(y,k_z,t)$. On the other hand, for each $k_z$ and $\theta$, ${\cal A}_\theta (k_z)$, ${\cal B}_\theta (k_z)$, and ${\cal C}_\theta (k_z)$ are bi-infinite matrices whose elements are one dimensional integro-differential operators in $y$. The structure of these operators depends on frequency representation of $E$, $F$, $G$, and $C$ in~(\ref{eq.LNSE}). In short, ${\cal B}_\theta (k_z)$ and ${\cal C}_\theta (k_z)$ are block-diagonal operators and
    \[
    {\cal A}_\theta
    \; = \;
    {\cal A}_{0\theta}
    \; + \,
    \sum_{l \, = \, 1}^{\infty}
    \alpha^l \, {\cal A}_{l\theta},
    \]
where ${\cal A}_{0\theta}$ and ${\cal A}_{l\theta}$ are structured operators (see Appendix~\ref{app.ABC-theta} for more details). The particular structure of $\ca_{0 \theta}$ and $\ca_{l \theta}$ is exploited in perturbation analysis of the energy amplification for small control amplitudes $\alpha$ in~\S~\ref{sec.pert}.


\subsection{Energy density of the linearized model}
    \label{sec.H2}

Frequency representation~(\ref{eq.FR}) contains a large amount of information about linearized dynamics. For example, it can be used to assess stability properties of the base flow. However, since the {\em early stages of transition\/} in wall-bounded shear flows are not appropriately described by the stability properties of the linearized equations~\citep[see, for example,][]{schhen01,sch07}, we perform receptivity analysis of stochastically forced model~(\ref{eq.FR}) to assess the effectiveness of the proposed control strategy. Namely, we set the initial conditions in~(\ref{eq.FR}) to zero and study the responses of the linearized dynamics to uncertain body forces. When the body forces are absent, the response of stable flows decays asymptotically to zero. However, in the presence of stochastic body forces, the linearized NS equations are capable of maintaining high levels of the steady-state variance~\citep{farioa93,bamdah01,jovbamJFM05}. Our analysis quantifies the effect of imposed streamwise traveling waves on the asymptotic levels of variance and describes how receptivity changes in the presence of control. We note that there are substantial differences between the problem considered here and in~\citet{jovbamJFM05}; these differences arise from lack of homogeneity in the streamwise direction which introduces significant computational challenges which we discus below. Furthermore, even though our study is similar in spirit to~\cite{jovPOF08}, current work studies dynamics of fluctuations around spatially periodic base velocity, whereas~\cite{jovPOF08} considered dynamics of fluctuations around time periodic base velocity. Theoretical framework for quantifying receptivity in these two conceptually different cases was developed by~\citet*{farjovbam08} and~\citet{jovfarAUT08}, respectively.

Let us assume that a stable system~(\ref{eq.FR}) is subject to a zero-mean white stochastic process (in $y$ and $t$), $\bd_\theta(y,k_z,t)$. Then, for each $k_z$ and $\theta$, the ensemble average energy density of the statistical steady-state is determined by
    \be
    \ba{rcl}
    \bmrE (\theta,k_z)
    \!\! & = & \!\!
    \ds{\lim_{t \, \rightarrow \, \infty}}
    \inprod{\bv_\theta (\cdot,k_z,t)}{\bv_\theta (\cdot,k_z,t)}
    \\[0.1cm]
    \!\! & = & \!\!
    \trace
    \left(
    \ds{\lim_{t \, \rightarrow \, \infty}}
    {\cal E} \left\{ \bv_\theta (\cdot,k_z,t) \otimes \bv_\theta (\cdot,k_z,t)
    \right\}
    \right),
    \ea
    \non
    \ee
where $\inprod{\cdot}{\cdot}$ denotes the $L_2 [-1,1]$ inner product and averaging in time, i.e.,
    \be
    \ba{rcl}
    \inprod{\bv_\theta}{\bv_\theta}
    \!\! & = & \!\!
    \ds{
    {\cal E}
    \left\{
    \int_{-1}^{1}{
    \bv^{*}_\theta (y,k_z,t) \, \bv_\theta (y,k_z,t)
    \, \mrd y}
    \right\}
    },
    \\[0.25cm]
    {\cal E}
    \left\{
    v(\cdot,t)
    \right\}
    \!\! & = & \!\!
    \ds{
    \lim_{T \, \rightarrow \, \infty}
    \dfrac{1}{T}
    \int_0^T{
    v(\cdot,t \, + \, \tau)
    \,
    \mrd \tau
    }
    },
    \ea
    \label{eq.L2ip}
    \ee
and $\bv_\theta \otimes \bv_\theta$ is the tensor product of $\bv_\theta$ with itself. We note that $\bmrE (\theta,k_z)$ determines the asymptotic level of energy (i.e., variance) maintained by a stochastic forcing in~(\ref{eq.FR}). Typically, this quantity is computed by running DNS of the NS equations until the statistical steady-state is reached. However, for linearized system~(\ref{eq.FR}), the energy density $\bmrE (\theta,k_z)$ can be determined using the solution to the following operator Lyapunov equation~\citep{farjovbam08}
    \be
    \ca_\theta (k_z) \cx_\theta (k_z)
    \; + \;
    \cx_\theta (k_z) \ca_\theta^{*} (k_z)
    \; = \,
    - \, \cb_\theta (k_z) \cb^{*}_\theta (k_z),
    \label{eq.LE}
    \ee
as
    \[
    \bmrE (\theta,k_z)
    \; = \;
    \trace
    \left(
    \cx_\theta (k_z)
    \,
    {\cal C}^*_\theta (k_z)
    \,
    {\cal C}_\theta (k_z)
    \right).
    \]
Here, $\cx_\theta (k_z)$ denotes the autocorrelation operator of $\bpsi_\theta$, that is
    \[
    \cx_\theta (k_z)
    \; = \;
    \lim_{t \, \rightarrow \, \infty}
    {\cal E}
    \left\{
    \bpsi_\theta (\cdot,k_z,t)
    \otimes
    \bpsi_\theta (\cdot,k_z,t) \right\}.
    \]
Since ${\cal C}_\theta^{*} (k_z) \, {\cal C}_\theta (k_z)$ is an identity operator, we have
    \be
    \bmrE (\theta,k_z)
    \, = \,
    \trace
    \left(
    \cx_\theta (k_z)
    \right)
    \, = \,
    \sum_{n \, = \, -\infty}^{\infty}
    \trace \left( X_d (\theta_n,k_z) \right),
    \label{eq.E-Xd}
    \ee
where $X_d (\theta_n,k_z)$ denotes the elements on the main diagonal of operator $\cx_\theta$. We note that $\bmrE$ also has an interesting deterministic interpretation; namely, if $\bv_\theta (\cdot,k_z,t)$ denotes the impulse response of~(\ref{eq.FR}), then
    \be
    \bmrE(\theta,k_z)
    \; = \;
    \ds{\int_{0}^{\infty}}
    \trace \left(
    \bv_\theta (\cdot,k_z,t) \otimes \bv_\theta (\cdot,k_z,t)
    \right) \,
    \mrd t.
    \non
    \ee
Thus, the same quantity can be used to assess receptivity of the linearized NS equations to exogenous disturbances of either stochastic or deterministic origin.


\subsection{Perturbation analysis of energy density}
    \label{sec.pert}

Solving~(\ref{eq.LE}) is computationally expensive; a discretization of the operators (in $y$) and truncation of the bi-infinite matrices convert~(\ref{eq.LE}) into a large-scale matrix Lyapunov equation. Our computations suggest that in order to obtain convergence of
    \[
    \bmrE(\theta,k_z)
    \; \approx \,
    \sum_{n \, = \, -N}^{N}
    \trace \left( X_d (\theta_n,k_z) \right),
    \]
a choice of $N$ between ten (for $\omega_x \sim {\cal O}(1)$) and a few thousands (for $\omega_x \sim {\cal O}(0.01)$) is required. Since we aim to conduct a detailed study of the influence of streamwise traveling waves on dynamics of velocity fluctuations, determining energy density for a broad range of traveling wave parameters, $k_z$ and $\theta$ still poses significant computational challenges.


Instead, we employ an efficient perturbation analysis based approach introduced by~\citet{farbam08} for solving equation~(\ref{eq.LE}). For our problem, this approach turns out to be at least $20$ times faster than the truncation approach. This method is well-suited for systems with small amplitude spatially periodic terms and it converts~(\ref{eq.LE}) into a set of conveniently coupled system of operator-valued Lyapunov and Sylvester equations. A finite dimensional approximation of these equations yields a set of algebraic matrix equations whose order is determined by the product between the number of fields in the evolution model (here $2$, the wall-normal velocity and vorticity) and the size of discretization in $y$. While consideration of small wave amplitudes simplifies analysis by providing an explicit expression for energy density, it is also motivated by our earlier observation that large values of $\alpha$ introduce high cost of control which is not desirable from a physical point of view.

It can be shown (see Appendix~\ref{app.htwo-pert-detail} for details) that the energy density of system~(\ref{eq.LNSE}) can be represented as
    \be
    \bmrE (\theta,k_z;R_c,\alpha,c,\omega_x)
    \, = \;
    \bmrE_{0} (\theta,k_z;R_c,\omega_x)
    \, +
    \sum_{l \, = \, 1}^{\infty}
    \alpha^{2 l} \,
    \bmrE_{2 l} (\theta,k_z;R_c,c,\omega_x),
    ~
    0 < \alpha \ll 1.
    \label{eq.ED}
    \ee
Thus, only terms with even powers in $\alpha$ contribute to $\bmrE$, which in controlled flow depends on six parameters. Since our objective is to identify trends in energy density, we confine our attention to a perturbation analysis up to a second order in $\alpha$. We briefly comment on the influence of higher order corrections in \S~\ref{sec.alpha} where it is shown that the essential trends are correctly predicted by the second order of correction.

\section{Energy amplification in Poiseuille flow with $R_c = 2000$}
    \label{sec.energy-amp}

In this section, we study energy amplification of stochastically forced linearized NS equations in Poiseuille flow controlled with streamwise traveling waves. Equation~(\ref{eq.ED}) reveals the dependence of the energy density on traveling wave amplitude $\alpha$, for $0 < \alpha \ll 1$. However, since the operators in~(\ref{eq.FR}) depend on the spatial wavenumbers ($\theta$ and $k_z$), $R_c$, $\omega_x$, and $c$, the energy density is also a function of these parameters. Finding the optimal triple $(\alpha, c, \omega_x)$ that maximally reduces the energy of the velocity fluctuations is outside the scope of the current study; instead, we identify the values of $c$ and $\omega_x$ that are capable of reducing receptivity in the presence of small amplitude streamwise traveling waves. Since we are interested in energy amplification of the transitional Poiseuille flow, we choose $R_c = 2000$ in all of our subsequent computations. This value is selected because it is between the critical Reynolds number at which linear instability takes place, $R_c = 5772$, and the value at which transition is observed in experiments and DNS, $R_c \approx 1000$. The same Reynolds number was used by~\citet{minsunspekim06} in their DNS study.


    \begin{figure}
    \begin{center}
    \includegraphics[height=1.9in]
    {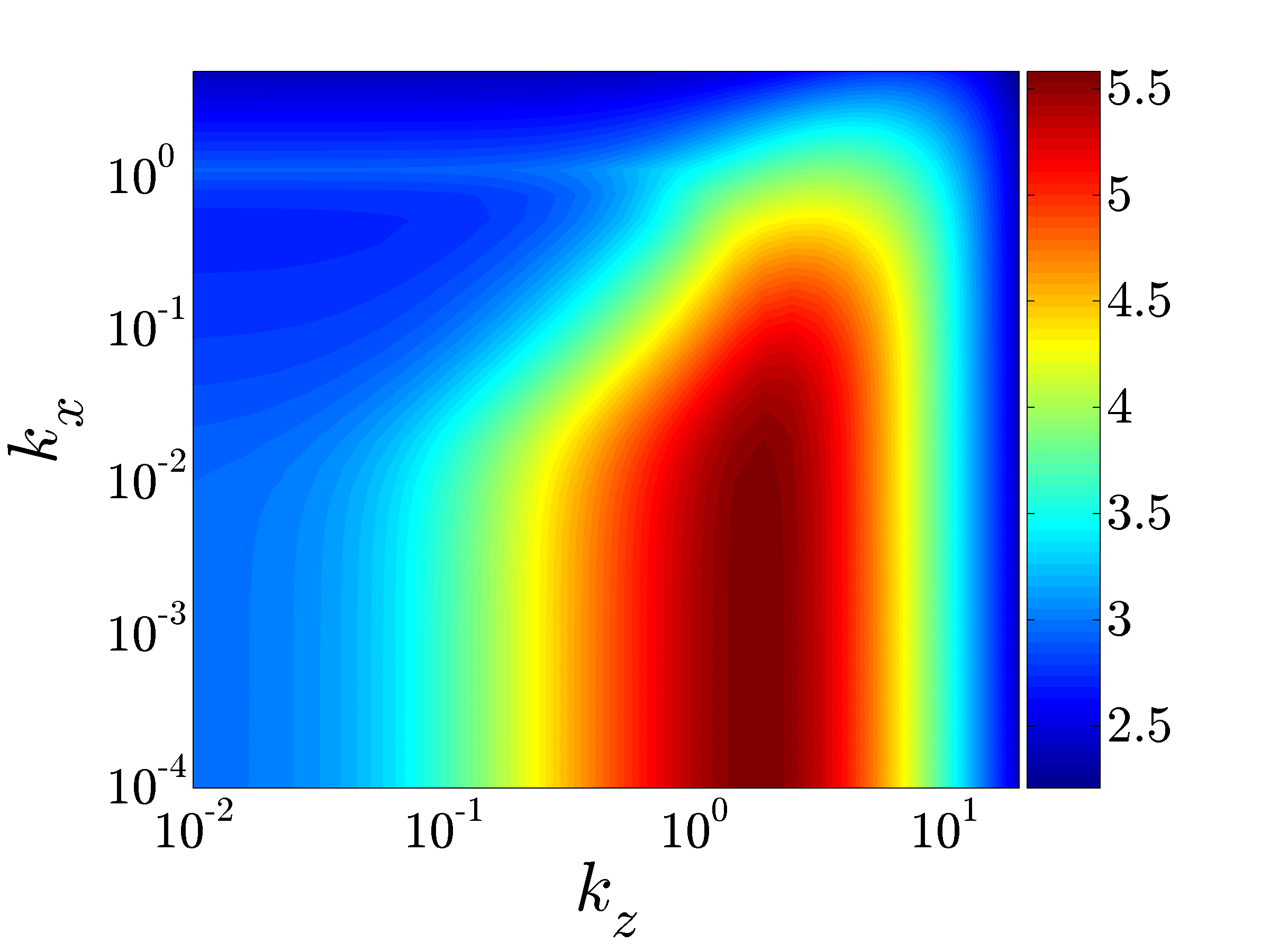}
    \end{center}
    \caption{
    Energy density $\tmrE_0 (k_x,k_z)$ of the uncontrolled Poiseuille flow with $R_c = 2000$. The plot is given in the log-log-log scale.
    }
    \label{fig.E0kxkz}
    \end{figure}

\subsection{Energy density of flow with no control}
    \label{sec.E0}

\begin{figure}
    \begin{center}
    \includegraphics[height=2.5in]
    {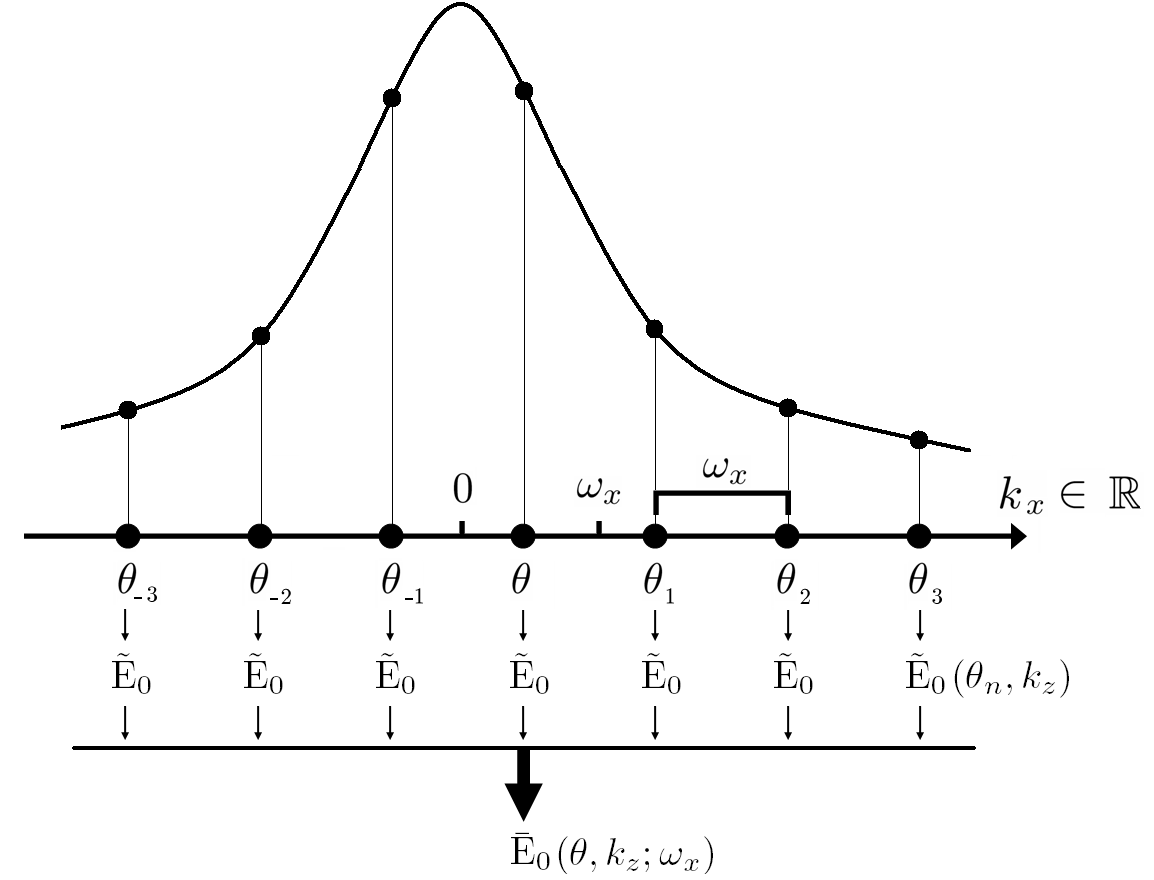}
    \end{center}
    \caption{
    For fixed $\omega_x$ and $\theta$, $\bmrE_0(\theta,k_z;\omega_x)$ represents energy density of fluctuations composed of all wavenumbers $\theta_n = \{\theta + n \omega_x \}_{n \, \in \, \bbZ}$;
    $
    \bmrE_0(\theta, k_z; \omega_x)
    \, = \,
    \sum_{n \,=\, -\infty}^{\infty}{\tmrE_0(\theta + n\omega_x,k_z)}.
    $
    }
    \label{fig.Energy-kx-vs-theta}
    \end{figure}

We briefly comment on the energy density in uncontrolled Poiseuille flow with $R_c = 2000$; for an in-depth treatment see~\citet{jovbamJFM05}. The appropriate normal modes in the uncontrolled flow are purely harmonic streamwise and spanwise waves, $\mre^{\mri k_x x} \, \mre^{\mri k_z z}$, where $k_x$ denotes the streamwise wavenumber. Figure~\ref{fig.E0kxkz} illustrates the energy density of the uncontrolled flow as a function of $k_x$ and $k_z$, which we denote by $\tmrE_0(k_x,k_z)$. The streamwise constant fluctuations with ${\cal O}(1)$ spanwise wavenumbers carry most energy in flow with no control. Namely, the largest value of $\tmrE_0 (k_x,k_z)$ occurs at ($k_x = 0$, $k_z \approx 1.78$), which means that the most amplified flow structures (the streamwise streaks) are infinitely elongated in the streamwise direction and have the spanwise length scale of approximately $3.5 \delta$, where $\delta$ is the channel half-height. We note that these input-output resonances do not correspond to the least-stable modes of the linearized NS equations. Rather, they arise because of the coupling from the wall-normal velocity $v$ to the wall-normal vorticity $\eta$. Physically, this coupling is a product of the vortex tilting (lift-up) mechanism~\citep{lan75}; the base shear is tilted in the wall-normal direction by the spanwise changes in $v$, which lead to a nonmodal amplification of $\eta$. This mechanism does not take place either when the base shear is zero (i.e., $U' = 0$), or when there are no spanwise variations in $v$ (i.e., $k_z = 0$). On the other hand, the least-stable modes (TS waves) of uncontrolled flow create a local peak in $\tmrE_0 (k_x,k_z)$ around ($k_z = 0$, $k_x \approx 1.2$), with a magnitude significantly lower compared to the magnitude achieved by the streamwise constant flow structures. Finally, we note that the uncontrolled energy density $\bmrE_0(\theta,k_z;\omega_x)$ as appeared in~(\ref{eq.ED}) can be obtained from $\tmrE_0(k_x,k_z)$ using the following expression
    \be
    \bmrE_0(\theta, k_z; \omega_x)
    \; = \,
    \ds{
    \sum_{n \,=\, -\infty}^{\infty}{\tmrE_0(\theta_n,k_z)}
    \; = \,
    \sum_{n \,=\, -\infty}^{\infty}{\tmrE_0(\theta + n\omega_x,k_z)}
    }.
    \non
    \ee
In other words, for fixed $\omega_x$ and $\theta$, $\bmrE_0(\theta,k_z;\omega_x)$ represents the energy density of velocity fluctuations that are composed of all wavenumbers $k_x = \{\theta + n \omega_x \}_{n \, \in \, \bbZ}$. In comparison, $\tmrE_0(k_x,k_z)$ is the energy density of velocity fluctuations composed of a single wavenumber $k_x$ (see figure~\ref{fig.Energy-kx-vs-theta} for an illustration).

\subsection{Energy amplification of flow with control}
    \label{sec.E2}

We next consider energy amplification of velocity fluctuations in Poiseuille flow with $R_c = 2000$ in the presence of both UTWs and DTWs. As shown in \S~\ref{sec.pert}, for small amplitude blowing and suction along the walls, the perturbation analysis yields an {\em explicit formula\/} for energy amplification (cf.~(\ref{eq.ED})),
    \be
    \frac{\bmrE (\theta,k_z;\alpha,c,\omega_x)}
    {\bmrE_{0} (\theta,k_z;\omega_x)}
    \; = \;
    1
    \, + \,
    \alpha^2
    \,
    g_2 (\theta,k_z;c,\omega_x)
    \, + \,
    {\cal O} (\alpha^4),
    ~~
    0 < \alpha \ll 1.
    \non
    \ee
Thus, for small wave amplitudes the influence of control can be assessed by evaluating function $g_2 = \bmrE_2/\bmrE_0$ that quantifies energy amplification up to a second order in $\alpha$. Sign of $g_2$ determines whether energy density is increased or decreased in the presence of control; positive (negative) values of $g_2$ identify wave speed and frequency that increase (decrease) receptivity.
Since function $g_2$ is sign-indefinite with vastly different magnitudes, it is advantageous to visualize $g_2$ using a sign-preserving logarithmic scale
    \be
    \hat{g}_2
    \, = \,
    {\rm sign}(g_2)\,
    {\rm log_{10}}(1 \, + \, |g_2|).
    \non
    \ee
For example, $\hat{g}_2 = 5$ or $\hat{g}_2 = -3$, respectively, signify  $\bmrE_2 = 10^5 \, \bmrE_0$ or $\bmrE_2 = -10^3 \, \bmrE_0$. Since $\hat{g}_2(\theta, k_z; c, \omega_x)$ depends on four parameters, for visualization purposes, we confine our attention to cross-sections of $\hat{g}_2$ by fixing two of the four parameters. We first study energy amplification of the modes with $k_z = 1.78$ and $k_z = 0$ as a function of $c$ and $\omega_x$; these spanwise wavenumbers are selected in order to capture influence of control on streamwise streaks and TS waves, respectively. Since, in uncontrolled flow, streamwise streaks (respectively, TS waves) occur at $k_x = 0$ (respectively, $k_x = 1.2$), fluctuations with $\theta = 0$ (respectively, $\theta (\omega_x) = 1.2 - \omega_x \lfloor 1.2/\omega_x \rfloor$) are considered; these values of $\theta$ are chosen to make sure that streamwise streaks (respectively, TS waves) represent modes of the controlled flow as well. (Here, $\lfloor a \rfloor$ denotes the largest integer not greater than $a$.) We then analyze the energy amplification of disturbances with different values of $\theta$ and $k_z$ for a fixed set of control parameters $c$ and $\omega_x$. Our analysis illustrates the ability of properly designed traveling waves to weaken the intensity of both most energetic and least stable modes of the uncontrolled flow. Direct numerical simulations of Part~2 show that this can be done with positive net efficiency.

\begin{figure}
    \begin{center}
    \begin{tabular}{cc}
    {\sc upstream:}
    &
    {\sc upstream:}
    \\[-0.18cm]
    \subfigure[]
    {
    \includegraphics[height=1.9in,width=2.5in]
    {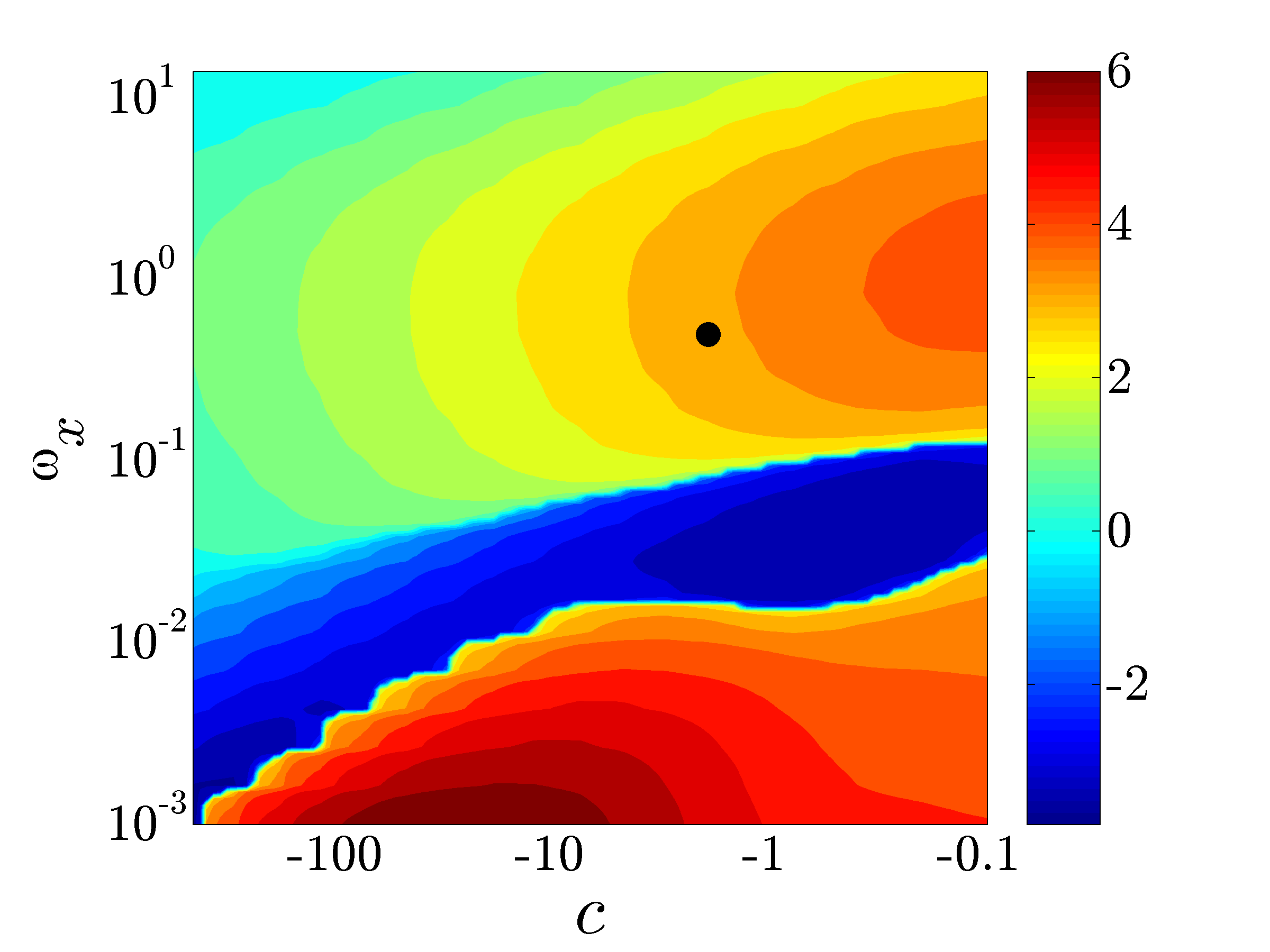}
    \label{fig.log10-E_2-corbc-vs-cm500tomp1-vs-wm3to2-theta0-kz1p78-R2000}}
    &
    \subfigure[]
    {
    \includegraphics[height=1.9in,width=2.5in]
    {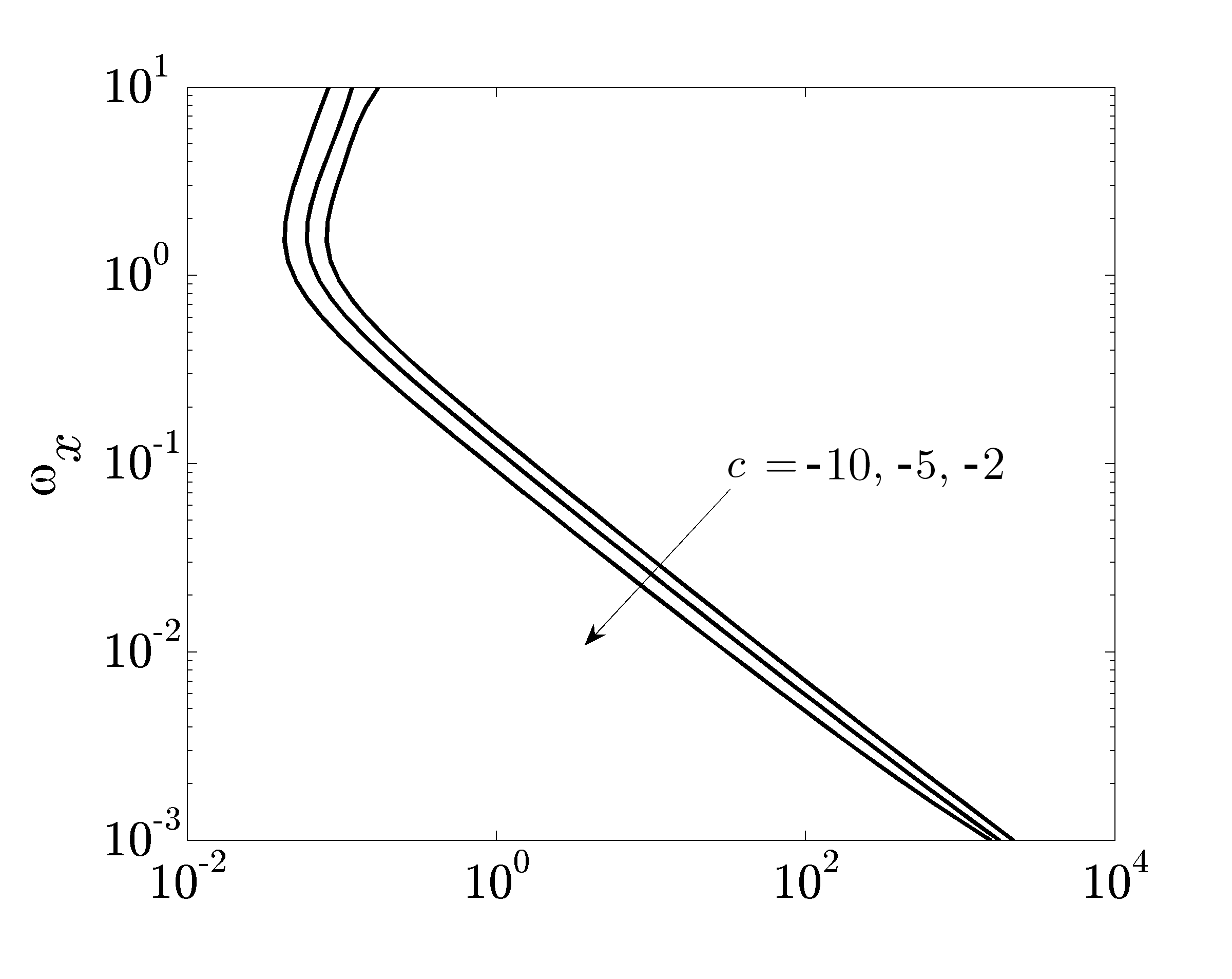}
    \label{fig.Preq2-wm3to1-cm10-m5-m2}}
    \\[0.2cm]
    {\sc downstream:}
    &
    {\sc downstream:}
    \\[-0.18cm]
    \subfigure[]
    {
    \includegraphics[height=1.9in,width=2.5in]
    {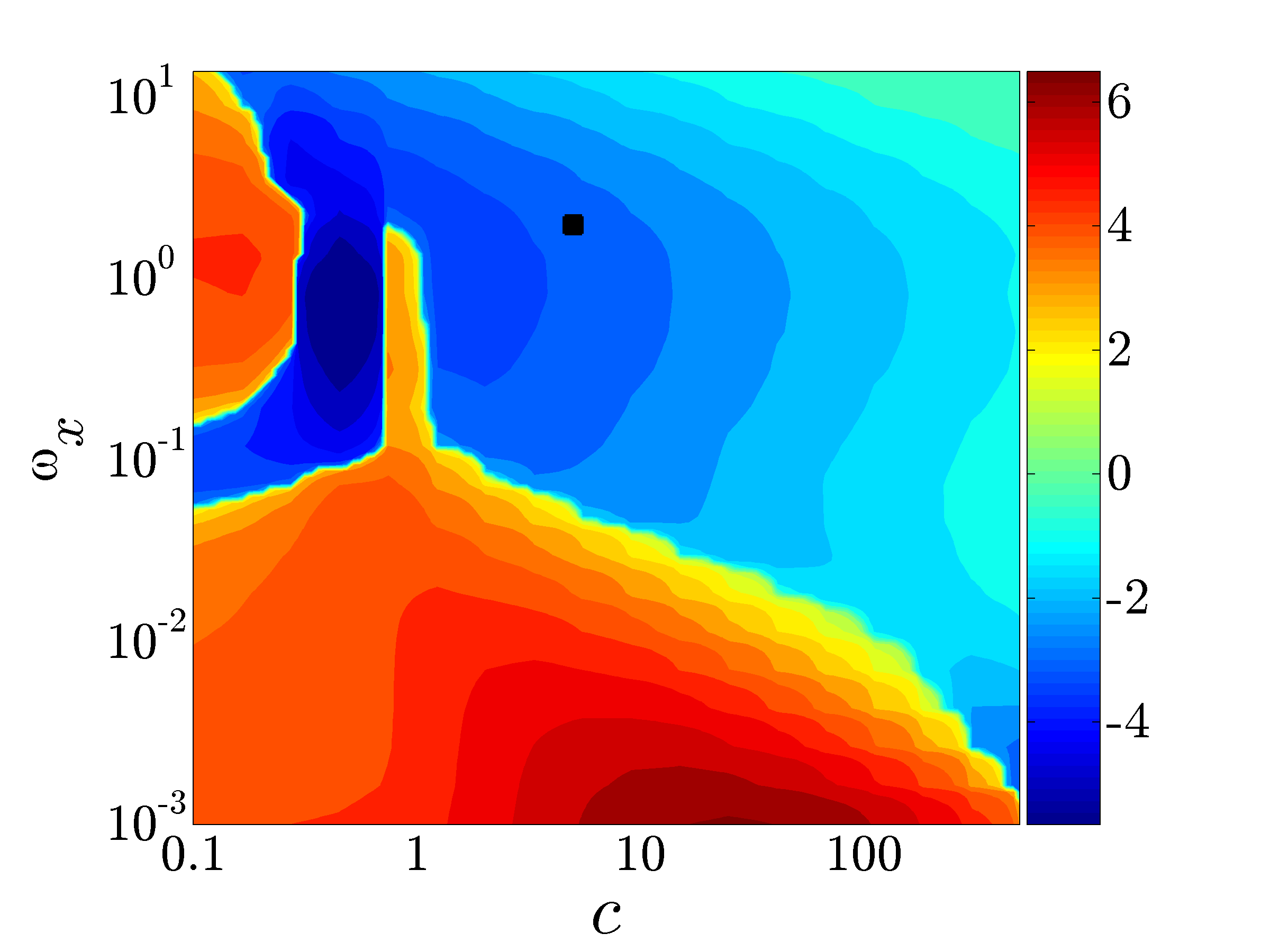}
    \label{fig.log10-E_2-corbc-vs-cpp1top500-vs-wm3to2-theta0-kz1p78-R2000}}
    &
    \subfigure[]
    {
    \includegraphics[height=1.9in,width=2.5in]
    {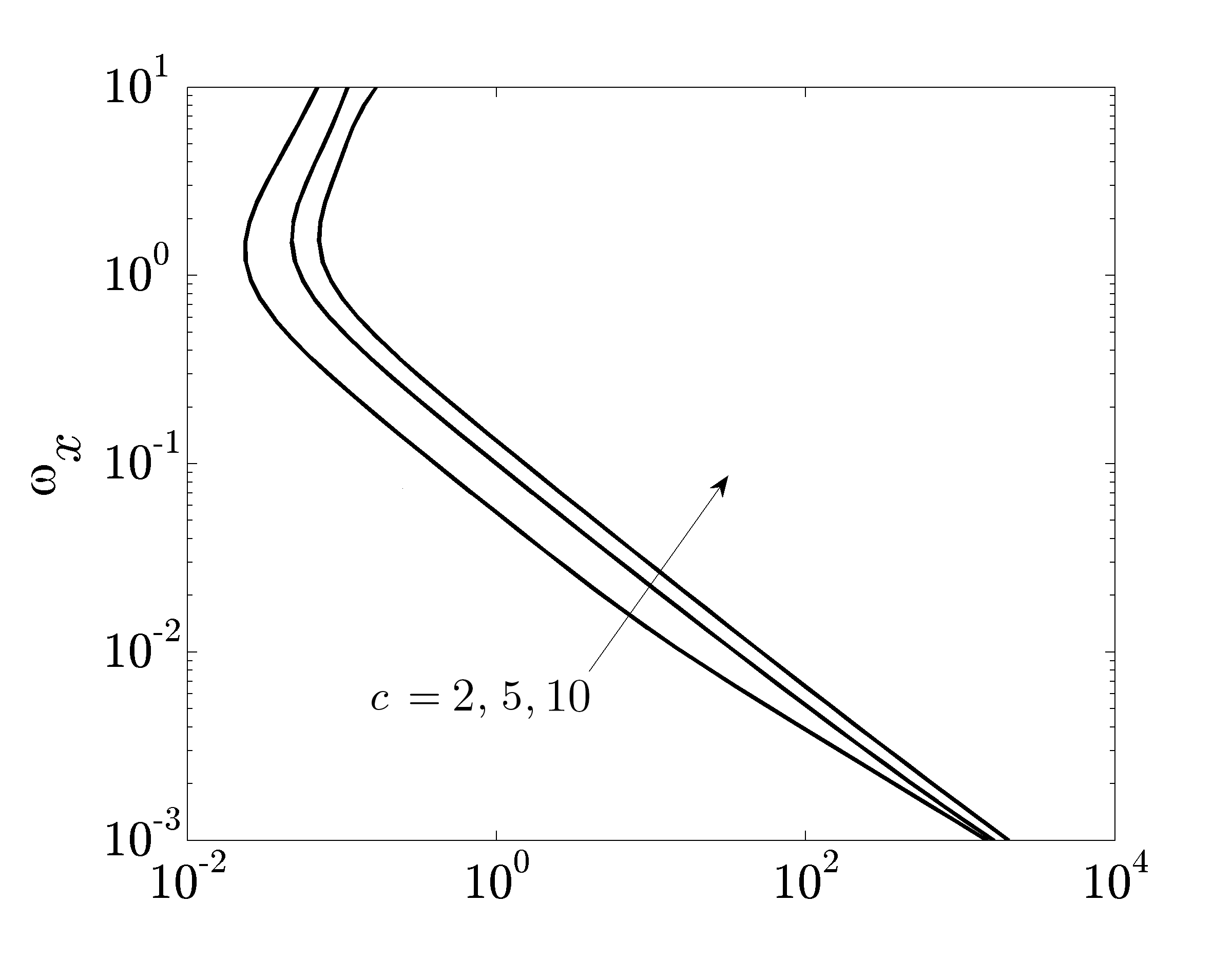}
    \label{fig.Preq2-wm3to1-c2-5-10}}
    \end{tabular}
    \end{center}
    \caption{
    {(a) and (c): Second order correction to the energy amplification, $\hat{g}_2 (c,\omega_x)$, of the modes with $(\theta, k_z) = (0, 1.78)$,     in the presence of (a) UTWs; and (c) DTWs in Poiseuille flow with $R_c =  2000$.
    (b) and (d): Second order correction to the nominal required power, $\Pi_{req,2} (\omega_x; c)$, for (b) UTWs; and (d) DTWs.
    The dot and the square, respectively, denote $( c = -2,$ $\omega_x = 0.5 )$ (as selected in~\cite{minsunspekim06}) and $( c = 5,$ $\omega_x = 2 )$.
    Note: the color plots are obtained using a sign-preserving logarithmic scale; e.g., $\hat{g}_2 = 5$ and $\hat{g}_2 = -3$ should be interpreted as $\bmrE_2 = 10^5 \, \bmrE_0$ and $\bmrE_2 = -10^3 \, \bmrE_0$, respectively.}
    }
    \label{fig.log10-E2-Preq-cpm}
    \end{figure}

Since most amplification in flow with no control occurs for fluctuations with $(k_x = 0$, $k_z = 1.78)$, it is relevant to first study the influence of controls on these most energetic modes. In flow with control, the streamwise-constant flow structures are imbedded in the fundamental mode, i.e.\ fluctuations with $\theta = 0$ (cf.\ \S~\ref{sec.freq3D3C}). As the plots of $\hat{g}_2(c,\omega_x)$ in figures~\ref{fig.log10-E_2-corbc-vs-cm500tomp1-vs-wm3to2-theta0-kz1p78-R2000} and~\ref{fig.log10-E_2-corbc-vs-cpp1top500-vs-wm3to2-theta0-kz1p78-R2000} reveal, the values of $c$ and $\omega_x$ determine whether these structures are amplified or attenuated by the traveling waves. Up to a second order in $\alpha$, the control parameters associated with the blue regions in these two figures reduce the energy amplification of the uncontrolled flow. As evident from figure~\ref{fig.log10-E_2-corbc-vs-cm500tomp1-vs-wm3to2-theta0-kz1p78-R2000}, only a narrow range of UTWs with $\omega_x \lesssim 0.1$ is capable of reducing the energy amplification. However, since the required power for maintaining the nominal flow for such low frequency controls is prohibitively large (cf.\ figure~\ref{fig.Preq2-wm3to1-cm10-m5-m2}), the choice of UTWs for transition control is not favorable from efficiency point of view (receptivity reduction by these UTWs is further discussed in~\S~\ref{sec.alpha}). On the other hand, a large range of DTW parameters with $c \gtrsim 1$ and $\omega_x \gtrsim 0.1$ is capable of making the controlled flow less sensitive to stochastic excitations (cf.\ figure~\ref{fig.log10-E_2-corbc-vs-cpp1top500-vs-wm3to2-theta0-kz1p78-R2000}). Moreover, figure~\ref{fig.Preq2-wm3to1-c2-5-10} shows that the $\omega_x \gtrsim 0.1$ region contains the smallest required power for sustaining the DTWs. These two features identify properly designed DTWs as suitable candidates for controlling the onset of turbulence with positive net efficiency (as confirmed by DNS in Part 2).

It is noteworthy that traveling waves with parameters considered in~\cite{minsunspekim06} (i.e., $\omega_x = \{0.5, 1, 1.5, 2\}$ and $-4 < c < 0$) increase amplification of the most energetic modes of the uncontrolled flow (cf.\ figure~\ref{fig.log10-E_2-corbc-vs-cm500tomp1-vs-wm3to2-theta0-kz1p78-R2000}). This is in agreement with a recent study of~\cite{leeminkim08} where a transient growth larger than that of the laminar uncontrolled flow was observed for UTWs with $c = \{-1, -2\}$ and $\omega_x = 1.5$. Furthermore, it is shown in Part 2 that such UTWs promote turbulence even for initial conditions for which the uncontrolled flow stays laminar.

\begin{figure}
    \begin{center}
    \begin{tabular}{cc}
    \subfigure[$(c = -2$, $\omega_x = 0.5)$]
    {
    \includegraphics[height=1.9in,width=2.5in]
    {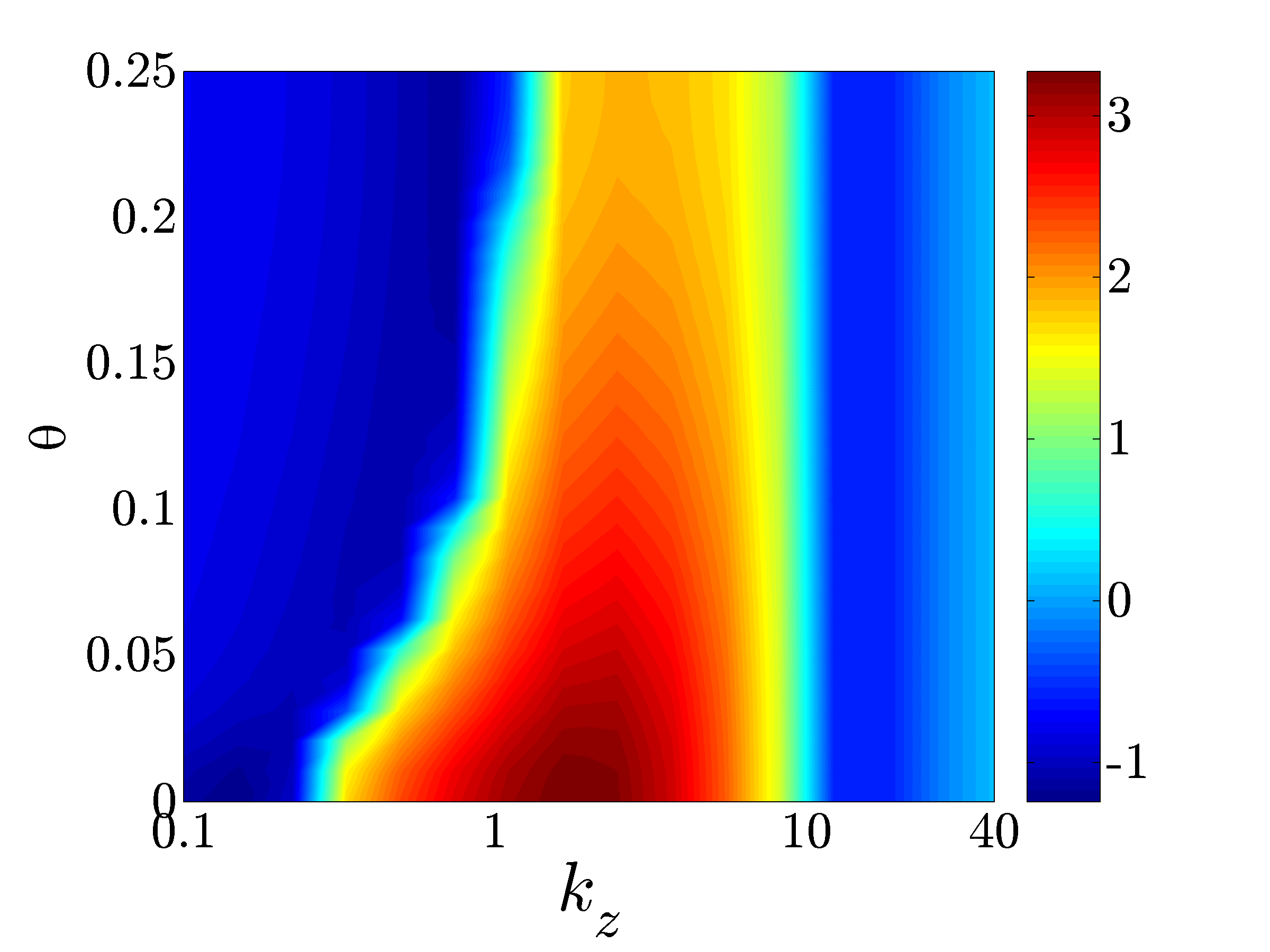}
    \label{fig.log10-E_2-vs-kz-vs-theta-cm2-w0p5-R2000}}
    &
    \subfigure[$(c = 3$, $\omega_x = 1.5)$]
    {
    \includegraphics[height=1.9in,width=2.5in]
    {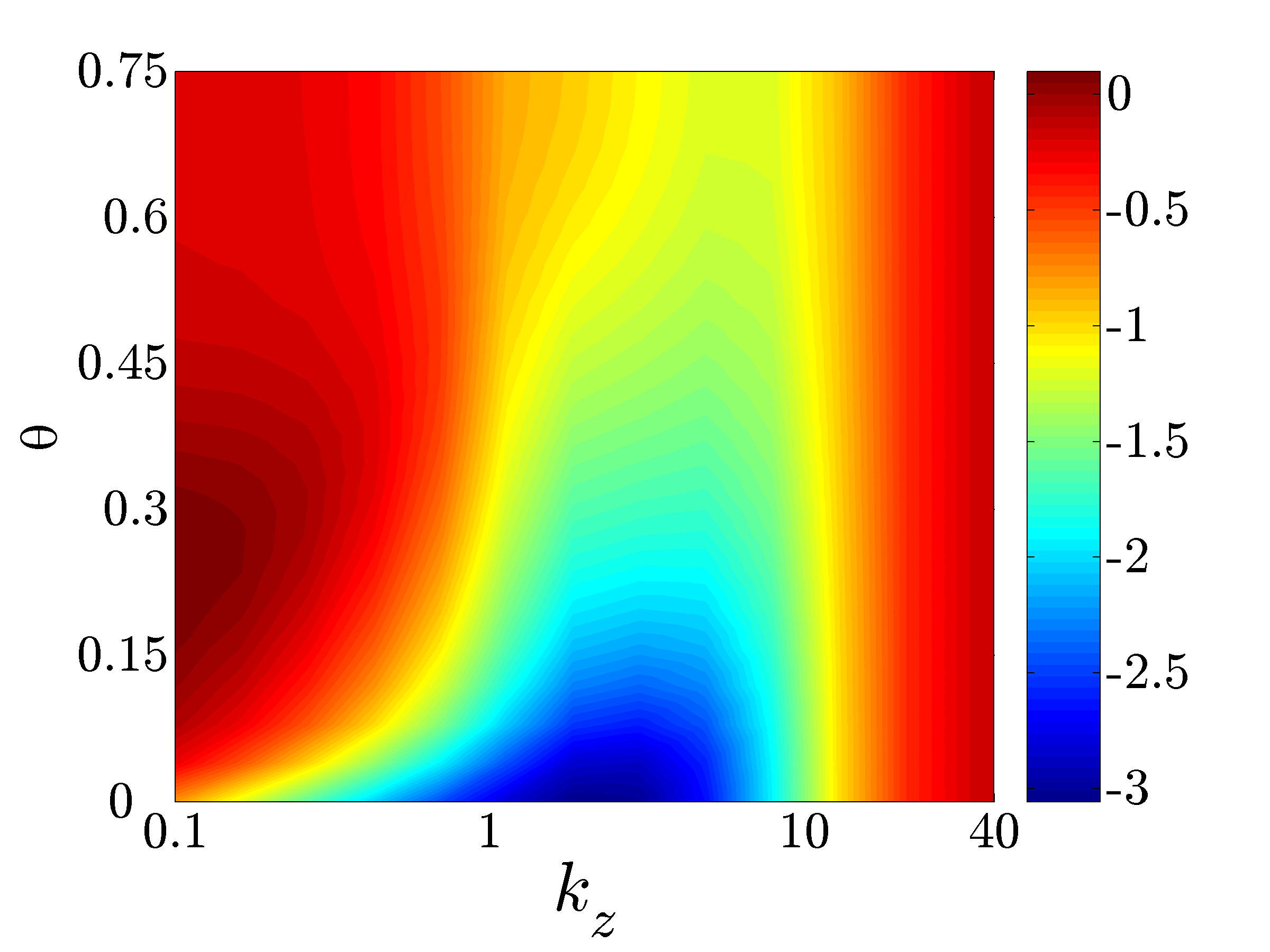}
    \label{fig.log10-E_2-vs-kz-vs-theta-c3-w1p5-R2000}}
    \\[-0.2cm]
    \subfigure[$(c = 5$, $\omega_x = 0.5)$]
    {
    \includegraphics[height=1.9in,width=2.5in]
    {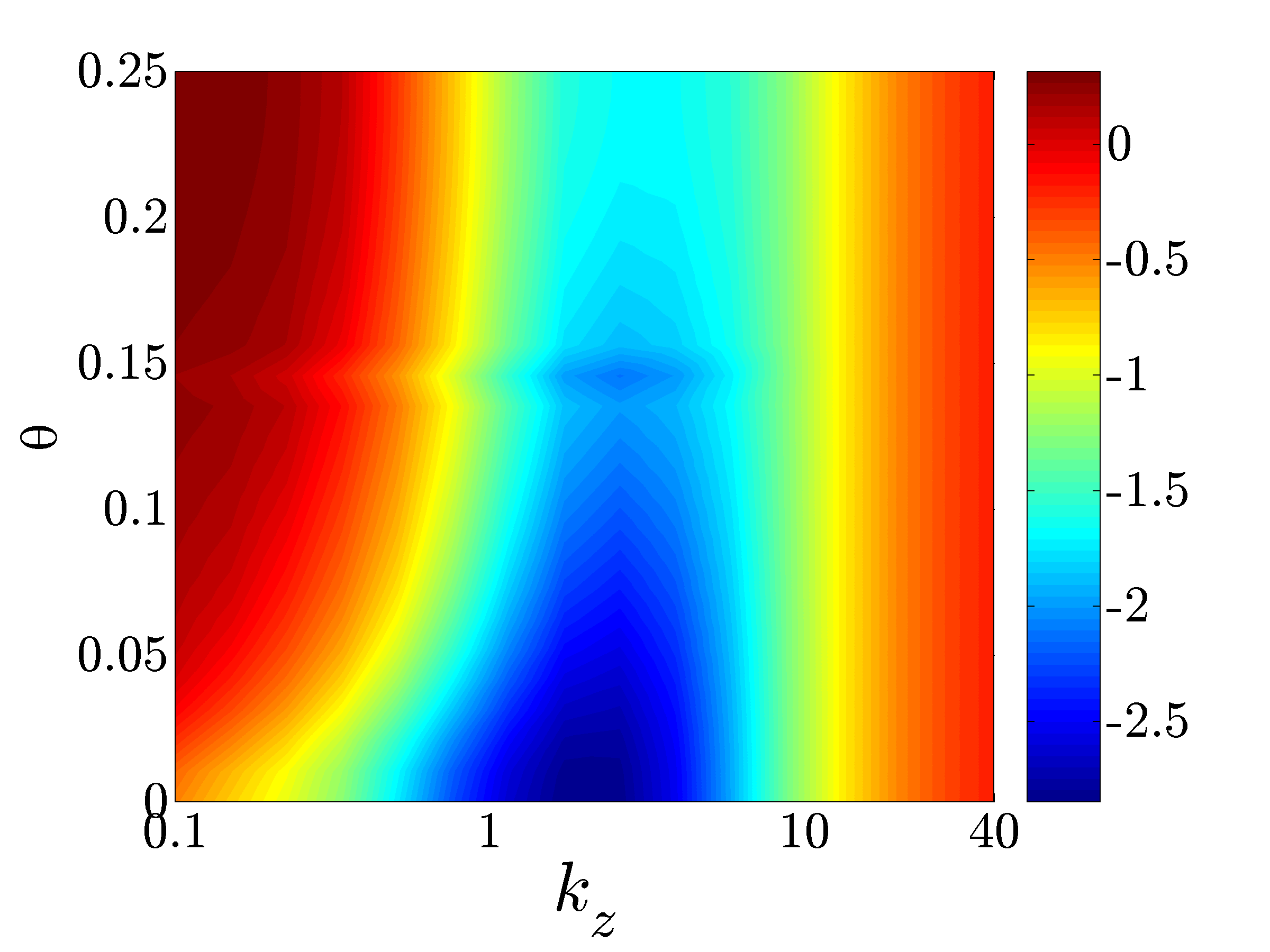}
    \label{fig.log10-E_2-vs-kz-vs-theta-c5-w0p5-R2000}}
    &
    \subfigure[$(c = 5$, $\omega_x = 2)$]
    {
    \includegraphics[height=1.9in,width=2.5in]
    {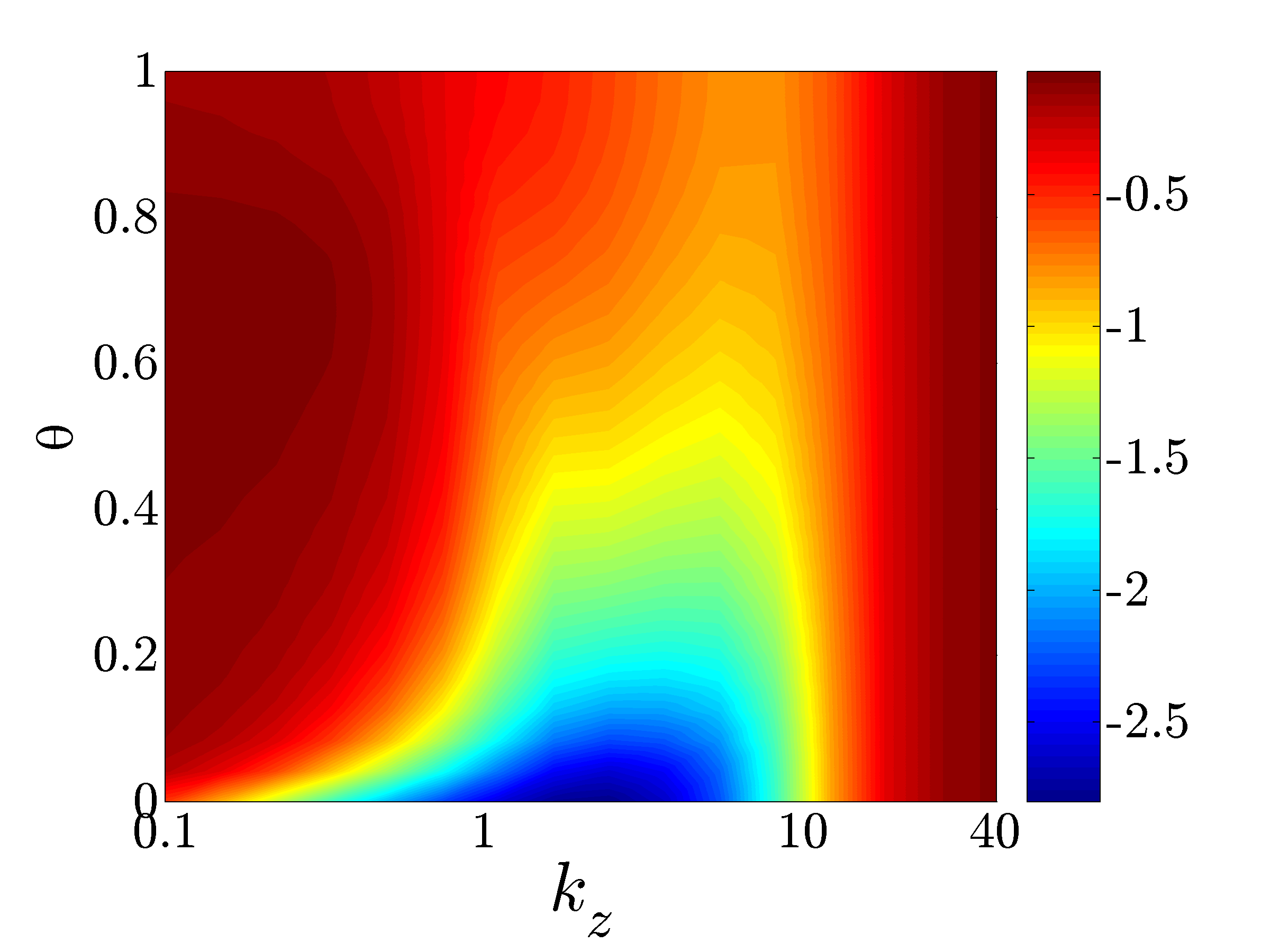}
    \label{fig.log10-E_2-vs-kz-vs-theta-c5-w2-R2000}}
    \end{tabular}
    \end{center}
    \caption{
    Second order correction to the energy amplification, $\hat{g}_2(\theta,k_z)$, for traveling waves with
    (a) $(c = -2$, $\omega_x = 0.5)$;
    (b) $(c = 3$, $\omega_x = 1.5)$;
    (c) $(c = 5$, $\omega_x = 0.5)$; and
    (d) $(c = 5$, $\omega_x = 2)$ in Poiseuille flow with $R_c = 2000$.
    }
    \label{fig.log10-E2-vs-kz-vs-theta}
    \end{figure}

The above analysis illustrates the ability of the DTWs to weaken the intensity of the most energetic modes of the uncontrolled flow; this is achieved by reducing receptivity to stochastic disturbances. However, an important aspect in the evaluation of any control strategy is to consider the influence of controls on all of the system's modes. In view of this, we next discuss how control affects the full three dimensional fluctuations. Since for a given $\omega_x$ the energy amplification is symmetric around $\theta = \omega_x/2$, it suffices to only consider the modes with $\theta \in [0, \omega_x/2]$. Figure~\ref{fig.log10-E2-vs-kz-vs-theta} shows $\hat{g}_2(\theta,k_z)$ for a UTW with $(c = -2$, $\omega_x = 0.5)$, and three DTWs with $(c = 3$, $\omega_x = 1.5)$, $(c = 5$, $\omega_x = 0.5)$, and $(c = 5$, $\omega_x = 2)$. As evident from figure~\ref{fig.log10-E2-Preq-cpm}, the selected UTW increases amplification of the fundamental mode with $k_z = 1.78$; on the other hand, all three DTWs reduce energy amplification of modes with $(\theta = 0$, $k_z = 1.78)$. Figure~\ref{fig.log10-E2-vs-kz-vs-theta} further reveals that the largest change in amplification for all of these traveling waves takes place at $(\theta = 0$, $k_z \approx 1.78)$, which is precisely where the uncontrolled flow contains most energy. This observation suggests presence of resonant interactions between the traveling waves and the most energetic modes of the uncontrolled flow. Additionally, as can be seen from figures~\ref{fig.log10-E_2-vs-kz-vs-theta-cm2-w0p5-R2000} and~\ref{fig.log10-E_2-vs-kz-vs-theta-c5-w2-R2000}, the energy of modes with $k_z \approx 0$ is reduced by a UTW with $(c = -2$, $\omega_x = 0.5)$ and a DTW with $(c = 5$, $\omega_x = 2)$ for all $\theta$. On the other hand, figure~\ref{fig.log10-E_2-vs-kz-vs-theta-c3-w1p5-R2000} shows that a DTW with $(c = 3$, $\omega_x = 1.5)$ increases amplification of fluctuations with $(0.1 \lesssim \theta \lesssim 0.4$, $k_z \approx 0)$; similarly, receptivity of fluctuations with $(0.05 \lesssim \theta \lesssim 0.45$, $k_z \approx 0)$ is increased by a DTW with $(c = 2$, $\omega_x = 0.5)$ (cf.\ figure~\ref{fig.log10-E_2-vs-kz-vs-theta-c5-w0p5-R2000}). Thus, from the four considered cases, only a DTW with $(c = 5$, $\omega_x = 0.5)$ can be used to inhibit intensity of full three dimensional velocity fluctuations (i.e., for all values of $\theta$ and $k_z$).

\begin{figure}
    \begin{center}
    \begin{tabular}{cc}
    {\sc upstream:}
    &
    {\sc downstream:}
    \\[-0.18cm]
    \subfigure[]
    {
    \includegraphics[height=1.9in,width=2.5in]
    {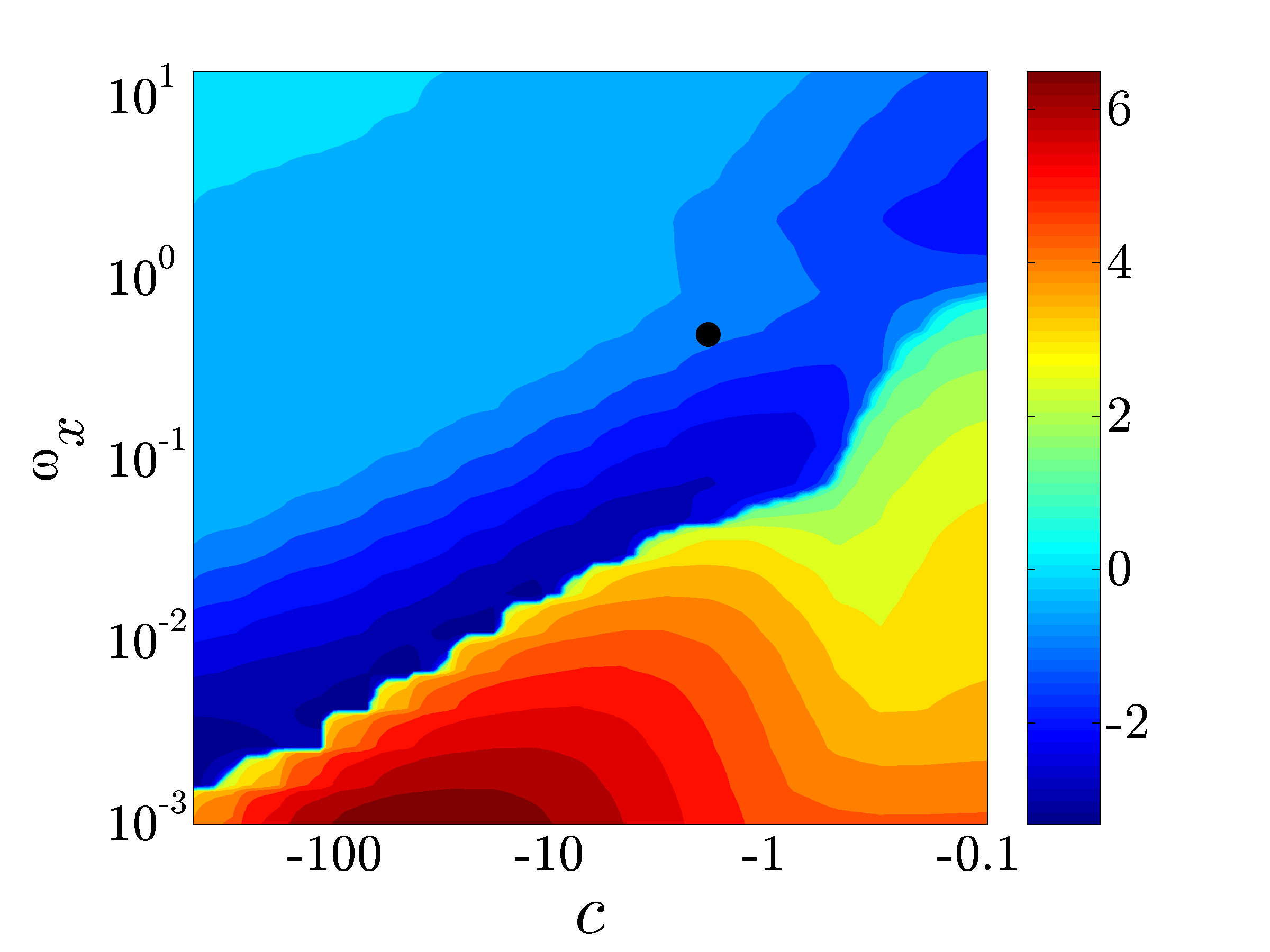}
    \label{fig.log10-E_2-corbc-vs-cm500tomp1-vs-wm3to2-kx1p2-kz0-R2000}}
    &
    \subfigure[]
    {
    \includegraphics[height=1.9in,width=2.5in]
    {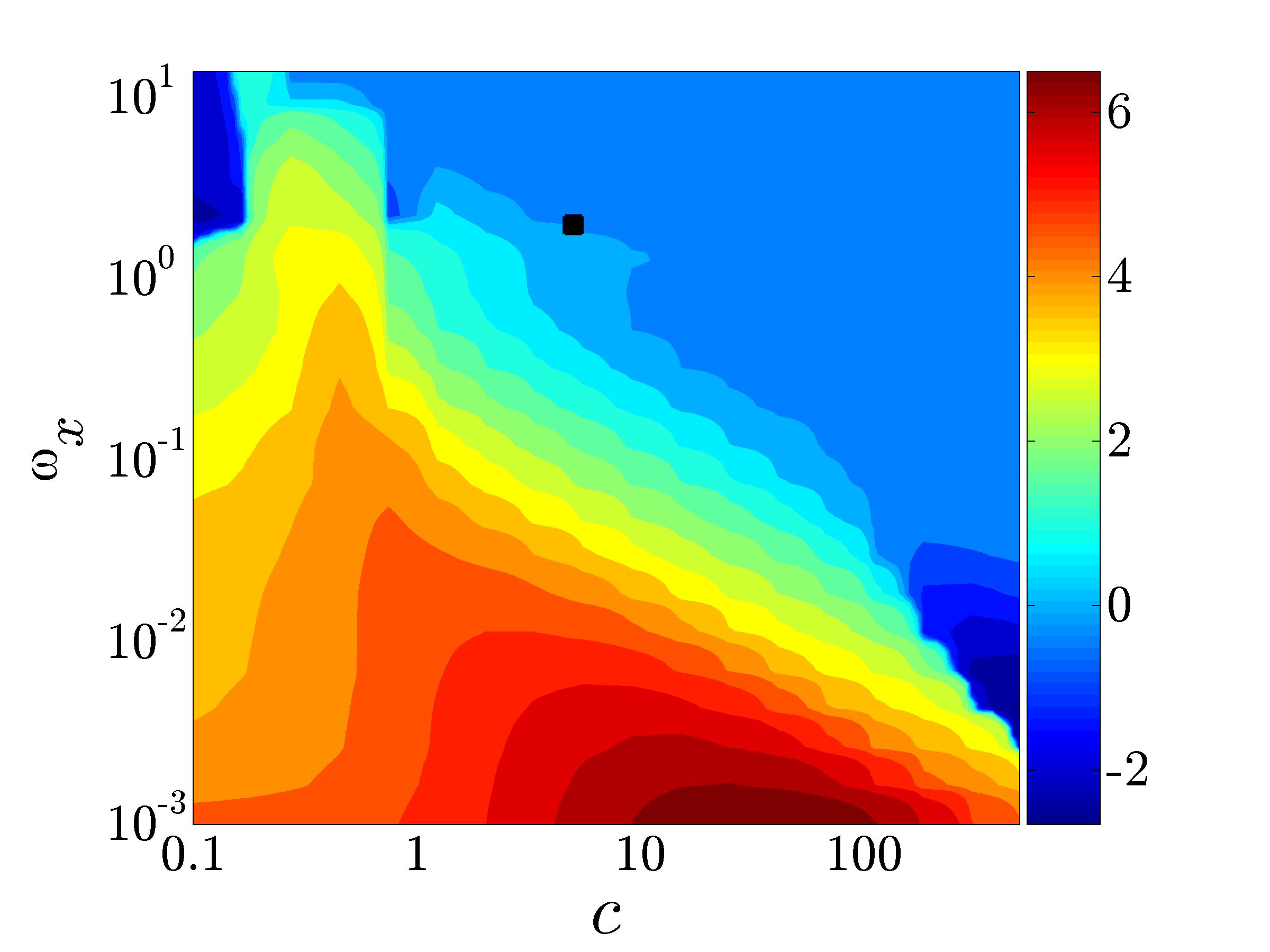}
    \label{fig.log10-E_2-corbc-vs-cpp1top500-vs-wm3to2-kx1p2-kz0-R2000}}
    \end{tabular}
    \end{center}
    \caption{
    Second order correction to the energy amplification, $\hat{g}_2 (c,\omega_x)$, of the modes with $(\theta(\omega_x),k_z) = (1.2 - \omega_x \lfloor \frac{1.2}{\omega_x} \rfloor, 0)$, in the presence of (a) UTWs; and (b) DTWs in Poiseuille flow with $R_c = 2000$.
    The dot and the square, respectively, denote $( c = -2,$ $\omega_x = 0.5 )$ and $( c = 5,$ $\omega_x = 2 )$.
    }
    \label{fig.log10-E2-stab-c-w}
    \end{figure}

\begin{figure}
    \begin{center}
    \begin{tabular}{cc}
    {\sc upstream:}
    &
    {\sc downstream:}
    \\[-0.18cm]
    \subfigure[]
    {
    \includegraphics[height=1.9in,width=2.5in]
    {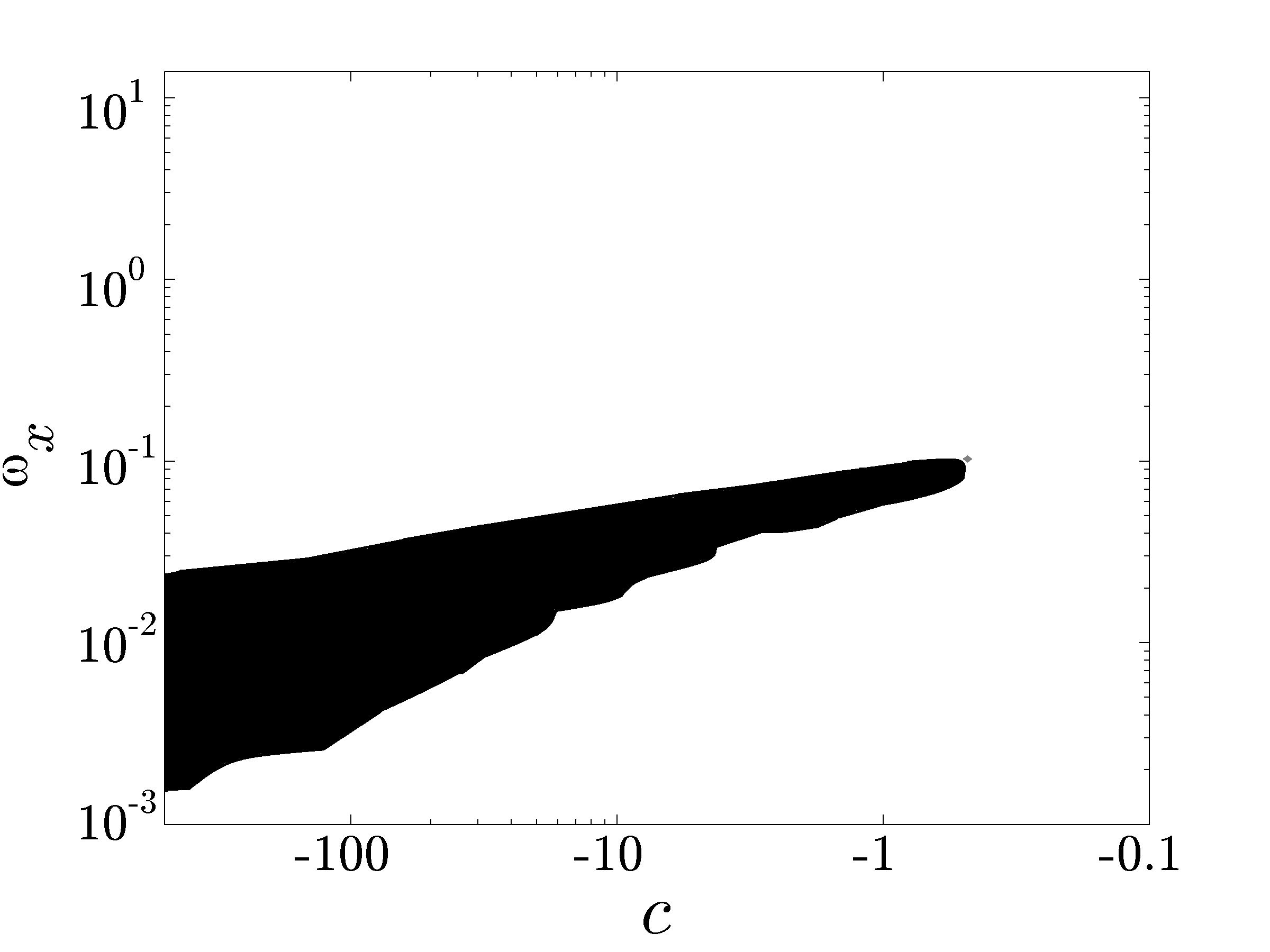}
    \label{fig.log10-blueregion-stab-ener-corbc-vs-cm500tomp1-vs-wm3to2-R2000}}
    &
    \subfigure[]
    {
    \includegraphics[height=1.9in,width=2.5in]
    {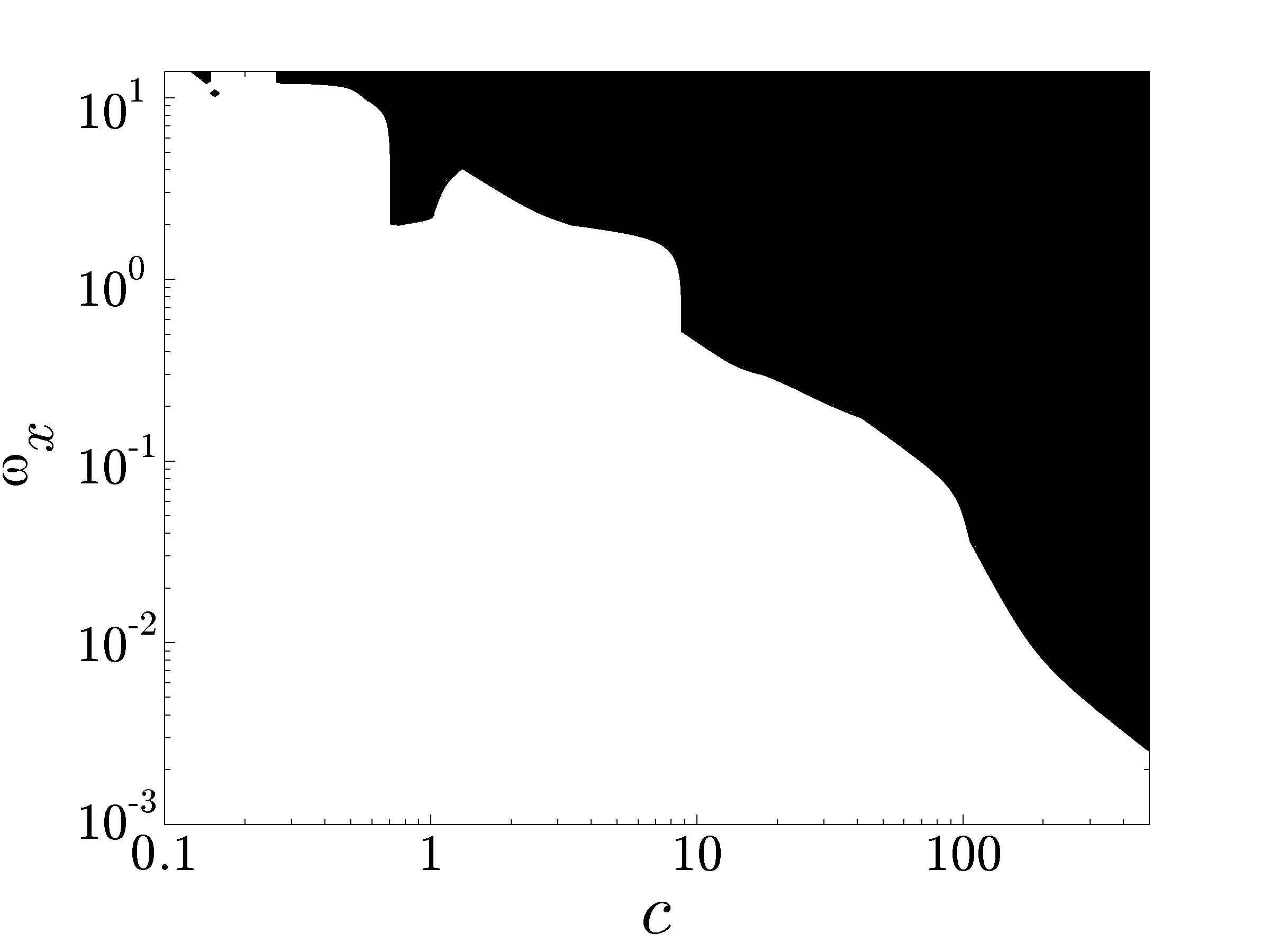}
    \label{fig.log10-blueregion-stab-ener-corbc-vs-cpp1top500-vs-wm3to2-R2000}}
    \end{tabular}
    \end{center}
    \caption{
    The dark regions identify values of wave speed and frequency that, up to a second order in $\alpha$, suppress the energy amplification of both most energetic and least stable modes in Poiseuille flow with $R_c = 2000$ subject to: (a) UTWs; and (b) DTWs.
    }
    \label{fig.log10-E2-ener-stab-c-w}
    \end{figure}

While the fundamental mode is most influential in determining the effect of control on the energy amplification, figures~\ref{fig.log10-E_2-vs-kz-vs-theta-c3-w1p5-R2000} and~\ref{fig.log10-E_2-vs-kz-vs-theta-c5-w0p5-R2000} indicate that the modes with $\theta \neq 0$ and large spanwise wavelengths (i.e., $k_z \approx 0$) can be significantly amplified by the traveling waves. We thus take a closer look at how control affects the spanwise constant fluctuations. The TS waves are characterized by $(k_x = 1.2$, $k_z = 0)$ and, for a given $\omega_x$, they are imbedded in the modes of the controlled flow for fluctuations with $\theta (\omega_x) = 1.2 - \omega_x \lfloor 1.2/\omega_x \rfloor$. Figure~\ref{fig.log10-E2-stab-c-w} shows the second order correction $\hat{g}_2(c,\omega_x)$ to the energy amplification of the modes with $k_z = 0$ subject to both UTWs and DTWs. Note that figure~\ref{fig.log10-E2-stab-c-w} correctly captures the increased intensity of the TS waves by DTWs with $(c = 3$, $\omega_x = 1.5)$ and $(c = 5$, $\omega_x = 0.5)$, as already observed in figures~\ref{fig.log10-E_2-vs-kz-vs-theta-c3-w1p5-R2000} and~\ref{fig.log10-E_2-vs-kz-vs-theta-c5-w0p5-R2000}. We also see that the traveling waves considered in~\cite{minsunspekim06} reduce energy of the TS waves (we recall that these promote amplification of the streamwise streaks; cf.\ figures~\ref{fig.log10-E_2-corbc-vs-cm500tomp1-vs-wm3to2-theta0-kz1p78-R2000} and~\ref{fig.log10-E_2-corbc-vs-cm500tomp1-vs-wm3to2-kx1p2-kz0-R2000}). On the other hand, the DTW with $c = 5$ and $\omega_x = 2$ decreases energy amplification of both streamwise streaks and TS waves (cf.\ figures~\ref{fig.log10-E_2-corbc-vs-cpp1top500-vs-wm3to2-theta0-kz1p78-R2000} and~\ref{fig.log10-E_2-corbc-vs-cpp1top500-vs-wm3to2-kx1p2-kz0-R2000}). The values of $c$ and $\omega_x$ capable of reducing the energy amplification (up to a second order in $\alpha$) of both most energetic and least stable modes of the uncontrolled flow are marked by the dark region in figure~\ref{fig.log10-E2-ener-stab-c-w}.


\subsection{Effect of control amplitude on energy amplification}
    \label{sec.alpha}

We next discuss influence of control amplitude on the energy amplification. We show that perturbation analysis (up to a second order in $\alpha$) correctly predicts the essential trends. This is done by comparing perturbation analysis results with computations obtained using large-scale truncation of the operators in Lyapunov equation~(\ref{eq.LE}).

\begin{figure}
    \begin{center}
    {
    \includegraphics[height=1.9in]
    {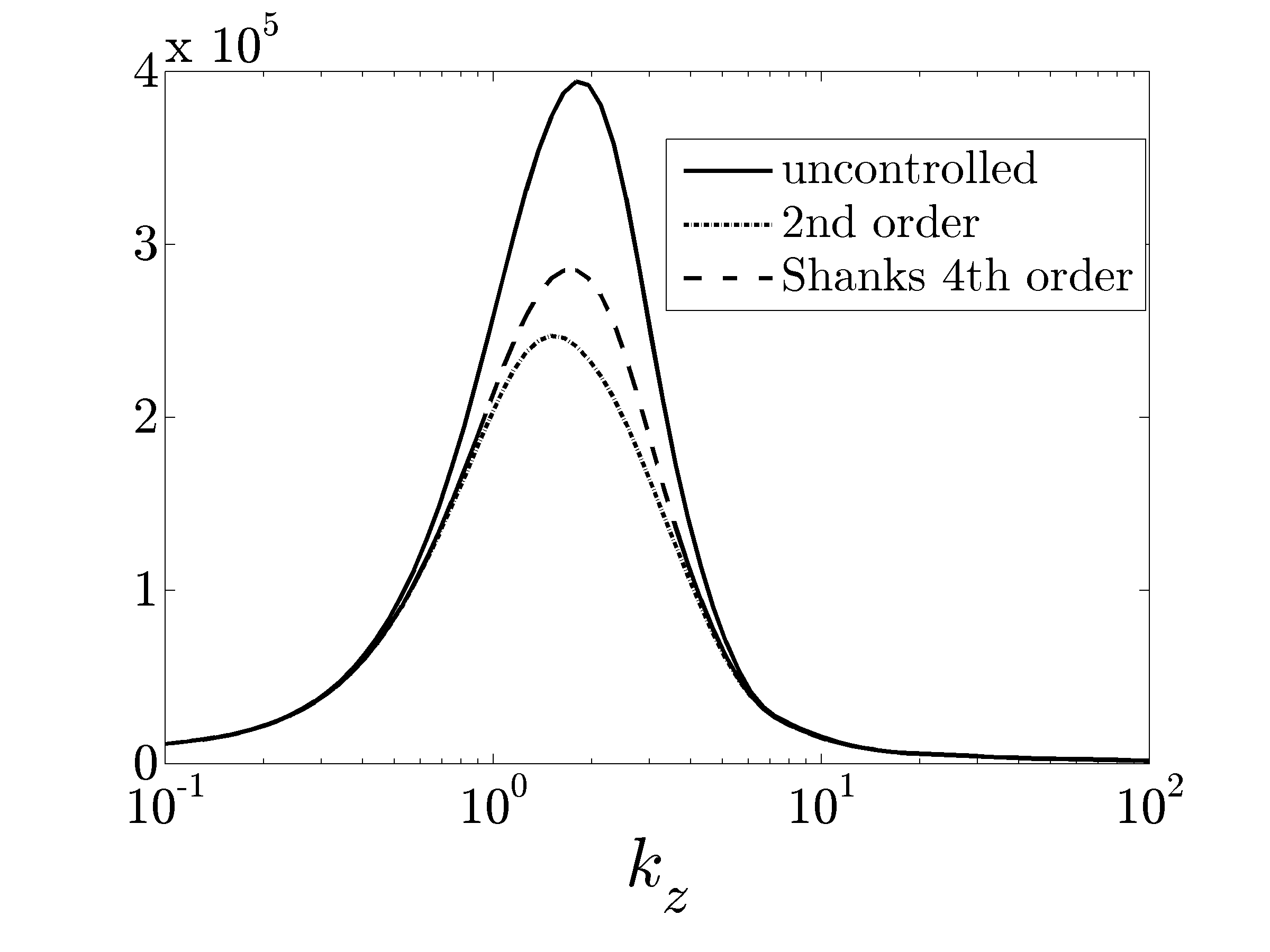}
    }
    \end{center}
    \caption{
    Energy density, $\bmrE (k_z)$, of the fundamental mode $\theta = 0$ in Poiseuille flow with $R_c = 2000$ and $(c = 5$, $\omega_x = 2$, $\alpha = 0.025)$. The controlled flow results are obtained using perturbation analysis up to a second order in $\alpha$, and Shanks transformation up to a fourth order in $\alpha$.
    }
    \label{fig.E0-E2-Eshanks-alpha50-w2-cp5}
    \end{figure}

The limit of the perturbation series~(\ref{eq.ED}) can be obtained by applying Shanks transformation~\citep{shanks55,vandyke64} on the perturbation-analysis-based correction coefficients in~(\ref{eq.ED}). This transformation represents an effective means for providing convergence (respectively, faster convergence) to a divergent (respectively, slowly convergent) series~\citep{sidi03}. It turns out that Shanks transformation significantly increases the maximum value of $\alpha$ for which series~(\ref{eq.ED}) converges. Figure~\ref{fig.E0-E2-Eshanks-alpha50-w2-cp5} shows the energy density of the fundamental mode, as a function of $k_z$, in the uncontrolled Poiseuille flow with $R_c = 2000$ and a flow subject to a DTW with $(c = 5$, $\omega_x = 2$, $\alpha = 0.025)$. The controlled flow results are obtained using truncation of series~(\ref{eq.ED}) up to a second order in $\alpha$, and Shanks transformation up to a fourth order in $\alpha$. Note that even though the second order correction overestimates the amount of receptivity reduction, it correctly captures the essential trends.

    \begin{figure}
    \begin{center}
    \begin{tabular}{cc}
    {\sc upstream:}
    &
    {\sc downstream:}
    \\[-0.18cm]
    \subfigure[]
    {
    \includegraphics[height=1.9in,width=2.5in]
    {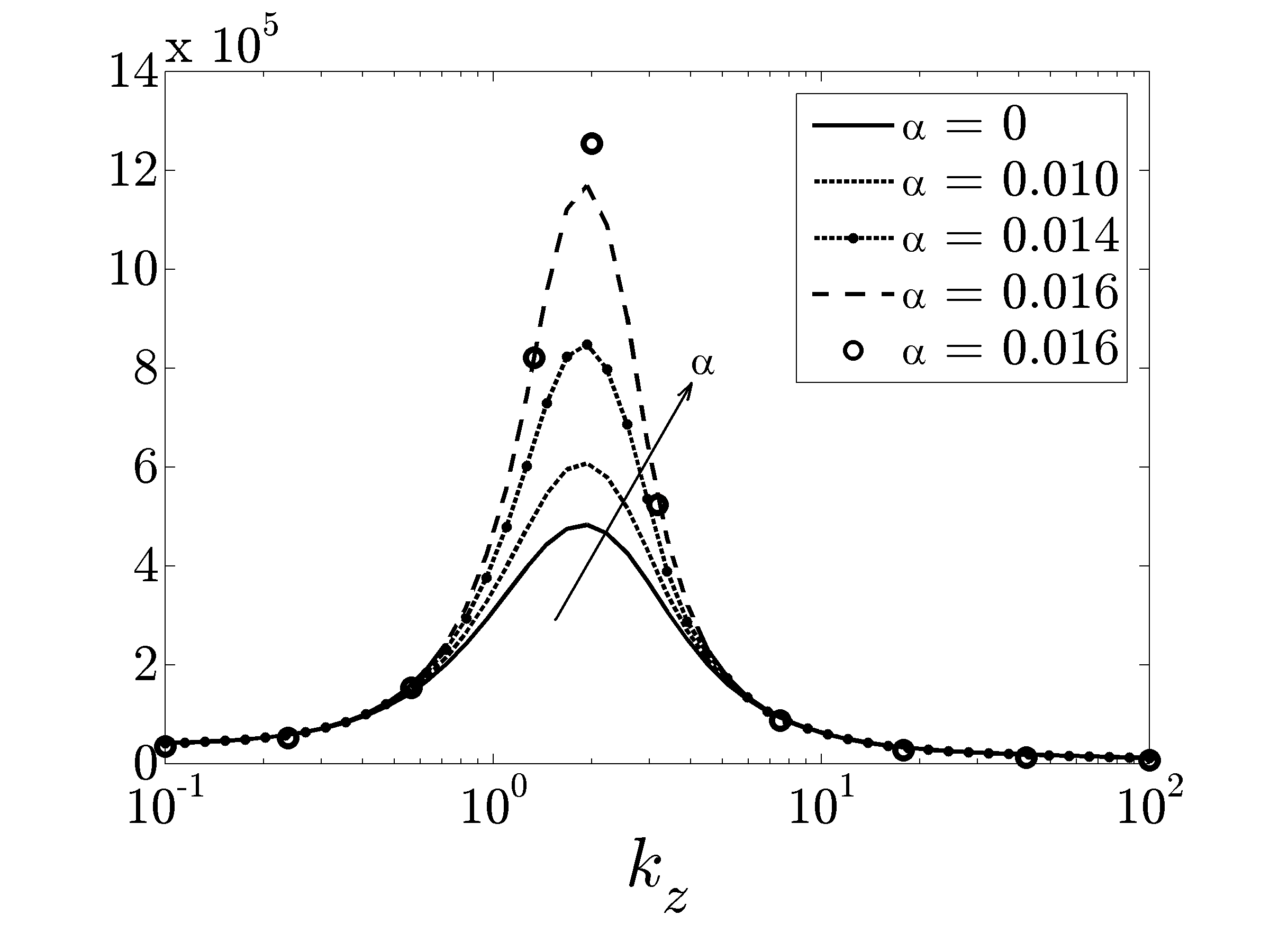}
    \label{fig.Esum-alpha0p01-0p014-0p016-Eshanks4-w0p5-cm2}}
    &
    \subfigure[]
    {
    \includegraphics[height=1.9in,width=2.5in]
    {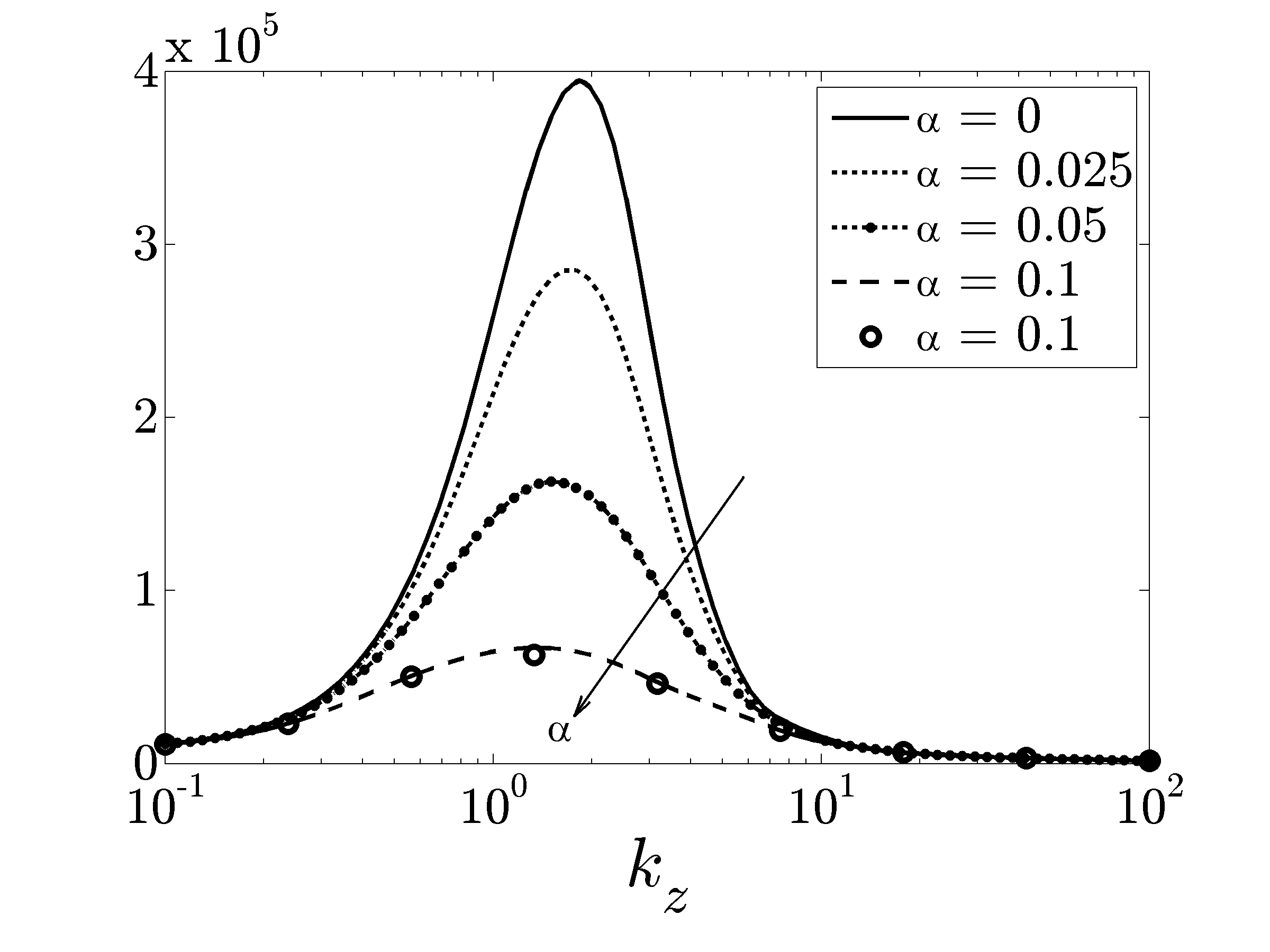}
    \label{fig.Esum-alpha0p025-0p05-0p1-Eshanks4-w2-cp5}}
    \end{tabular}
    \end{center}
    \caption{
    Energy density, $\bmrE (k_z)$, of the fundamental mode $\theta = 0$ in Poiseuille flow with $R_c = 2000$ subject to:
    (a) a UTW with $c = -2$ and $\omega_x = 0.5$;
    and
    (b) a DTW with $c = 5$ and $\omega_x = 2$.
    Shanks transformation up to a fourth order in $\alpha$ is used in computations. The truncation results (hollow circles) are obtained for (a) $\alpha = 0.016$; and (b) $\alpha = 0.1$.
    }
    \label{fig.Esum-alpha}
    \end{figure}

    \begin{figure}
    \begin{center}
    \includegraphics[height=1.9in]
    {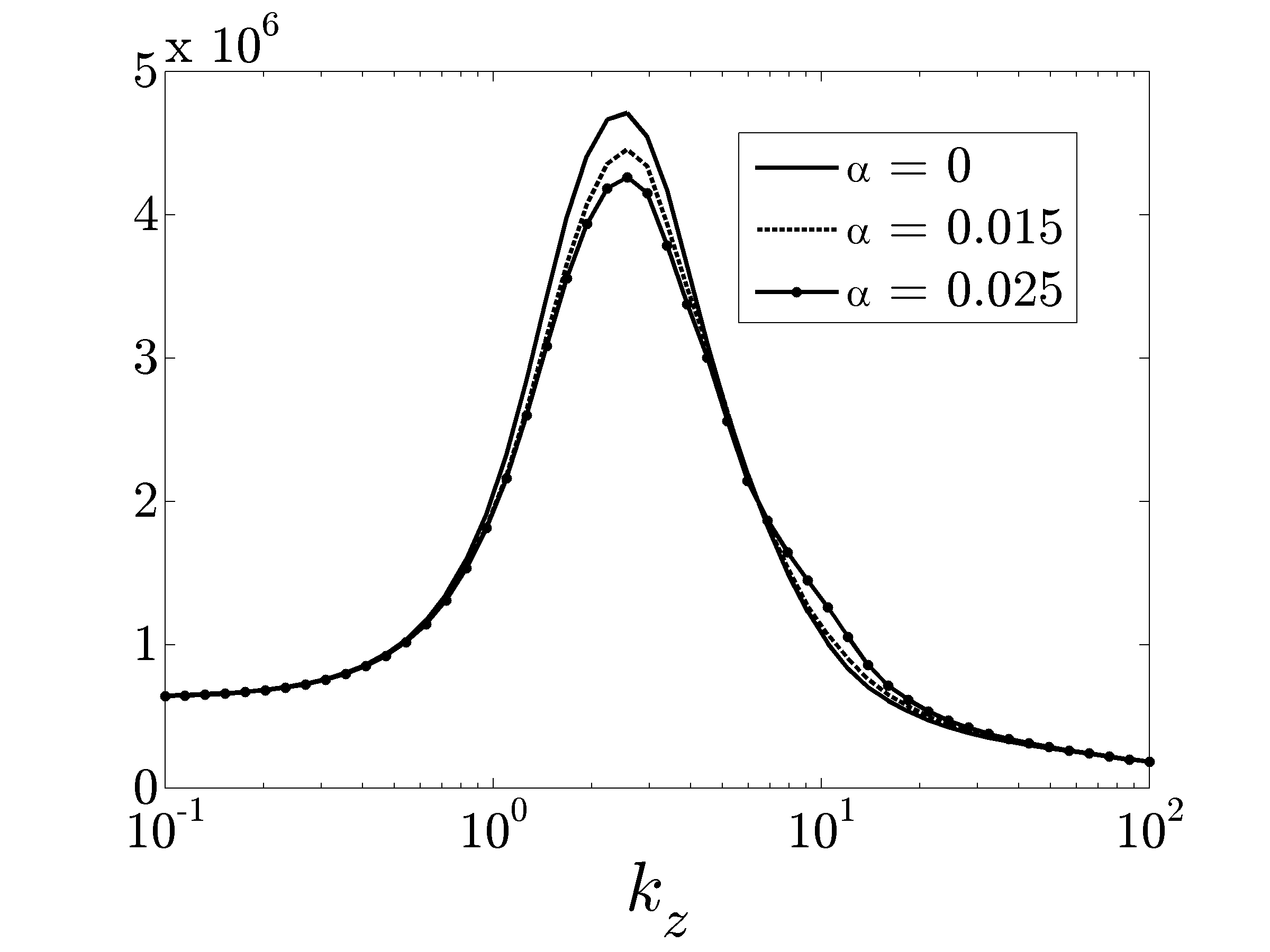}
    \end{center}
    \caption{
    Energy density, $\bmrE (k_z)$, of the fundamental mode $\theta = 0$ in Poiseuille flow with $R_c = 2000$ subject to a UTW with $c = -5$ and $\omega_x = 0.03$. Shanks transformation up to a fourth order in $\alpha$ is used in computations.
    }
    \label{fig.Esum-alpha0p015-0p025-0p1-Eshanks4-w0p03-cm5}
    \end{figure}

Figure~\ref{fig.Esum-alpha} compares energy density of the fundamental mode in uncontrolled Poiseuille flow with $R_c = 2000$, and in the controlled flows subject to: (a) a UTW with $c = -2$ and $\omega_x = 0.5$, figure~\ref{fig.Esum-alpha0p01-0p014-0p016-Eshanks4-w0p5-cm2}; and (b) a DTW with $c = 5$ and $\omega_x = 2$, figure~\ref{fig.Esum-alpha0p025-0p05-0p1-Eshanks4-w2-cp5}. The controlled flow results are obtained using Shanks transformation up to a fourth order in $\alpha$, and they closely match the large-scale truncation results (hollow circles). Figure~\ref{fig.Esum-alpha0p025-0p05-0p1-Eshanks4-w2-cp5} shows that the properly designed DTWs with amplitudes equal to $5 \, \%$, $10 \, \%$, and $20 \, \%$ of the base centerline velocity reduce the largest energy density of the uncontrolled flow by approximately $28 \, \%$, $60 \, \%$, and $80 \, \%$, respectively. It is noteworthy that substantial reduction is obtained at the expense of relatively small increase (compared to the laminar flow) in the nominal drag coefficient, which approximately increases by $1 \, \%$, $4 \, \%$, and $13 \, \%$. Further increase in the amplitude of a DTW with $c = 5$ and $\omega_x = 2$ results even in larger receptivity reduction. Part~2 demonstrates that this approach can be successfully used for controlling the onset of turbulence in flows subject to large initial disturbances. However, the power required for maintaining laminar flow under these conditions is prohibitively large, which limits the advantage of using DTWs for transition control from efficiency point of view.

In contrast to DTWs, figure~\ref{fig.Esum-alpha0p01-0p014-0p016-Eshanks4-w0p5-cm2} demonstrates that the UTW with $c = -2$ and $\omega_x = 0.5$ increases receptivity. We note that all of these trends are correctly captured by the second order correction (in $\alpha$) to the energy amplification and that our results agree with the transient growth study of~\cite{leeminkim08}. Furthermore, large energy amplification of the UTWs may be thought of as a precursor to flow instability; namely, it turns out that the UTWs destabilize the flow for $\alpha > 0.03$ which is a smaller value compared to the amplitudes chosen in~\cite{minsunspekim06} ($\alpha = 0.05$ and $\alpha = 0.125$, respectively).

As described in \S~\ref{sec.E2}, figure~\ref{fig.log10-E_2-corbc-vs-cm500tomp1-vs-wm3to2-theta0-kz1p78-R2000} suggests that the UTWs with $\omega_x \lesssim 0.1$ can reduce the intensity of the most energetic modes of the uncontrolled flow. Here, we demonstrate that such UTWs lead to a very modest receptivity reduction. Figure~\ref{fig.Esum-alpha0p015-0p025-0p1-Eshanks4-w0p03-cm5} illustrates that a UTW with $(c = -5,$ $\omega_x = 0.03,$ $\alpha = 0.025)$ reduces energy amplification by about $8 \, \%$. On the other hand, modal stability analysis can be used to show that amplitudes as small as $\alpha \approx 0.03$ make the flow linearly unstable. Therefore, relative to flow with no control, the UTWs at best exhibit similar receptivity to disturbances.

For control amplitudes shown in figures~\ref{fig.Esum-alpha} and~\ref{fig.Esum-alpha0p015-0p025-0p1-Eshanks4-w0p03-cm5}, we have verified stability of fluctuations around base velocities in both UTWs and DTWs by computing the eigenvalues of the large-scale truncation of operator $\ca_\theta(k_z)$ in~(\ref{eq.FR}). Compared to solving the truncated version of Lyapunov equation~(\ref{eq.LE}), perturbation analysis in conjunction with Shanks transformation provides much more efficient way for determining energy amplification. For example, while it takes four days on a PC to obtain the truncated results (hollow circles) in figure~\ref{fig.Esum-alpha0p025-0p05-0p1-Eshanks4-w2-cp5}, the Shanks approximation is computed in four hours on the same PC. Moreover, once the correction coefficients in~(\ref{eq.ED}) have been determined, the energy amplification for a reasonably wide range of control amplitudes can be obtained at no further cost.

\subsection{Energy amplification mechanisms}
    \label{sec.mech-energy}

\begin{figure}
    \begin{center}
    \begin{tabular}{ccc}
    $+,\bmrE_{0,p}$; $\circ,\bmrE_{0,d}$; $-,\bmrE_{0}$:
    &
    $+,\bmrE_{2,p}$; $\circ,\bmrE_{2,d}$; $-,\bmrE_{2}$:
    &
    $-,\bmrE_{p},\bmrE_{d}$; $+,\bmrE_{0,p}$; $\circ,\bmrE_{0,d}$:
    \\[-0.1cm]
    \subfigure[{\small uncontrolled, $\omega_x = 2$}]
    {
    \includegraphics[height=1.6in,width=1.6in]
    {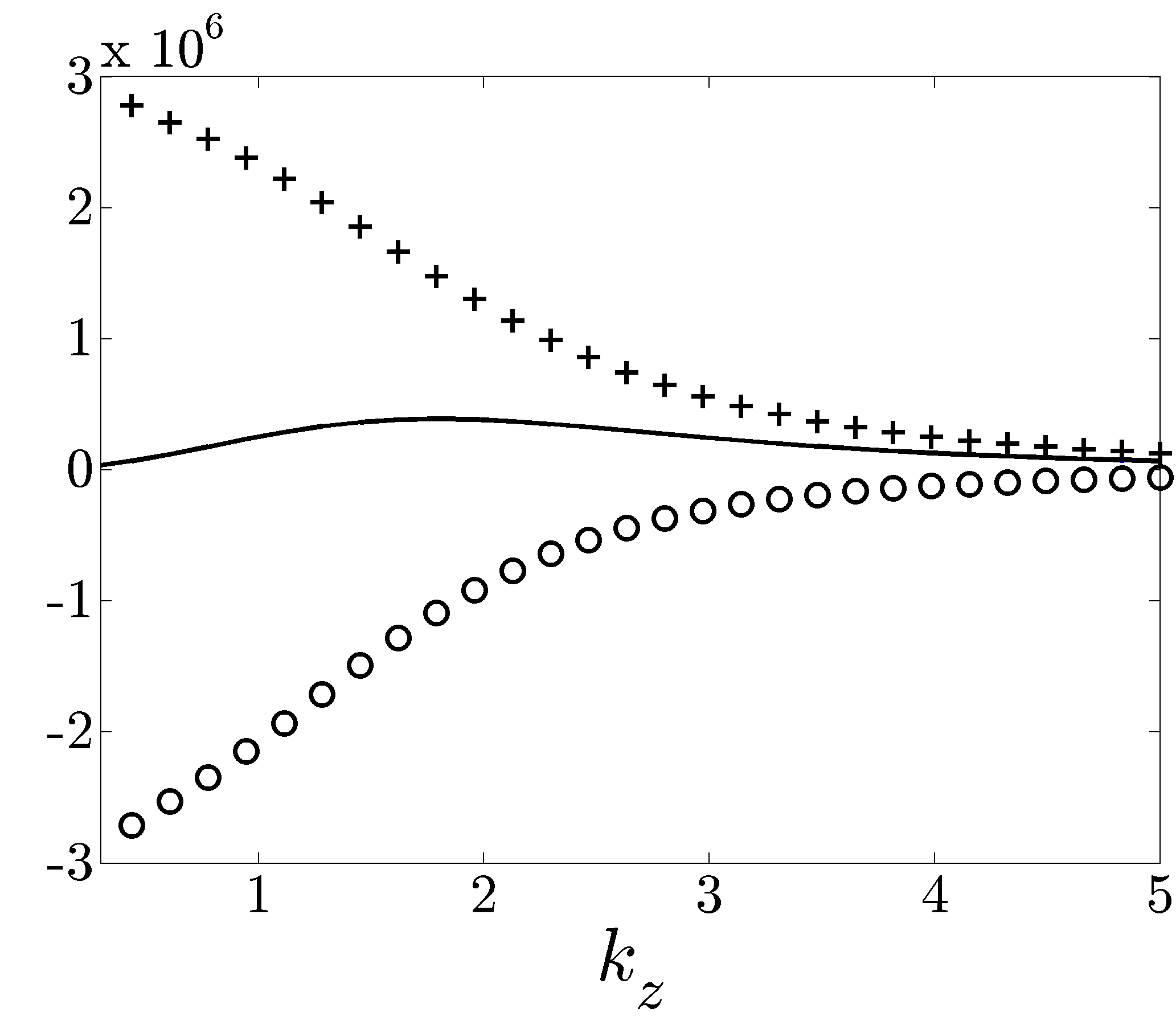}
    \label{fig.RO-c5-w2-kz-proddiff0}}
    &
    \subfigure[{\small DTW, $c = 5, \omega_x = 2$}]
    {
    \includegraphics[height=1.6in,width=1.6in]
    {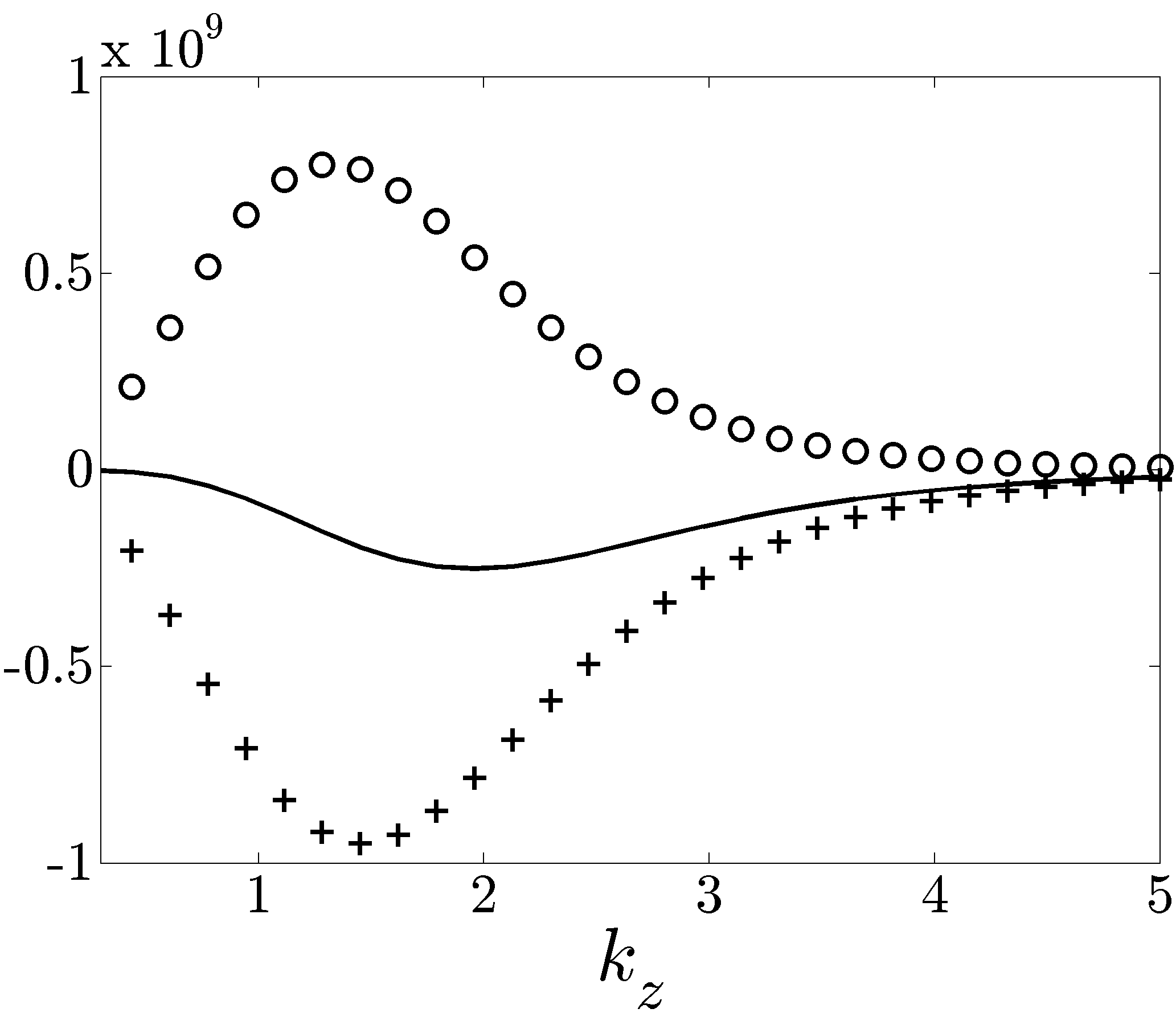}
    \label{fig.RO-c5-w2-kz-proddiff2}}
    &
    \subfigure[{\small DTW, $\alpha = 0.025$}]
    {
    \includegraphics[height=1.6in,width=1.6in]
    {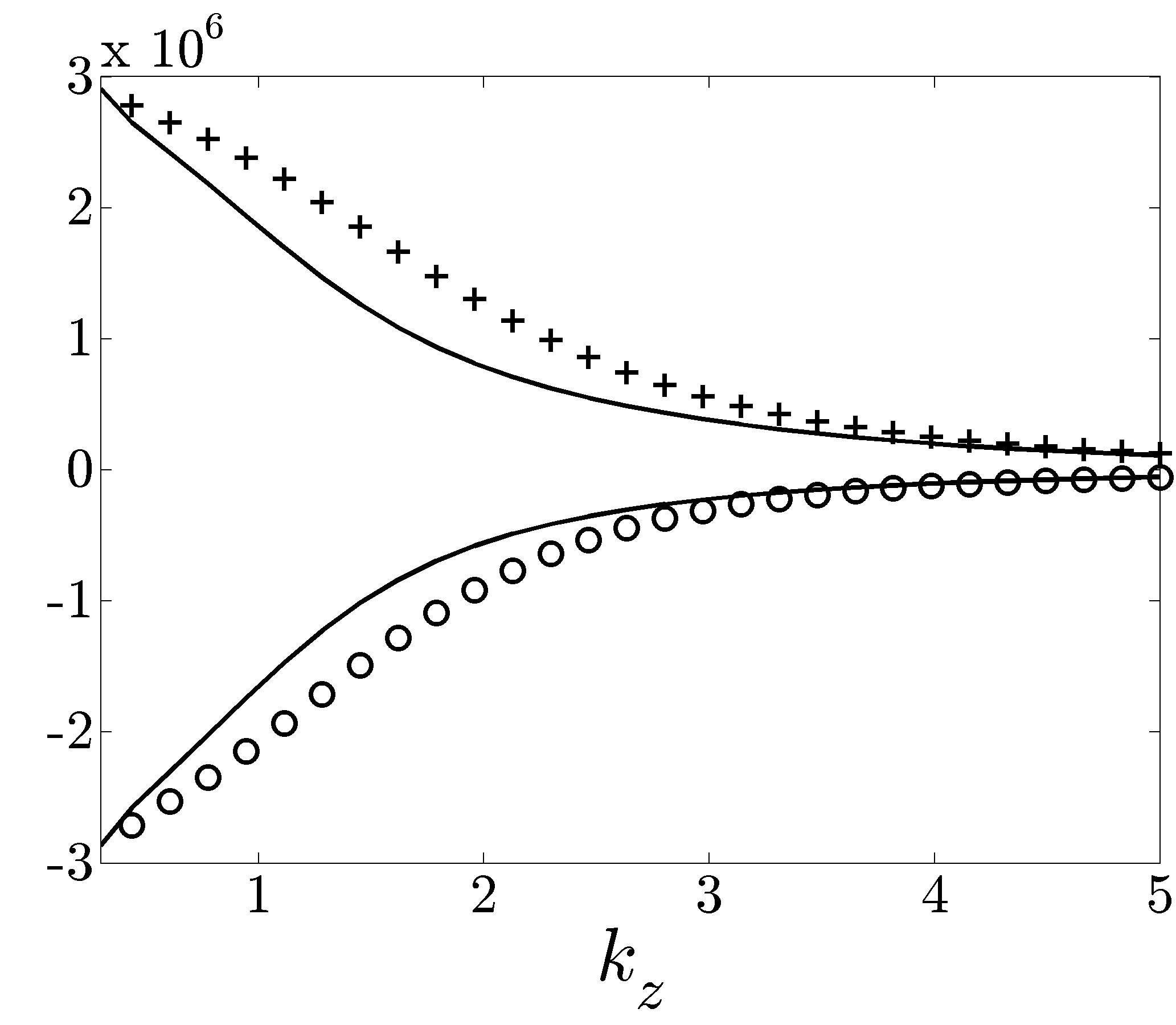}
    \label{fig.RO-c5-w2-alpha0p025-kz-proddiff}}
    \\[0.1cm]
    \subfigure[{\small uncontrolled, $\omega_x = 0.5$}]
    {
    \includegraphics[height=1.6in,width=1.6in]
    {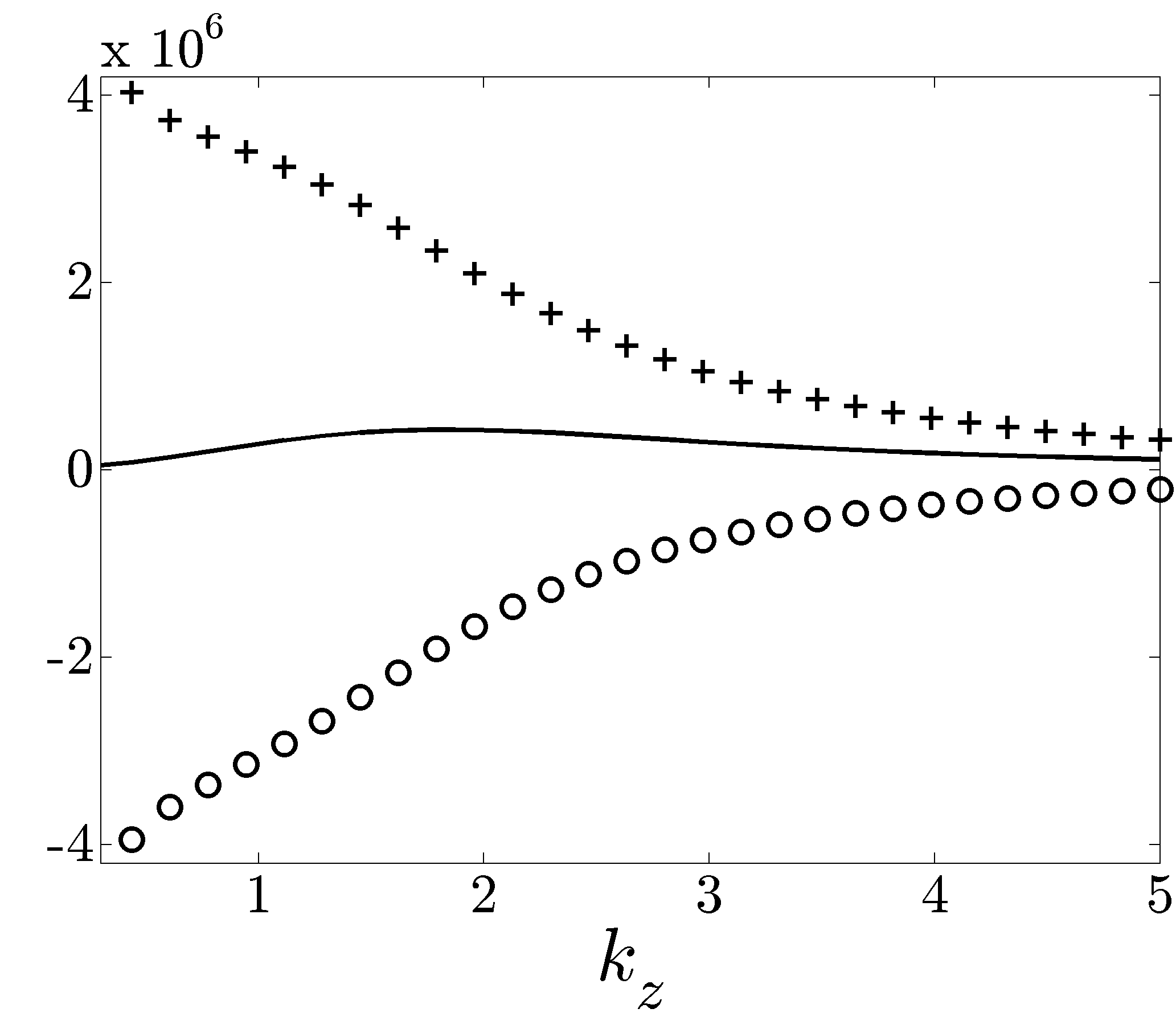}
    \label{fig.RO-cm2-w0p5-kz-proddiff0}}
    &
    \subfigure[{\small UTW, $c = -2, \omega_x = 0.5$}]
    {
    \includegraphics[height=1.6in,width=1.6in]
    {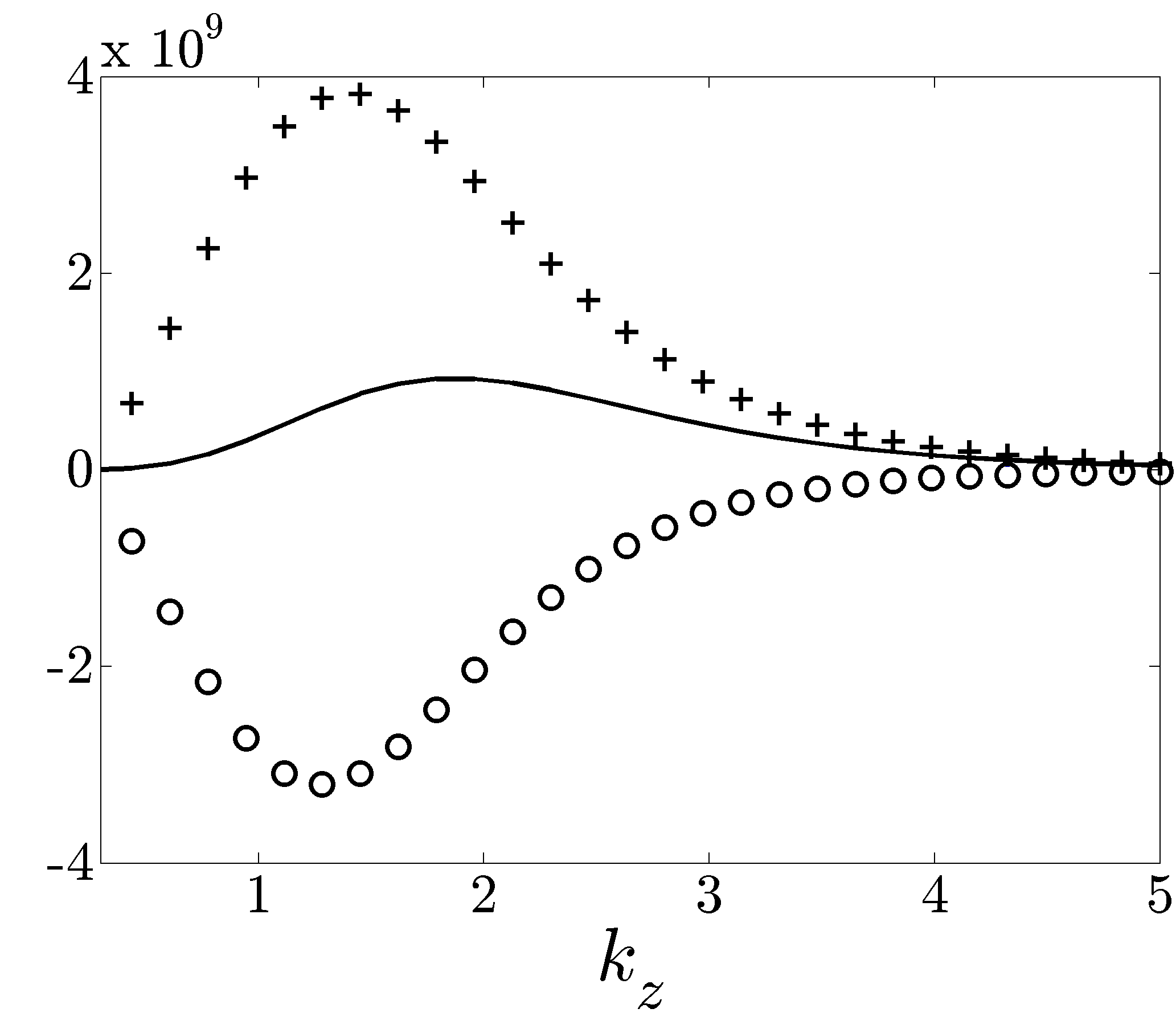}
    \label{fig.RO-cm2-w0p5-kz-proddiff2}}
    &
    \subfigure[{\small UTW, $\alpha = 0.015$}]
    {
    \includegraphics[height=1.6in,width=1.6in]
    {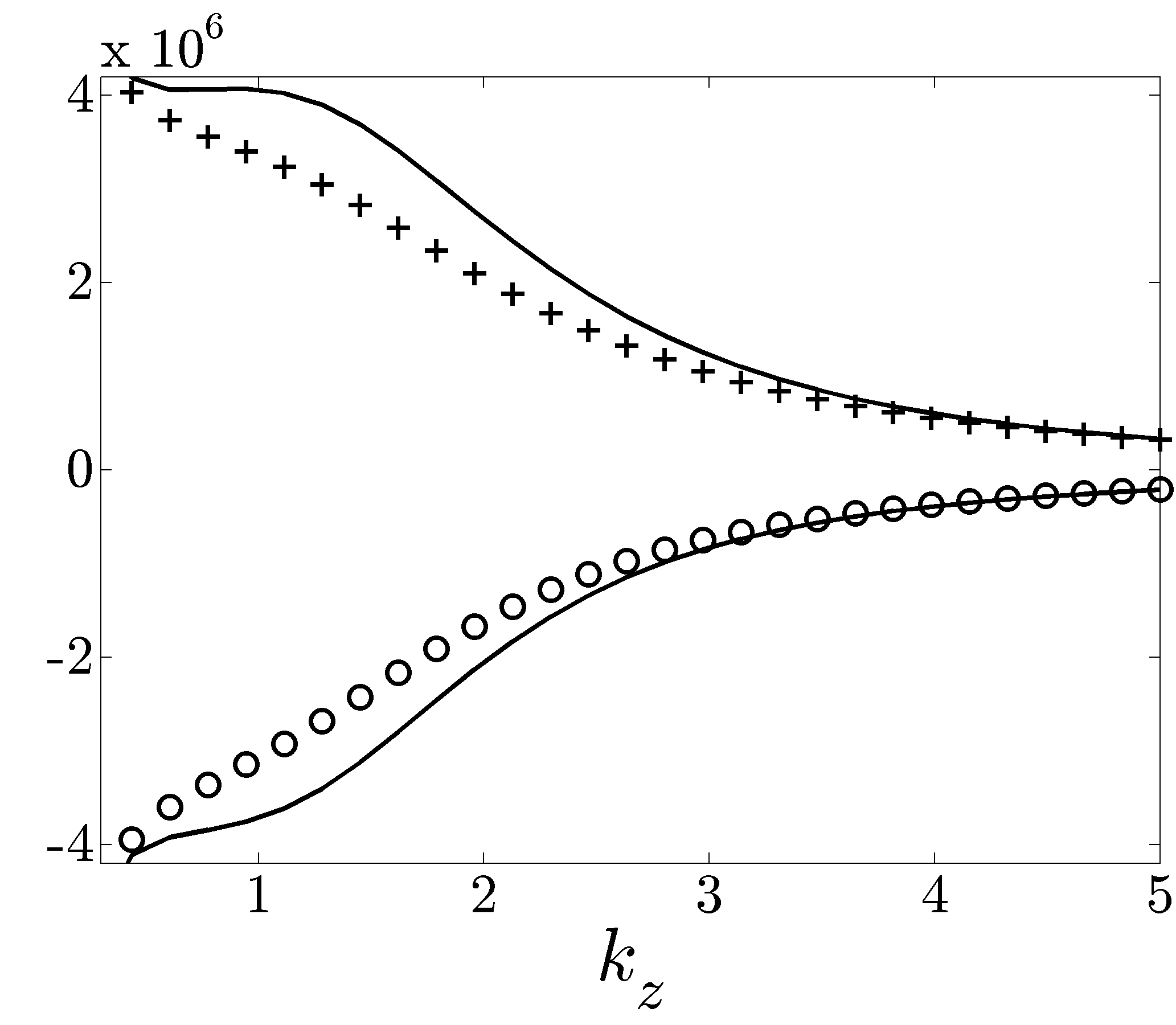}
    \label{fig.RO-cm2-w0p5-alpha0p015-kz-proddiff}}
    \end{tabular}
    \end{center}
    \caption{
    Contribution of production and dissipation terms to energy density of $\theta =  0$ mode in Poiseuille flow with $R_c = 2000$ subject to
    (a)-(c) a DTW with $(c = 5,$ $\omega_x = 2)$;
    and
    (d)-(f) a UTW with $(c = -2,$ $\omega_x = 0.5)$.
    (a,d) uncontrolled flow;
    (b,e) second order corrections;
    and (c,f) controlled flows.
    In (c,f), the controlled flow results are obtained using approximation up to a second order in $\alpha$, and the uncontrolled flow results are shown for comparison.
    }
    \label{fig.RO-DTW-UTW}
    \end{figure}

The energy of velocity fluctuations around a given base flow can also be obtained from the Reynolds-Orr equation~\citep{schhen01}. This equation can be used to elucidate the energy amplification mechanisms and facilitate better understanding of the influence of UTWs and DTWs on transitional channel flows. In this section, we consider the Reynolds-Orr equation for the fundamental modes (i.e., modes with $\theta = 0$; cf.\ equation~(\ref{eq.NM})). Our results reveal that, relative to the uncontrolled flow, the DTWs reduce the production of kinetic energy, thereby enabling the smaller receptivity to disturbances. As opposed to the DTWs, the UTWs increase the production of kinetic energy. For the streamwise-periodic base flow, $\bu_b = (U(x,y),$ $V(x,y),$ $0)$, the time evolution of the kinetic energy of the fundamental modes,
    $
    \bmrE(\theta = 0, k_z; t)
    =
    \inprod{\bv_\theta (\cdot,k_z,t)}{\bv_\theta (\cdot,k_z,t)}|_{\theta \, = \, 0}
    $,
is governed by
    \be
    \ba{rcl}
    \dfrac{1}{2}\,\dfrac{\mrd \bmrE}{\mrd t}
    & \!\!\! = \!\!\! &
    -\inprod{u_{\theta}}{{\cal U}_y v_{\theta}}
    \, - \,
    \inprod{v_{\theta}}{{\cal V}_y v_{\theta}}
    \, - \,
    \inprod{u_{\theta}}{{\cal U}_x u_{\theta}}
    \, - \,
    \inprod{v_{\theta}}{{\cal V}_x u_{\theta}}
    \, + \,
    \\[0.15cm]
    & \!\! \!\! \!\! &
    (1/R_c)
    \left(
    \inprod{\bv_{\theta}}{{\cal D}_{xx} \bv_{\theta}}
    \, + \,
    \inprod{\bv_{\theta}}{\partial_{yy} \bv_{\theta}}
    \, - \,
    k_z^2 \, \inprod{\bv_{\theta}}{\bv_{\theta}}
    \right)
    \, + \,
    \inprod{\bv_{\theta}}{\bd_{\theta}},
    ~
    \theta
    =
    0,
    \label{eq.RO}
    \ea
    \ee
where, for example, $\bv_\theta = \col \, \{\bv (n \omega_x,y,k_z,t)\}_{n \, \in \, \bbZ}$ for the fundamental modes. In~(\ref{eq.RO}), ${\cal D}_{xx}$ is a diagonal operator with $\{(-n \omega_x)^2 I\}_{n \, \in \, \bbZ}$ on its main diagonal, $I$ is the identity operator, $\inprod{\cdot}{\cdot}$ denotes the $L_2 [-1,1]$ inner product and averaging in time (cf.\ equation~(\ref{eq.L2ip})), and ${\cal U}_y$, ${\cal V}_y$, ${\cal U}_x$, and ${\cal V}_x$ are block-Toeplitz operators whose $r$th sub-diagonals are determined by the $r$th harmonic in the Fourier series representation of $U_y (x,y)$, $V_y (x,y)$, $U_x (x,y)$, and $V_x (x,y)$ (see Appendix~\ref{app.ABC-theta} for details). The first four terms on the right-hand-side of~(\ref{eq.RO}) denote the work of Reynolds stresses on the base shear and they contribute to production of the kinetic energy. The next group of terms represents viscous dissipation and the last term accounts for the direct work of the forcing on the velocity fluctuations. It can be shown that the direct work of $\bd$ on $\bv$ is balanced by a fixed portion of the viscous dissipation, and that the difference between the production terms and the remaining dissipation terms determines the energy density (cf.\ figure~\ref{fig.RO-DTW-UTW}).

In the steady-state limit,~(\ref{eq.RO}) can be used to obtain the following expression for the energy density of the fundamental mode, $\bmrE (0, k_z) = \lim_{t \, \rightarrow \, \infty} \bmrE (0, k_z; t)$,
    \be
    \bmrE (0, k_z)
    \; = \;
    \bmrE_p (0, k_z)
    \, + \,
    \bmrE_d (0, k_z).
    \non
    \ee
Here, $\bmrE_p (0, k_z)$ denotes the contribution of production terms to the energy density and $\bmrE_d (0, k_z)$ represents the joint contribution of viscous dissipation and the work of disturbances
    \be
    \ba{rcl}
    \bmrE_p (0, k_z)
    \!\! & = & \!\!
    - (R_c/k_z^2)
    \left(
    \inprod{u_{\theta}}{{\cal U}_y v_{\theta}}
    \, + \,
    \inprod{v_{\theta}}{{\cal V}_y v_{\theta}}
    \, + \,
    \inprod{u_{\theta}}{{\cal U}_x u_{\theta}}
    \, + \,
    \inprod{v_{\theta}}{{\cal V}_x u_{\theta}}
    \right),
    \\[0.15cm]
    \bmrE_d (0, k_z)
    \!\! & = & \!\!
    (1/k_z^2)
    \left(
    \inprod{\bv_{\theta}}{{\cal D}_{xx} \bv_{\theta}}
    \, + \,
    \inprod{\bv_{\theta}}{\partial_{yy} \bv_{\theta}}
    \right)
    \, + \,
    (R_c/k_z^2) \,
    \inprod{\bv_{\theta}}{\bd_{\theta}},
    ~~
    \theta
    =
    0.
    \ea
    \label{eq.RO-ss}
    \ee
In flows subject to small amplitude traveling waves, a perturbation analysis can be employed to study the effect of each term on the right-hand-side of~(\ref{eq.RO-ss}) on the energy density
    \be
    \ba{rcl}
    \bmrE (0, k_z)
    \!\! & = & \!\!
    \bmrE_{0} (0, k_z)
    \, + \,
    \alpha^2 \, \bmrE_{2} (0, k_z)
    \, + \,
    {\cal O} (\alpha^4),
    \\[0.1cm]
    \bmrE_0 (0, k_z)
    \!\! & = & \!\!
    \bmrE_{0,p} (0, k_z)
    \, + \,
    \bmrE_{0,d} (0, k_z),
    \\[0.1cm]
    \bmrE_2 (0, k_z)
    \!\! & = & \!\!
    \bmrE_{2,p} (0, k_z)
    \, + \,
    \bmrE_{2,d} (0, k_z),
    \ea
    \label{eq.RO-ss-pert}
    \ee
where all the above terms can be readily determined from the solution to the Lyapunov equation~(\ref{eq.LE}).

Figure~\ref{fig.RO-DTW-UTW} illustrates $\bmrE_p (0,k_z)$ and $\bmrE_d (0,k_z)$ in the uncontrolled flow and in flows subject to a DTW with $(c = 5,$ $\omega_x = 2)$ and a UTW with $(c = -2,$ $\omega_x = 0.5)$. As expected, in the uncontrolled flow the joint contribution of dissipation and forcing is negative while the contribution of production is positive (see figures~\ref{fig.RO-c5-w2-kz-proddiff0} and~\ref{fig.RO-cm2-w0p5-kz-proddiff0}). The energy density (solid curve) is determined by the sum of $\bmrE_{0,p}$ and $\bmrE_{0,d}$, and it peaks at $k_z \approx 1.78$. The second order corrections (in $\alpha$) to $\bmrE_p$ and $\bmrE_d$ are shown in figures~\ref{fig.RO-c5-w2-kz-proddiff2} (for the DTW) and~\ref{fig.RO-cm2-w0p5-kz-proddiff2} (for the UTW). In flows subject to a DTW, the correction to $\bmrE_p$ is negative while the correction to $\bmrE_d$ is positive. Furthermore, the effect of $\bmrE_{2,p}$ dominates that of $\bmrE_{2,d}$ which implies that the DTW reduces the energy density of the uncontrolled flow (solid curve in figure~\ref{fig.RO-c5-w2-kz-proddiff2} shows that a DTW introduces a negative correction to $\bmrE_0$). On the other hand, flows subject to a UTW exhibit opposite trends; the correction to $\bmrE_p$ is positive, the correction to $\bmrE_d$ is negative, and since $\bmrE_{2,p}$ has the dominant effect, the UTW increases the energy density of the uncontrolled flow (solid curve in figure~\ref{fig.RO-cm2-w0p5-kz-proddiff2} shows that a UTW introduces a positive correction to $\bmrE_0$). In figures~\ref{fig.RO-c5-w2-alpha0p025-kz-proddiff} and~\ref{fig.RO-cm2-w0p5-alpha0p015-kz-proddiff} perturbation analysis up to a second order in $\alpha$ is used to show $\bmrE_{p}$ and $\bmrE_{d}$ (solid curves) for a DTW with $\alpha = 0.025$ and for a UTW with $\alpha = 0.015$. Relative to the uncontrolled flow (symbols), the DTW decreases both production and dissipation terms. On the contrary, the UTW increases both of these terms. For both UTWs and DTWs, production dominates dissipation and determines whether the energy is increased or decreased. In addition, our computations show that $\inprod{u_{\theta}}{{\cal U}_y v_{\theta}}$ is orders of magnitude larger than the other production terms. Moreover, $\inprod{u}{\partial_{yy} u}$ completely dominates other dissipation terms. Therefore, the work of the Reynolds stress $u v$ against the base shear $U_y$ is responsible for almost all of the energy production and the maximum viscous dissipation is associated with the wall-normal diffusion of the streamwise velocity fluctuation. These results are confirmed by DNS of the NS equations in Part~2.

\section{Concluding remarks}
    \label{sec.concl}

This paper disentangles three distinct effects of blowing and suction along the walls on pumping action, required control power, and kinetic energy reduction. We have shown that analysis of dynamics is paramount to designing the streamwise traveling waves. If velocity fluctuations are well-behaved then the pumping action and required control power can be ascertained from the steady-state analysis. The proposed method uses receptivity analysis of the linearized NS equations to study the fluctuations' energy in transitional channel flows. Motivated by our observation that a positive net efficiency can be achieved by preventing transition, we develop a framework for design of the traveling waves that reduce receptivity to three dimensional body force fluctuations. Direct numerical simulations of the NS equations, conducted in Part~2 of this study~\citep{liemoajov10}, verify that the traveling waves identified here are indeed an effective means for controlling the onset of turbulence. This demonstrates the predictive power of model-based approach to sensorless flow control; our simulation-free approach captures the essential trends in a computationally efficient manner and avoids the need for DNS and experiments in the early design stages.

Our perturbation analysis has revealed that properly designed DTWs can significantly reduce energy amplification of three dimensional fluctuations, including the streamwise streaks and the TS waves, which makes them well-suited for preventing transition. The DNS of Part~2 confirm that transient response of fluctuations' kinetic energy can be maintained at low levels using the values of wave frequency and speed that reduce receptivity of the linearized NS equations. This facilitates maintenance of laminar flow; positive net efficiency can be achieved if the wave amplitude necessary for controlling the onset of turbulence is not prohibitively large~\citep{liemoajov10}. On the other hand, we show that the UTWs are poor candidates for preventing transition for they, at best, exhibit similar receptivity to background disturbances as the uncontrolled flow. In particular, the UTWs considered by~\citet{minsunspekim06} largely amplify the most energetic modes of the uncontrolled flow, thereby promoting turbulence even when the uncontrolled flow stays laminar~\citep{liemoajov10}.

In spite of promoting turbulence, the UTWs may offer a viable strategy for reducing skin-friction drag in fully developed turbulent flows. The DNS of~\citet{minsunspekim06} and~\citet{liemoajov10} suggest that the UTWs alter the dynamics of velocity fluctuations favorably in the turbulent regime, e.g., by reducing skin-friction drag coefficient compared to the uncontrolled flow. However, since the UTWs induce turbulent flow that departs from base flow obtained in the absence of velocity fluctuations, the model used in this work cannot be employed to explain utility of UTWs. This would require development of control-oriented models that contain essential physics of turbulent flows and, at the same time, are convenient for control design. For example, turbulent viscosity models have been successfully used to determine the turbulent mean velocity~\citep{reytie67,reyhus72-3} and to identify the dominant turbulent flow structures~\citep{alajim06,cospujdep09,pajgarcosdep09}. Motivated by these successes, we intend to explore development of new control-oriented models that are capable of capturing the essential features of turbulent flow dynamics.

The contribution of this paper goes beyond the problem of designing transpiration-induced streamwise traveling waves. The techniques presented here may also find use in designing periodic geometries and waveforms for maintaining the laminar flow or skin-friction drag reduction. Our work (i) suggests that strategies capable of reducing high flow sensitivity represent viable approach to controlling the onset of turbulence; and (ii) offers a computationally attractive
(and simulation-free) method to determine the energy amplification of the linearized flow equations in the presence of periodic controls.




\section*{Acknowledgments}
    \label{sec.Ack}

Financial support from the National Science Foundation under CAREER Award CMMI-06-44793 and 3M Science and Technology Fellowship (to R.\ M.) is gratefully acknowledged. The University of Minnesota Supercomputing Institute is acknowledged for providing computing resources. This work was initiated during the 2006 Center for Turbulence Research Summer Program with financial support from Stanford University and NASA Ames Research Center. M.\ R.\ J.\ would like to thank Prof.\ P.\ Moin for creating an inspiring intellectual atmosphere and Dr.\ D.\ You for his hospitality during the stay.

\appendix

\section{Base velocity}
    \label{app.detail-nom-velocity}

In order to determine the corrections to base parabolic profile in flows subject to small amplitude traveling waves, we use a weakly nonlinear analysis to solve~(\ref{eq.NScts-trans}) subject to~(\ref{eq.BCorig}). We only present the equations for corrections up to a second order in $\alpha$; similar equations can be obtained for higher order corrections. Stream functions, $\Psi_{1,\pm 1}(y)$, can be used to determine the first harmonic in Fourier series representation of the base velocity (cf.~(\ref{eq.base-vel-pert}))
    \be
    U_{1,\pm1} (y) \, = \, \Psi_{1,\pm 1}' (y),
    ~~
    V_{1,\pm1} (y) \, = \, \mp \mri \omega_x \Psi_{1,\pm 1} (y),
    \non
    \ee
where $\Psi_{1,\pm 1}(y)$ are solutions to
    \be
    \ba{c}
    (1/R_c)
    \Delta_{\omega_x}^2 \, \Psi_{1,\pm 1}
    \, \pm \,
    \mri \omega_x
    \left(
    (c \, - \, U_{0}) \,
    \Delta_{\omega_x} \, \Psi_{1,\pm 1}
    \, + \,
    U_{0}'' \, \Psi_{1,\pm 1}
    \right)
    \, = \,
    0,
    \\[0.1cm]
    \Psi_{1,-1}(\pm 1)
    \, = \,
    \pm \mri / \omega_x,
    ~
    \Psi_{1,1}(\pm 1)
    \, = \,
    \mp \mri / \omega_x,
    ~
    \Psi_{1,\pm 1}'(\pm 1)
    \, = \,
    0.
    \ea
    \non
    \ee
Here, $\Delta_{\omega_x} = \partial_{yy} - \omega_x^2$ with Dirichlet boundary conditions and $\Delta_{\omega_x}^2 = \partial_{yyyy} - 2 \omega_x^2 \partial_{yy} + \omega_x^4$ with Cauchy boundary conditions. Moreover, $U_{2,0}$ is obtained by equating terms of order $\alpha^2$ in the streamwise averaged $x$-momentum equation
    \be
    (1/R_c) U_{2,0}''
    \, = \,
    V_{1, 1} \, U_{1, -1}'
    \, - \,
    U_{1, 1} \, V_{1, -1}'
    \, + \,
    V_{1, -1} \, U_{1, 1}'
    \, - \,
    U_{1, -1} \, V_{1, 1}',
    ~~
    U_{2,0}(\pm 1)
    \, = \,
    0.
    \non
    \ee

\section{Frequency representation of the evolution model}
    \label{app.ABC-theta}

We first describe how base velocity modified by the traveling waves enters in evolution model~(\ref{eq.LNSE}). Frequency representation of the evolution model is discussed next. It turns out that the components of base velocity determine coefficients of operator $F$ in~(\ref{eq.LNSE}). For base velocity,
    $
    {\bf u}_b
    =
    ({U(x,y)}, \, {V(x,y)}, \, {0}),
    $
$F$ is a $2 \times 2$ block-operator with components
    \be
    \ba{rcl}
    F^{11}
    & \!\! = \!\! &
    ({1}/{R_c}) \Delta^2
    \, + \,
    ((\Delta U) - (U - c I) \Delta)\partial_x
    \, - \,
    (\Delta V)\partial_y
    \, - \,
    V \Delta \partial_y
    \, - \,
    \\[0.1cm]
    & &
    2 V_x \partial_{xy}
    \, + \,
    U_x (\Delta - 2 \partial_{xx})
    \, - \,
    (\Delta V_y)
    \, + \,
    \left( 2 (\Delta V) \partial_x
    \, + \,
    \Delta V_x
    \, + \,
    \right.
    \\[0.1cm]
    & &
    \left.
    V_x (\Delta - 2 \partial_{yy})
    \, - \,
    2 U_x \partial_{xy} \right)
    (\partial_{xx} + \partial_{zz})^{-1} \partial_{xy},
    \\[0.1cm]
    F^{12}
    & \!\! = \!\! &
    - \left( 2 (\Delta V)
    \partial_x
    \, + \,
    \Delta V_x
    \, + \,
    V_x (\Delta - 2 \partial_{yy})
    \, - \,
    2 U_x \partial_{xy}
    \right)
    (\partial_{xx} + \partial_{zz})^{-1}
    \partial_{z},
    \\[0.1cm]
    F^{21}
    & \!\! = \!\! &
    -\left( U_y \partial_z + V_x (\partial_{xx} + \partial_{zz})^{-1}
    \partial_{yyz} \right),
    \\[0.1cm]
    F^{22}
    & \!\! = \!\! &
    ({1}/{R_c}) \Delta
    \, - \,
    (U_x + (U - c I) \partial_x
    \, + \,
    V \partial_y)
    \, - \,
    V_x (\partial_{xx} + \partial_{zz})^{-1} \partial_{xy},
    \ea
    \non
    \ee
where $(\partial_{xx} + \partial_{zz})^{-1}$ is defined by
    \be
    (\partial_{xx} \, + \, \partial_{zz})^{-1}:
    \, f \, \mapsto \, g
    ~ \Leftrightarrow ~
    \left\{
    \ba{rll}
    f
    & \!\! = \!\! &
    (\partial_{xx} \, + \, \partial_{zz}) g
    \\
    & \!\! = \!\! &
    g_{xx} \, + \, g_{zz}.
    \ea
    \right.
    \non
    \ee

Frequency representation~(\ref{eq.FR}) of the linearized evolution model~(\ref{eq.LNSE}) can be determined using the following simple rules~\citep{farjovbam08}:
    \bi
\item[(a)] A {\em spatially invariant\/} operator $L$ with Fourier symbol $L(k_x)$ has a block-diagonal representation
    $
    {\cal L}_\theta
    =
    \diag \,
    \{ L(\theta_n) \}_{n \, \in \, \bbZ}.
    $
For example, if $L \, = \, \partial_x$, then
    $
    {\cal L}_\theta
    =
    \diag \,
    \{
    \mri (\theta + n \omega_x) I
    \}_{n \, \in \, \bbZ}.
    $
Operators $E$, $G$, $C$, $F_0$, and $F_{l,r}$ in~(\ref{eq.LNSE}) are spatially invariant and, thus, their representations are block-diagonal.

\item[(b)] A {\em spatially periodic function\/} $T(x)$ with Fourier series coefficients $\{ T_n \}_{n \, \in \, \bbZ}$ has a $\theta$-independent block-Toeplitz representation
    \be
    {\cal T}
    \; = \;
    \toep \,
    \left\{
    \cdots, T_2, T_1, \fbox{$T_0$} \,, T_{-1}, T_{-2}, \cdots
    \right\}
    \; = \;
    \left[
      \begin{array}{ccccc}
        \ddots &       &        &        &  \\
         & T_{0} & T_{-1} & \!\!T_{-2} &  \\
         & T_{1} & T_{0}  & \!\!T_{-1} &  \\
         & T_{2} & T_{1}  & \!\!T_{0}  &  \\
         &       &        &        & \ddots \\
      \end{array}
    \right],
    \non
    \ee
where the box denotes the element on the main diagonal of ${\cal T}$. For example, $T(x) = \mre^{- \mri r x}$ has a block-Toeplitz representation ${\cal T} = {\cal S}_r$ with the only non-zero element $T_{-r} = I$.

\item[(c)] A representation of the sums and cascades of spatially periodic functions and spatially invariant operators is readily determined from these special cases. For example, a matrix representation of operator $\mre^{- \mri r x} \partial_x$ is given by ${\cal S}_r \, \diag \, \{ \mri (\theta + n \omega_x) I \}_{n \, \in \, \bbZ}$.
    \ei

Based on these, we get the following representations for ${\cal A}_\theta$, ${\cal B}_\theta$, and ${\cal C}_\theta$ in~(\ref{eq.FR})
    \beq
    \ba{c}
    {\cal A}_\theta
    \; = \;
    {\cal E}_\theta^{-1} {\cal F}_\theta
    \; = \;
    {\cal E}_\theta^{-1}
    {\cal F}_{0\theta}
    \; + \;
    \ds{ \sum_{l \, = \, 1}^{\infty} }
    \alpha^l
    \sum_{r \, \overset{2}{=} \, -l}^{l}
    {\cal E}_\theta^{-1}
    {\cal S}_{-r}
    {\cal F}_{l,r\theta}
    \; = \;
    {\cal A}_{0\theta}
    \; + \;
    \ds{ \sum_{l \, = \, 1}^{\infty} }
    \alpha^l \, {\cal A}_{l\theta},
    \\[0.1cm]
    {\cal B}_\theta
    \; = \;
    {\cal E}_\theta^{-1} \, {\cal G}_\theta,
    ~~
    {\cal G}_\theta
    \; = \;
    \diag \, \{ G(\theta_n) \}_{n \, \in \, \bbZ},
    ~~
    {\cal C}_\theta
    \; = \;
    \diag \, \{ C(\theta_n) \}_{n \, \in \, \bbZ},
    \ea
    \non
    \eeq
where we have used the fact that ${\cal E}_\theta = \diag \, \{ E(\theta_n) \}_{n \, \in \, \bbZ}$ is an invertible operator. For convenience of later algebraic manipulations, we rewrite ${\cal A}_{l\theta}$ as
    $
    {\cal A}_{l\theta}
    =
    { \sum_{r \, \overset{2}{=} \, {-l}}^{l}
    {{\cal S}_{-r} \, {\cal A}_{l,r\theta}} }
    $
where
    $
    {\cal A}_{l,r\theta}
    =
    \diag \, \{ A_{l,r}(\theta_n) \}_{n \, \in \, \bbZ}
    =
    \diag \, \{ E^{-1} (\theta_{n + r}) \, F_{l,r}(\theta_{n}) \}_{n \, \in \,
    \bbZ}.
    $
In other words, for a given $l \geq 1$ operator $\ca_{l \theta}$ has non-zero blocks only on $r$th sub-diagonals with $r \in \{-l, -l+2, \ldots, l-2, l\}$. The frequency symbols of the operators $E(\theta_{n}), G(\theta_{n}), C(\theta_{n})$, and $F_{l,r}(\theta_{n})$ are given by
    \be
    \ba{l}
    F^{11}_0 (\theta_n,k_z)
    \, = \,
    ({1}/{R_c}) \Delta^2
    \, + \,
    \mri \theta_n (U_0'' \, - \, (U_0 \, - \, c) \Delta),
    ~
    F^{12}_0 (\theta_n,k_z)
    \, = \,
    0,
    \\[0.15cm]
    F^{21}_0 (\theta_n,k_z)
    \, = \,
    - \mri k_z U_0',
    ~
    F^{22}_0 (\theta_n,k_z)
    \, = \,
    ({1}/{R_c}) \Delta
    \, - \,
    \mri \theta_n (U_0 \, - \, c),
    \non
    \ea
    \ee
and
    \be
    \ba{rcl}
    F^{11}_{l,r} (\theta_n,k_z)
    & \!\! = \!\! &
    \mri \theta_n
    \left(
    (\Delta_{r\omega_x} \, U_{l,r})
    \, - \,
    U_{l,r} \, \Delta
    \, - \,
    2 \mri (r \omega_x)
    V_{l,r} \py
    \right)
    \, - \,
    \\[0.15cm]
    & \!\!  \!\! &
    \left(
    (\Delta_{r\omega_x} \, V_{l,r})
    \, + \,
    V \, \Delta
    \right)
    \py
    \, - \,
    \mri r \omega_x
    U_{l,r} (\Delta \, + \, 2 \theta_n^2)
    \, - \,
    \mri r \omega_x
    (\Delta_{r\omega_x} \, U_{l,r})
    \, - \,
    \\[0.15cm]
    & \!\!  \!\! &
    (\theta_n / k^2)
    \left(
    2 \theta_n
    \left(
    (-\Delta_{r\omega_x} \, V_{l,r}) \py
    \, + \,
    \mri r \omega_x U_{l,r} \pyy
    \right)
    \, - \,
    \right.
    \\[0.15cm]
    & \!\!  \!\! &
    \left.
    r \omega_x
    \left(
    (\Delta_{r\omega_x} \, V_{l,r})
    \, + \,
    V_{l,r} ( \Delta \, - \, 2 \pyy )
    \right)
    \py
    \right),
    \\[0.15cm]
    F^{12}_{l,r} (\theta_n,k_z)
    & \!\! = \!\! &
    (k_z/k^2)
    \left(
    2 \theta_n
    \left(
    (-\Delta_{r \omega_x} \, V_{l,r} )
    \, + \,
    \mri r \omega_x
    U_{l,r} \py
    \right)
    \, - \,
    \right.
    \\[0.15cm]
    & &
    \left.
    r \omega_x
    \left(
    ( \Delta_{r \omega_x} \,V_{l,r} )
    \, + \,
    V_{l,r} (\Delta - 2 \partial_{yy})
    \right)
    \right),
    \\[0.15cm]
    F^{21}_{l,r} (\theta_n,k_z)
    & \! = \! &
    - \, \mri k_z \left( U'_{l,r} - (\mri r \omega_x/k^2) \, V_{l,r} \partial_{yy} \right),
    \\[0.15cm]
    F^{22}_{l,r} (\theta_n,k_z)
    & \! = \! &
    -V_{l,r} \py
    \, - \,
    \mri r \omega_x
    U_{l,r}
    \, + \,
    \mri \theta_n
    \left(
    (\mri r \omega_x/k^2) V_{l,r} \py
    \, - \,
    U_{l,r}
    \right),
    \ea
    \non
    \ee
where $k^2 = \theta_n^2 + k_z^2$, $\Delta = \partial_{yy} - k^2$ and $\Delta_{r \omega_x} = \partial_{yy} - (r \omega_x)^2$ with Dirichlet boundary conditions, and $\Delta^2 = \partial_{yyyy} - 2 k^2 \partial_{yy} + k^4$ with Cauchy boundary conditions. Operators $E$, $G$, and $C$ are given by
    \be
    \ba{c}
    E(\theta_n,k_z)
    \; = \;
    \tbt{\Delta}{0}{0}{I}, ~
    G(\theta_n,k_z)
    \; = \;
    \tbth
    {-\mri \theta_n \partial_y}{-k^2 I}{-\mri k_z \partial_y}
    {\mri k_z I}{0}{-\mri \partial_y },
    \\[0.35cm]
    C(\theta_n,k_z)
    \; = \;
    \thbt
    {\mri (\theta_n/k^2) \partial_y}{-\mri k_z / k^2}
    {I}{0}
    {\mri (k_z/k^2) \partial_y }{\mri \theta_n / k^2}.
    \non
    \ea
    \ee

\section{Perturbation analysis of energy density}
    \label{app.htwo-pert-detail}

As discussed in~\S~\ref{sec.H2}, steady-state energy density, $\bmrE (\theta,k_z)$, of the linearized system~(\ref{eq.FR}), can be determined using the solution to the operator Lyapunov equation~(\ref{eq.LE}). For sufficiently small values of $\alpha$, the solution of~(\ref{eq.LE}) can be expressed as a perturbation series
    $
    \cx_\theta
    \, = \,
    \sum_{m \, = \, 0}^{\infty}
    {\alpha^m \, \cx_{m \theta}}
    $.
After substituting into~(\ref{eq.LE}) and factoring out the terms with equal power in $\alpha$, we have
    \be
    \ba{rl}
    \alpha^0:
    &
    \ca_{0 \theta} \, \cx_{0 \theta}
    \, + \,
    \cx_{0 \theta} \, \ca_{0 \theta}^*
    \; = \;
    - \cb_\theta \, \cb_\theta^*,
    \\[0.1cm]
    \alpha^m:
    &
    \ca_{0 \theta} \, \cx_{m \theta}
    \, + \,
    \cx_{m \theta} \, \ca_{0 \theta}^*
    \; = \;
    -
    \ds{ \sum_{l \, = \, 1}^{m}}
    \left(
    \ca_{l \theta} \, \cx_{m-l \theta}
    \, + \,
    \cx_{m-l \theta} \, \ca_{l \theta}^*
    \right),
    ~~
    m \geq 1
    \ea
    \label{eq.lyap-n}
    \ee
Since operator $\ca_{0 \theta}$ is block-diagonal, $\cx_{m \theta}$ inherits the same structure as the right-hand-side of~(\ref{eq.lyap-n}). One can show that $\cx_{m \theta}$ has non-zero blocks only on the first $s \leq m$ odd (for odd $m$) or even (for even $m$) upper and lower sub-diagonals. Up to a second order in $\alpha$, we have
    \be
    \ba{rcl}
    \cx_{0 \theta}
    \!\! & = & \!\!
    \cx_{0,0 \theta},
    \\[0.1cm]
    \cx_{1 \theta}
    \!\! & = & \!\!
    \csq_{1} \, \cx_{1,1 \theta}
    \; + \;
    \cx_{1,1 \theta}^* \, \csq_{-1},
    \\[0.1cm]
    \cx_{2 \theta}
    \!\! & = & \!\!
    \csq_{2} \, \cx_{2,2 \theta}
    \; + \;
    \cx_{2,0 \theta},
    \; + \;
    \cx_{2,2 \theta}^* \, \csq_{-2},
    \ea
    \label{eq.P012}
    \ee
where ${\cx}_{m,s \theta} \, = \, {\rm diag} \, \{ X_{m,s}(\theta_n) \} _{n \, \in \, \bbZ}$ and $\csq_{r}$ is defined in Appendix~\ref{app.ABC-theta}. Substituting into~(\ref{eq.lyap-n}) yields
    \be
    \ba{rcl}
    \ca_{0 \theta} \, {\cx}_{0,0 \theta}
    \, + \,
    \cx_{0,0 \theta} \, \ca_{0 \theta}^*
    \!\! & = & \!\!
    - \cb_\theta \, \cb_\theta^*,
    \\[0.1cm]
    \ca_{0 \theta} \, \csq_{1} {\cx}_{1,1 \theta}
    \, + \,
    \csq_{1} \, \cx_{1,1 \theta} \, \ca_{0 \theta}^*
    \!\! & = & \!\!
    - \left(
    \csq_{1} \, \ca_{1,-1 \theta} \, \cx_{0,0 \theta}
    \, + \,
    \cx_{0,0 \theta} \, \ca_{1,1 \theta}^* \, \csq_{1}
    \right),
    \\[0.1cm]
    \ca_{0 \theta} \, {\cx}_{2,0 \theta}
    \, + \,
    \cx_{2,0 \theta} \, \ca_{0 \theta}^*
    \!\! & = & \!\!
    - \left(
    \ca_{2,0 \theta} \, \cx_{0,0 \theta}
    \, + \,
    \cx_{0,0 \theta} \, \ca_{2,0 \theta}^*
    \, + \,
    \csq_{-1} \, \ca_{1,1 \theta} \, \csq_{1} \, \cx_{1,1 \theta}
    \, + \,
    \right.
    \\[0.2cm]
    & &
    \hskip-2cm
    \left.
    \csq_{1} \, \ca_{1,-1 \theta} \, \cx_{1,1 \theta}^* \, \csq_{-1}
    \, + \,
    \csq_{1} \, \cx_{1,1 \theta} \, \ca_{1,-1 \theta}^* \, \csq_{-1}
    \, + \,
    \cx_{1,1 \theta}^* \, \csq_{-1} \, \ca_{1,1 \theta}^* \, \csq_{1}
    \right).
    \ea
    \label{eq.lyap012}
    \ee
Finally, each block on the main diagonal of $\cx_{m,s \theta}$ in~(\ref{eq.lyap012}) is obtained from
    \be
    \ba{rcl}
    A_{0}(\theta_{n}) X_{0,0}(\theta_{n})
    +
    X_{0,0}(\theta_{n}) A_{0}^*(\theta_{n})
    \!\! & = & \!\!
    - B(\theta_{n}) B^*(\theta_{n}),
    \\[0.1cm]
    A_{0}(\theta_{n-1}) X_{1,1}(\theta_{n})
    +
    X_{1,1}(\theta_{n}) A_{0}^*(\theta_{n})
    \!\! & = & \!\!
    - \left(
    A_{1,-1}(\theta_{n}) X_{0,0}(\theta_{n})
    +
    X_{0,0}(\theta_{n-1}) A_{1,1}^*(\theta_{n-1})
    \right),
    \\[0.1cm]
    A_{0}(\theta_{n}) X_{2,0}(\theta_{n})
    +
    X_{2,0}(\theta_{n}) A_{0}^*(\theta_{n})
    \!\! & = & \!\!
    - \left(
    A_{2,0}(\theta_{n}) X_{0,0}(\theta_{n})
    +
    X_{0,0}(\theta_{n}) A_{2,0}^*(\theta_{n})
    \, +
    \right.
    \\[0.1cm]
    & &
    \left.
    A_{1,1}(\theta_{n-1}) X_{1,1}(\theta_{n})
    +
    A_{1,-1}(\theta_{n+1}) X_{1,1}^*(\theta_{n+1})
    \, +
    \right.
    \\[0.1cm]
    & &
    \left.
    X_{1,1}(\theta_{n+1}) A_{1,-1}^*(\theta_{n+1})
    +
    X_{1,1}^*(\theta_{n}) A_{1,1}^*(\theta_{n-1})
    \right).
    \ea
    \non
    \ee

\end{document}